\pdfoutput=1
\documentclass[acmsmall,screen]{acmart}
\newif\ifARXIV
\ARXIVfalse
\newif\ifShowRevisions
\ShowRevisionsfalse
\ifShowRevisions
\usepackage{soul}
\newcommand{\add}[1]{\textcolor{blue}{#1}}
\newcommand{\replace}[2]{\del{#1}\add{#2}}
\else
\newcommand{\add}[1]{{#1}}
\newcommand{\replace}[2]{\add{#2}}
\fi

\newif\ifPLDI
\PLDItrue

\AtBeginDocument{%
  }

\setcopyright{cc}
\setcctype{by-nd}
\acmDOI{10.1145/3763134}
\acmYear{2025}
\acmJournal{PACMPL}
\acmVolume{9}
\acmNumber{OOPSLA2}
\acmArticle{356}
\acmMonth{10}
\received{2025-03-25}
\received[accepted]{2025-08-12}

\usepackage[skins,theorems]{tcolorbox}
\tcbset{highlight math style={enhanced jigsaw,
  colframe=red,colback=white,arc=0pt,boxrule=0.25pt,
  borderline={0.5mm}{0mm}{blue!70!white,dashed}}}
\usepackage{subfigure}
\usepackage{hyperref}
\usepackage{array}
\usepackage{proof}
\usepackage{stmaryrd}
\usepackage{xspace}
\usepackage{graphicx}

\usepackage{lipsum}
\usepackage[utf8]{inputenc}
\usepackage{blindtext}
\usepackage{listings}
\usepackage{wrapfig}

\usepackage{amsmath}

\usepackage{cleveref}
\crefformat{footnote}{#2\footnotemark[#1]#3}
\usepackage[T1]{fontenc}
\usepackage{fontawesome}
\usepackage[utf8]{inputenc}
\usepackage{iris}
\usepackage{mathpartir}
\renewcommand{\TirNameStyle}[1]{\hypertarget{#1}{\textsc{#1}}}

\usepackage{xspace}
\def\etal.{\emph{et al.}}

\newcommand{\qfrac}{\kw{q}}

\newcommand{\vaddr}{\kw{va}}

\newcommand{\mydagger}{\mathord{\dagger}}
\renewcommand{\ddag}{\mydagger\kern-0.8mm\mydagger}

\newcommand{\paddr}{\textsf{pa}}

\newcommand{\vpage}{\textsf{v}}
\newcommand{\kw}[1]{\mathsf{#1}}

\newcommand{\Loctw}{\mathcal{W}_{12}}

\newcommand{\Locsf}{\mathcal{W}_{64}}

\newcommand{\Locft}{\mathcal{W}_{52}}

\newcommand{\rv}{\kw{rv}}
\newcommand{\rg}{\kw{r}}

\newcommand{\Loc}{\mathcal{W}_n}
\newcommand{\loc}{\kw{w}_n}
\newcommand{\regset}{\kw{greg}}

\newcommand{\reg}{\kw{r}}

\newcommand{\instr}{i}

\newcommand{\instrs}{\ensuremath{\vec\instr}}
\newcommand{\iskip}{\ensuremath{\kw{skip}}}
\newcommand{\iseq}[2]{\ensuremath{#1; #2}}

\newcommand{\varray}[1]{\begin{array}{@{}c@{}}#1\end{array}}
\newcommand{\starvarray}[1]{\bgroup\renewcommand{\star}{\\}\varray{#1}\egroup}

\renewcommand{\wp}[2]{\mathit{wp}\;#1\;#2}

\newcommand{\bigast}[2]{\mathop{\textstyle\Sep}\limits_{#1}\,#2}

\newcommand{\singletonMap}[2]{[#1 := #2]}
\newcommand{\coq}{\textsc{Rocq}\xspace}
\newcommand{\rocq}{\textsc{Rocq}\xspace}

\newcommand{\iris}{\textsc{Iris}\xspace}

\begin{document}

\title{Modal Abstractions for Virtualizing Memory Addresses}
\author{Ismail Kuru}
\email{ik335@drexel.edu}
\author{Colin S. Gordon}
\email{csgordon@drexel.edu}
\affiliation{%
  \institution{Drexel University}
  \city{Philadelphia, PA}
  \country{USA}
}

\renewcommand{\shortauthors}{Kuru and Gordon}

\begin{abstract}
\ifPLDI
\else
Operating system kernels employ virtual memory subsystems, which use a CPU's memory management units (MMUs) to virtualize the addresses of memory regions:
a logical (virtual) address is translated to a physical address in memory by the MMU based on
kernel-controlled page tables -- a hardware-defined sparse tree-map structure -- stored in memory, which itself is 
accessed by the kernel through virtual addresses.
Operating systems manipulate these virtualized memory mappings to isolate untrusted processes,
 restrict which memory is accessible to different processes, 
hide memory limits from user programs, 
ensure process isolation, implement demand-paging and copy-on-write behaviors for performance
and resource controls.
At the same time, misuse of MMU hardware can lead to kernel crashes.
\fi

Virtual memory management (VMM) code is a critical piece of general-purpose OS kernels, but verification of this functionality
is challenging due to the complexity of the hardware interface (the page tables are updated via writes to those
memory locations, using addresses which are themselves virtualized).
Prior work on verification of VMM code has either only handled a single address space, or trusted significant
pieces of assembly code.

In this paper, we introduce a modal abstraction to describe
the truth of assertions relative to a specific virtual address space: [r]P indicating that P holds in the
virtual address space rooted at r. Such modal assertions 
allow different address spaces to refer to each other, enabling complete verification of instruction sequences
manipulating multiple address spaces. Using them effectively requires working with other assertions,
% such as points-to assertions in our separation logic,  relative to a given address space.
such as points-to assertions about memory contents --- which implicitly depend on the address space
they are used in. 
We therefore define virtual points-to assertions to definitionally mimic hardware address translation,
relative to a page table root.
We demonstrate our approach with challenging fragments of VMM code showing that our approach
handles examples beyond what prior work can address, including reasoning about
a sequence of instructions as it changes address spaces.
\ifPLDI
Our results are formalized for a RISC-like fragment of x86-64 assembly in Rocq.
\looseness=-1
\else
All definitions and theorems mentioned in this paper including the operational model of a RISC-like fragment of x86-64, 
a simple language run on this operational model, and a logic as an instantiation of the Iris framework are mechanized 
inside Rocq.
\fi
\end{abstract}


\begin{CCSXML}
<ccs2012>
   <concept>
       <concept_id>10011007.10010940.10010992.10010998.10010999</concept_id>
       <concept_desc>Software and its engineering~Software verification</concept_desc>
       <concept_significance>500</concept_significance>
       </concept>
   <concept>
       <concept_id>10011007.10010940.10010941.10010949.10010950.10010951</concept_id>
       <concept_desc>Software and its engineering~Virtual memory</concept_desc>
       <concept_significance>500</concept_significance>
       </concept>
 </ccs2012>
\end{CCSXML}

\ccsdesc[500]{Software and its engineering~Software verification}
% \ccsdesc[500]{Theory of computation~Modal and temporal logics}
\ccsdesc[500]{Software and its engineering~Virtual memory}



\keywords{program verification, virtual memory, modal logic}

\maketitle

{
\theoremstyle{acmdefinition}
\newtheorem{assumption}[theorem]{Assumption}
}

\section{Introduction}
\label{sec:intro}
Virtual memory management lies at the core of modern OS kernel implementation. It is deeply intertwined with most other parts of a typical general-purpose OS kernel design, including scheduling, hardware drivers, and even the filesystem buffer cache. In writing the authoritative reference on the internals of the Solaris kernel, McDougall and Mauro went so far as to claim that ``\emph{the virtual memory sub-system can be considered the core of a Solaris instance, and the implementation of Solaris virtual memory affects just about every other subsystem in the operating system}''~\cite{mcdougall2006solaris}.
This makes support for verification the virtual memory management subsystem of an OS kernel critical to the correctness of every other piece of an OS or any software running atop it.

At its core, the virtual memory functionality of modern CPUs is about \emph{location virtualization}: the memory locations
(addresses) seen by most code are not, in fact, the exact location in physical memory where data reside. Instead these 
are \emph{virtual} addresses, which are mapped to actual physical resources by the cooperation of the hardware and OS. 
This is what enables separation of process memory resources:
the OS manipulates hardware functionality to ensure 
program memory accesses only succeed if the kernel has granted the program access to the accessed address.
But this is complicated by the fact that 
%the OS and hardware can also enable, shared (overlapping) access to physical memory regions; the fact that the kernel data structures themselves are accessed via virtual memory addresses; 
%and the fact that 
control over these mappings of virtual to physical addresses is itself mediated by \emph{in-memory data structures}, 
which the kernel also accesses via virtual address, leading to indirect cycles.
\looseness=-1

Further complicating matters, addresses themselves bear no information about which address space they originate 
from. 
%For user processes this is of little concern, as these have access to only their own address space. 
This is not a problem for usermode code, which only accesses its own address space.
But the kernel has
(or can grant itself) access to all address spaces. Mixing up addresses from different address spaces leads to severe bugs.
Proofs of kernel correctness must also track which \emph{assertions} hold in different address spaces:
some assertions should hold across all address spaces, while others hold in only one, and others may hold in 
multiple but still not all.

This kind of context-dependent assertion, where a fact may be true in one address space but not others, has a modal flavor. 
We propose tackling the verification of virtual memory subsystems (and kernels more broadly) by adapting ideas from hybrid
modal logic, which can label assertions true under \emph{other, named} circumstances (i.e., in another address space) with a 
modality indexed by a name for that space (in our case, the root of the page tables for an address space). This offers a 
\textit{convenient} and \textit{powerful} way to \emph{modularly}
isolate assertions specific to a particular address space,
explicitly state when an assertion is true across address spaces,
manipulate address spaces from within other address spaces, and
reason about change in address spaces.
This approach to reasoning about virtual memory is more flexible than prior program logic techniques~\cite{kolanski08vstte,kolanski09tphols}, 
which were only able to work with a single address space (the current address space on the CPU) because they were unable
to speak directly \emph{within the logic} about other address spaces, in addition to handling
the \emph{non-local} effects of page table updates whether within the current address space or across address spaces.
\looseness=-1

\paragraph{Contributions}
We develop these ideas in the form of a logic for working with virtual-address-space-relative assertions,
implemented as an embedded separation logic within {\textsc{Iris}}~\cite{jung2018iris}.
The result is a separation logic that lifts a number of major semantic restrictions present in the few
prior logics tackling virtual address translation.
The logic we develop covers core reasoning principles for reasoning about memory configurations and code
reliant upon or manipulating those memory configurations in the presence of in-memory page tables, the primary
memory protection mechanism across Intel/AMD's x86-64 processors; ARM's application class processors including
AArch64 CPUs; as well as POWER, RISC-V, and other architectures.
We prove the soundness of our \textsf{vProp} logic with respect to a RISC-like fragment of \textsf{AMD64} instructions.
We verify simplified versions of several critical virtual-memory-related pieces of OS functionality, 
including:
switching address spaces, converting physical addresses to virtual addresses for
software page table walks, and \add{a} complete \add{software} page table walk  --- the backbone of other essential functionalities of
virtual-memory which we discuss for mapping and unmapping pages. 
These examples demonstrate the suitability and flexibility of
a modal treatment of address spaces: each has a concise specification, each example either goes beyond the technical
capabilities of prior logics, or revisits an example from prior work
  in much greater detail (e.g., verifying critical and challenging supporting code that was
  axiomatized or treated as out-of-scope in prior work).
Our proofs are available~\cite{artifact}.
\looseness=-1

% One aspect of virtual memory management we do not address in this paper is updates to the translation lookaside
% buffer (TLB) present in most CPUs. Like CertiKOS~\cite{?}, seL4~\cite{sel4tocs}, and most other OS verification
% efforts we currently trust insertion of TLB flush operations, whose locations are critical but also few and
% well-known (after unmapping memory, or changing address spaces). Tackling correctness of TLB invalidation
% rigorously is an additional challenging project unto itself~\cite{...} with similarities to but
% significant divergences from reasoning about weak memory models, which future work should combine
% with the present work.

%\input{motivation.tex}
\section{Background}
\label{sec:background}
This section briefly recalls both fundamentals of page-table-based address translation in
general-puspose OS kernels, 
high-level background on separation logics and the \iris~\cite{jung2018iris} framework
we build on,
and some material on modal logic that informs our development.

\subsection{Machine Model}
\label{sec:backgroundonmachinemodel}

In typical system configurations, all memory addresses seen by programs running on modern computers are \emph{virtualized}: the address observed by a running
program generally will not correspond directly to the physical location in memory, and may not even correspond to a physical location that \emph{exists} 
in the machine. Instead, these \emph{virtual} addresses are translated to \emph{physical} addresses that correspond directly to locations in RAM. On most 
modern architectures, this translation is performed through cooperation of the hardware and OS kernel: while executing an instruction that dereferences a 
(virtual) address, the CPU's \emph{memory management unit} (\textsc{MMU}) hardware performs \emph{address translation},
resulting in a physical address used to access the cache\footnote{Technically, for performance reasons most caches are indexed with parts of the virtual 
address, but tagged with the physical data addresses, so cache lookups and address translations can proceed in parallel.} and/or memory-bus.
\setlength{\columnseprule}{4pt}
\begin{figure}\scriptsize
%    \begin{framed}
%\column{.59\textwidth}    
\begin{subfigure}\scriptsize

\includegraphics[width=\textwidth]{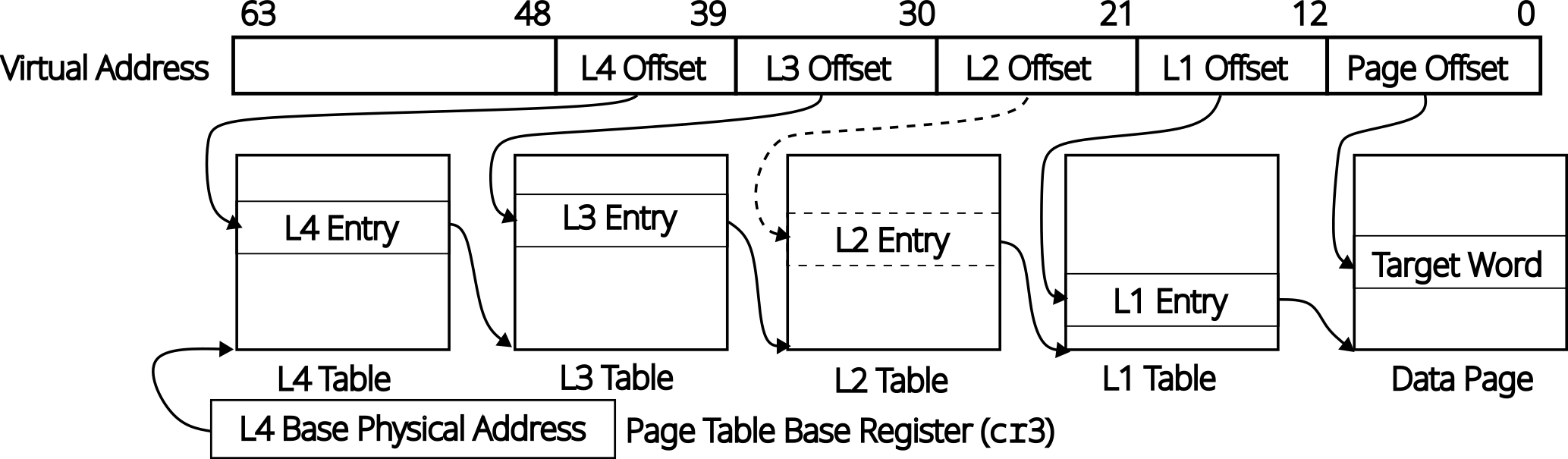}
\vspace{-2em}
          \caption{x86-64 hardware address translation.}
        \label{fig:pagetables}
    \end{subfigure}
    \vfill
\hspace{5em}
\begin{minipage}{.5\textwidth}
%    \begin{figure}
\begin{lstlisting}[mathescape,escapeinside={(*}{*)},basicstyle=\scriptsize]
pte_t *pte_get_next_table(pte_t *entry) {
  pte_t *next;
  if (!entry->present) { // Check if entry needs initialization
    pte_initialize(entry); // Alloc empty table for next level(*\label{line:ptc_alloc}*)
    entry->present = 1; // Mark valid(*\label{line:ptc_mark}*)
  }
  /* Convert phys. addr. from PTE to virt. addr. for access */
  uintptr_t next_phys_addr = PTE_PFN_TO_ADDR(entry->pfn)
  uintptr_t next_virt_addr = (uintptr_t) P2V(next_phys_addr);(*\label{line:ptc_p2v}*)
  next = (pte_t *) next_virt_addr;
  return next;
}
\end{lstlisting}
\end{minipage}
\hfill\hfill
\begin{minipage}{.4\textwidth}
\hspace{3em}
\begin{lstlisting}[mathescape,escapeinside={(*}{*)}, basicstyle=\scriptsize]
pte_t *walkpgdir(pte_t *root, void *va){ 
  pte_t *l4_entry = &root[L4Offset(va)];
  pte_t *l3 = pte_next_table(l4_entry);
  pte_t *l3_entry = &l3[L3Offset(va)];
  pte_t *l2 = pte_next_table(l3_entry);
  pte_t *l2_entry = &l2[L2Offset(va)];  
  pte_t *l1 = pte_next_table(l2_entry);
  pte_t *l1_entry = &l1[L1Offset(va)];  
  return l1_entry
}
\end{lstlisting}
%          \caption{Walking page-table tree to locate L1 entry C-source code.}
%        \label{fig:enter-label}
 %   \end{figure}
\end{minipage}
%\end{framed}
    \vspace{-2.5em}
    \caption{Most of a software page table walk in C, as used to map new pages.}
    \label{fig:pagetablescode}
    \vspace{-1em}
\end{figure}

On the x86-64 architecture, the \textsc{MMU}'s address translation uses a sparse hierarchical set of tables:
\emph{page tables} (referring to pages of memory). As Figure \ref{fig:pagetables} (based on Figure 5-17 of the 
AMD64 architecture manual~\cite{amd64_manual_vol2}\footnote{While x86 up through its 32-bit incarnation were due to Intel,
the x86-64 architecture as a 64-bit extension to x86 was originally due to AMD. As a result, it is sometimes also referred to as the \texttt{amd64} architecture.})
shows,
address translation proceeds by repeatedly taking designated slices of the virtual address and indexing into {successive} table\add{s} \add{of 512 8-byte entries (making each table 4KB in size)}.
The final lookup in the page tables gives the base physical address of a 4KB page of physical memory \add{accessible to the running
program}, to which the low-order
\add{12} bits of the accessed virtual address are added to determine the actual physical address retrieved.
On x86-64, standard configurations use 4 levels of page tables, labelled levels 4 through 1, with lookups in the level 
1 page table resulting in the actual page of physical memory holding the requested data, and the low-order 12 bits 
being used to index into this page.\footnote{Technically levels 1--3 have explicit historical names,
but for brevity and consistency, we simply number them, in keeping with the newer 5th level. Our formalization only deals with 4-level page tables, but is straightforwardly extensible to 5.} 
The translation process or algorithm is {often} referred to as a \emph{page-table walk}.
While Figure \ref{fig:pagetables} and most of our constants (how many levels, which virtual address bits index which
table levels) are specific to {common} x86-64 {CPUs},
  it is straightforward to adapt our approach to ARMv8/\texttt{aarch64}, RISC-V, or PowerPC, which use nearly-identical
  page table structures (only the order of flag bits in the entries differ),
  or to 5-level paging present in newer x86-64 or RISC-V designs (one step would be iterated an additional time).

The entries \add{(e.g., L4 Entry in Figure \ref{fig:pagetables})} of each table are 64 bits wide,
but each points to a physical address aligned to 4KB (4096 byte) boundaries, leaving 12 bits to spare to 
control a validity bit (called the \emph{present} bit), a read-write bit (enabling write access),
and a range of additional bits which can be used to control caching, write-through, and more.
This paper will only consider the present bit (0).
\looseness=-1

The page tables are managed by the {OS kernel's} \emph{virtual memory manager} (VMM).\footnote{Not to be confused with Virtual 
Machine Monitor. We focus on non-hypervisor scenarios, but hardware virtualization extensions for both 
x86-64 and ARM make use of an additional set of page tables translating what a \emph{guest} considers to be its 
(virtualized) physical memory to actual physical memory. Our contributions should offer value in this scenario as well.}
Typically each process has its own 
page tables, which the OS registers with the CPU by storing the (page-aligned)
physical address of the root of the page table tree (the start of the L4 table shown as $\mathsf{root}$ in Figure \ref{fig:pagetables}) in a specific register (\texttt{cr3}) as part of switching to a new process. 
Using different mappings, which map only disjoint portions of physical memory (with some exceptions in the next section) 
is how the OS ensures memory isolation between processes.
If an instruction is executed that accesses a virtual address that either has no mapping, or does not have a mapping permitting 
the kind of access that was performed (e.g., the instruction was a memory write, but the relevant address range was marked read-only in
the relevant page table entry), the hardware triggers a \emph{page fault}, transferring control to a \emph{page fault handler} registered 
with the hardware by the OS, allowing it to take corrective action.
If no mapping was supposed to exist, this is a program bug (e.g., dereferencing virtual address 0 / NULL)
and the faulting program should be terminated. But this can also be used for
other purposes such as demand paging to save on IO and better-manage physical memory~\cite{doeppner2010operating}.
\looseness=-1

The key pieces of VMM functionality are
adding a new page mapping (whether the mapped page contains zeros, file data, or swap data), and removing an existing 
page mapping.
While this initially sounds like relatively modest functionality whose implementation may be complicated by hardware 
subtleties, correctness of even these basic operations are actually quite intrictate.
Notably, updates to the page tables are performed as writes to memory --- \emph{which are themselves subject to address translation},
and finding the correct page table to update requires converting between physical and virtual addresses.
In the case of changing the mappings for the currently-active set of page tables, 
\emph{the OS kernel is modifying the tables involved in its
own access of the tables}.
\add{
  To get a sense of how subtle the required reasoning is, we can consider code such as Figure \ref{fig:pagetablescode},
  used in our evaluation to locate the appropriate L1 entry to map a page into the current address space.
  % Adding or removing mappings requires the VMM code to locate the correct page table entry to update,
  % using code such as that in Figure \ref{fig:pagetablescode}.
  \lstinline|walkpgdir| (right) essentially mimics the hardware address translation up to the L1 entry (its caller will modify
  the entry to map a new page),
  indexing into each successive table (e.g., with \lstinline|L4Offset(va)| retrieving the L4 offset of Figure \ref{fig:pagetables}),
  with \lstinline|pte_get_next_table| (left) fetching the base address of the next table
  from each entry.
  If, as shown in Figure \ref{fig:pagetables}, an entry such as the L2 entry is uninitialized (the \emph{present} flag is not set),
  \lstinline|pte_get_next_table| allocates the next table on Line \ref{line:ptc_alloc} and marks it valid on Line \ref{line:ptc_mark}.
  But there is a critical difference from hardware: every memory access in Figure \ref{fig:pagetablescode} uses \emph{virtual} addresses, as opposed
  to hardware's direct physical memory access.
  Thus \lstinline|pte_get_next_table| uses the \lstinline|P2V| (physical-to-virtual) macro on line \ref{line:ptc_p2v}
  to convert physical addresses stored in page table entries into virtual addresses which the kernel can use to access
  the corresponding physical location.
  Proving it correct requires proving that it yields
  an address that is not only mapped, but known to map back to the original physical address!
  Verifying this code is beyond the reach of prior work, which either
  does not model address translation for the kernel~\cite{gu15,gu2018certikos,Klein2009seL4,seL4TOCS,alkassar2008formal}
  and would thus reason unsoundly about this code,
  or models address translation but lacks features required to reason about this code~\cite{kolanski08vstte,kolanski09tphols}.
  We describe our proof of correctness for this code,
  based on the first formalization of critical VMM kernel invariants in an MMU-aware logic, in Section \ref{sec:experiment}.
}
\looseness=-1

Virtual memory affects the OS scheduler, which deals with multiple 
address spaces, so must track which virtual addresses are valid in which address spaces. 
Some virtual addresses are valid in only a single address space (e.g., a code address for a particular usermode 
process), while others are valid in all address spaces (e.g., kernel data structure pointers). 
The VMM must maintain some of these assumptions on behalf of the rest of the kernel, for example by guaranteeing that 
a certain range of virtual addresses (corresponding to the kernel's code and data) are valid in every address space.
\looseness=-1

% \paragraph{\add{Mapping a New Page}}
% \add{Mapping a new page into a virtual address space is a key piece fo functionality, and also a key challenge for verification.
% This operation takes a chunk of physical memory to install, and a (page-aligned) virtual address at which to make it
% visible to the process.
% Starting from the page table root (\lstinline|cr3| in Figure \ref{fig:pagetables}), it must traverse the page tables until it locates
% the correct location of the L1 entry which should point to the physical page (installing new L3, L2, and L1 tables as needed),
% and update the L1 entry to point to the physical page. This is complicated by the fact that the kernel accesses the page
% tables through \emph{virtual} addresses, but the page table root and the addresses in the L4--L1 entries are all \emph{physical}
% addresses. So the mapping process must repeatedly perform \emph{backwards} mappings (not supported by hardware) from
% physical addresses of each level's entry to a virtual address that is then mapped back to the corresponding physical address!
% And for modularity, the translation process (and possible allocation of missing tables) is typically implemented in a function
% used for every level of translation (i.e., a level-polymorphic function).
% These compounding subtleties are one reason \textsc{CertiKOS} and \textsc{seL4} \emph{trust} their page table
% manipulations; both the reasoning to prove correctness, and the kernel invariants that ensure correctness of typical implementations,
% are remarkably subtle.
% }
% \looseness=-1

\paragraph{Out of Scope: Translation Lookaside Buffers (TLBs)}
CPUs with MMUs typically have an additional \emph{translation lookaside buffer}
(TLB), to cache the (successful) results of page table walks, rather than transforming every virtual
memory access into 5 physical accesses. 
Any time a virtual address that was accessible becomes inaccessible (or has downgraded permissions),
the TLB (or at least entries in affected virtual address ranges) should be flushed.
In most kernels, this occurs in only a few well-known places, which is why existing verified kernels
\textsc{seL4}~\cite{Klein2009seL4,seL4TOCS} and \textsc{CertiKOS}~\cite{gu15,gu2016certikos} currently
trust that TLB flushes are handled correctly rather than actually modeling TLB hardware and verifying.
Eventual verification of TLB handling is a worthwhile long-term goal, but it is a challenging pursuit in its
own right. Based on others' early progress on verifying TLB operations in isolation~\cite{syeda2020formal}, we expect
it to be possible to combine this paper's insights with that support. 
Section \ref{sec:relwork} elaborates briefly on the challenges involved.
Even without TLB modeling,
our logic already enables verification of virtual memory management functionality that prior verified kernel work
either trusts completely (\textsc{seL4}, \textsc{CertiKOS}) or is incapable of reasoning about.
% Appendix \ref{apdx:tlb} gives more details on the challenges involved for the interested reader.
\looseness=-1

\subsection{Separation Logic}
\label{sec:seplogic}
% \todo{Colin says: I added this b/c prior systems reviewers wanted more verification background...
% not sure how I feel about it}
Separation logic~\cite{reynolds02} is a descendant of classic Hoare logic~\cite{hoare69},
where in addition to pre- and postcondition assertions, assertions themselves pick up a
\emph{separating conjunction} operator $\ast$, such that an assertion $A\ast B$ means $A$ and $B$ are true
of disjoint pieces of state. This allows for local reasoning about updates, because it articulates
that updates to the state backing $A$ do not affect the truth of $B$. This is classically demonstrated
through the \emph{points-to} assertion: $l\mapsto v$ asserts that the memory cell at address $l$ holds value $v$:
knowing $x\mapsto 3\ast y\mapsto 4$ and writing through $x$ means information about $y$ is preserved:
$\{x\mapsto 3\ast P\}\;x\mathrel{:=}5\;\{x\mapsto 5 \ast P\}$ can be derived for any $P$.

We build on the \iris~\cite{jung2018iris} separation logic framework,
an abstract separation logic embedded in Rocq, which is useful for both metatheoretical work
and interactive correctness proofs using the logic. Given an operational
semantics structured a certain way (in our case, semantics for
a fragment of x86-64 assembly including address translation),
if a small number of ``glue'' lemmas are proven, \iris
provides a ready-made separation logic with a number of advanced features, including
higher-order ghost state and impredicative invariants, for no additional work.

We suppress {some technical \iris details} for brevity, but briefly note
a few recurring details used heavily in \iris but not necessarily in traditional
separation logics.
Because \iris is an embedded framework in \rocq, proofs in \iris-derived
logics often encapsulate raw \rocq assertions: $\ulcorner P \urcorner$ is an embedding
of the \rocq assertion $P$ into an \iris assertion (used for things like equality and other
general pure logical assertions), similar to prior \rocq-embedded program logics~\cite{Chlipala2013Bedrock}. Newly-defined \iris assertions are in fact
\rocq terms of a particular type, rather than being drawn from a fixed vocabulary.
\looseness=-1

\iris includes two forms of implication. The magic wand operator $\wand$ is an affine implication:
$A\wand B$ describes a resource which, if combined with a resource satisfying $A$, will satisfy $B$.
Notably, this implication involves no changes to ghost state. \iris, building on the Views framework~\cite{Dinsdale-Young2013Views},
also includes a \emph{view shift} operator $\vs$ which models updates to ghost state: $A \vs B$
means resources satisfying $A$ may be transformed into resources satisfying $B$, intuitively by updating only ghost state
(a slight simplification of \iris's update modalities, but adequate intuition for non-\iris-experts).
% .\footnote{\iris
% experts may note that this is \emph{technically} mildly misleading given how \iris's update modalities and weakest precondition
% are defined, but
% this is adequate intuition for non-experts in \iris to follow their use.
% }
View shifts are essentially logical entailment plus ghost updates.
\looseness=-1

% Of particular interest to us, \iris's machinery is largely agnostic to the particular choice of
% \emph{resource algebra}\footnote{A modern descendant of the venerable partial commutative monoid used
% to abstractly model earlier separation logics~\cite{calcagno2007local}, 
% with additions to support the step-indexing required to solve the recursive domain equations
% that arise with higher-order ghost state and impredicativity~\cite{birkedal2011step,hobor2010theory}.
% }
% used to give algebraic semantics to the connectives of the assertion language.
% This means we can drop in an alternative model that naturally supports working with respect to
% a locally-fixed context, like an address space.

% Also relevant is that \textsc{Iris} naturally decomposes the traditional Hoare triple $\{P\}\;C\;\{Q\}$
% into the precondition implying the weakest precondition of the program with respect to the postcondition ---
% $P \wand \textsf{wp}\;C\;\{v\ldotp Q\}$.\footnote{An idea with long history~\cite{pratt1976semantical}.}
% This is a natural fit for assembly-level verification, where the \emph{Hoare double}

An important limitation is that to date, every separation logic
has assumed that all pointer addresses are meant for use in a single address space, which avoided the
problem of tracking that certain points-to assertions are true only for certain address spaces, but not others.

\subsection{Modal Logic}
\label{sec:backgroundonmodallogic}
The problem of needing to keep track of things being true in some contexts and not in others is hardly unique to virtual 
memory management, and is the general insight behind most flavors of modal logic, which use
unary operators to express that a logical claim $P$ is \emph{contingently} true 
in certain other circumstances, such as in other times~\cite{pnueli1977temporal} or places~\cite{murphy2008type,gordon2019modal}.

%   A unifying concept across any modality is that they behave as applicative functors, 
% typically satisfying (directly, or as a derived law, depending on the modality):
% \[ (P\rightarrow Q) \rightarrow M(P) \rightarrow M(Q)\]
% %\todo[inline]{Ismail, I think this means we have pure intro, $\square P \mathrel{-\ast} [r](\square P)$ if I'm using the right symbol for pure assertions }
% %\todo[inline,color=red]{The above todo comment isn't quite right.
% %Purity is about being able to duplicate. You had defined another typeclass/property-of-assertions that meant it didn't
% %care what address space it was in (like physical pointstos). That's orthogonal to purity.
% %}
% Many modalities, so-called \emph{normal} modalities also possess introduction rules of the form $P\rightarrow M(P)$, 
% the classic example being that if $P$ is true, then $P$ is \emph{necessarily} true with the contingency picked up.%($\square P$).

Of particular interest for reasoning about virtual memory are modalities that permit \emph{naming} the alternate 
circumstances, prominently \emph{hybrid} modal logics~\cite{blackburn1995hybrid,areces2001hybrid}, which come equipped 
with assertions of the form $[\ell](P)$ indicating that $P$ is true in the specific alternate circumstance (Kripke world)
 named by the {term} $\ell$. Note that a distinctive property of hybrid logics is that, rather than hiding
the points at which a modal assertion is evaluated inside the modality's definition, the choice of what world a modalized
assertion should be true in is chosen \emph{in the assertion itself}. This allows assertions to talk about not simply whether some other assertion
is true in some possible future or past world related in a fixed way to the current world, but to talk about \emph{arbitrary}
other worlds.
\add{This is somewhat different from the accessibility relation modalities more common in verification, but also well-established,
tracing its origin back to Arthur Prior~\cite{areces2006hybrid}, just as many temporal logics do.}

This explicit naming of alternate worlds increases the power of \add{propositional} modal logics~\cite{blackburn1995hybrid}, and is 
necessary for completeness in classical separation logics~\cite{brotherston2014parametric}.
\add{However typically modal logics can always be ``compiled'' into logics with suitable quantifiers:
the \emph{standard translation}~\cite{blackburn2006modal} of propositional modal logics into first-order logic
has been thoroughly studied. \textsc{Iris}'s support for higher-order impredicative quantification means
this applies to our modalities as well --- they do not strictly speaking make \textsc{Iris} more expressive.
But while propositional modal logics are no more expressive than first-order logic (with just two variables!),
the primary goal of a modal logic is not raw expressive power but intuitive specifications and proof rules.
Temporal logics are used because they simplify specifications (vs. having explicit time variables
in every base asertion) and reasoning principles (because the modal correspondents of quantification over
time is used in a highly structured way).
Our logic offers a similar value proposition: natural specifications and reasoning principles which
are powerful enough for reasoning about virtual memory management, but without requiring pervasive tracking of
and explicit abstraction over address spaces.
}

For our purposes, these are natural candidates to adapt for virtual memory management. We can reinterpret the notion of 
naming an alternate world slightly more loosely, and instead name \emph{address spaces} by the physical address of the 
page table root, since these structures are the physical representations of page tables. Thus in this paper we develop 
the notion that we can represent contingent truth of an assertion via $[r](P)$ indicating that $P$ holds in the address 
space rooted at physical address $r$. Because OS kernels create and destroy address spaces, it is sensible to use
a hybrid-style logic that is not specialized to a fixed set of modalities, but this introduces
some subtleties from the fact that the existence of certain modalities (address spaces) can change.

Interaction of hybrid modalities and substructural reasoning is relatively unexplored (see Section \ref{sec:relwork}).
% A relatively under-explored space of modal logics is the interaction of modal and substructural logics, 
% in particular hybrid-style modalities in substructural logics, which has seen only minimal exploration~\cite{dovsen1992modal,restall1993modalities,d1997grafting,kamide2002kripke,licata2017fibrational} and no prior 
% application. 
Our development atop \iris~\cite{jung2018iris} needs to explore some additional subtleties 
that arise where the modality itself may entail ownership of resources, 
as well as interactions between our hybrid-style 
modality and substructural rules.  
% For example, Iris contains a number of modalities that distribute over separating 
% conjunction, or for which resources can freely move into the modality 
% (e.g., $\blacktriangleright(P)\ast Q \wand \blacktriangleright(P\ast Q)$). In our setting some of these rules 
% apply while others do not. For example, 
% In our setting, an assertion that involves no modalities is interpreted as 
% holding in the current (active-on-the-CPU) address space, so clearly cannot move into arbitrary other address spaces,
% while address-space-relative assertions 
% --- unless guarded by another address space modality.
Some prior \iris-based work~\cite{dang2019rustbelt,dang2022compass} has constructed derived modalities in the style we propose, indexed
by thread IDs. However their intepretation of those modalities was fully fixed ahead of time (to refer to essentially buffers in operationalized versions of C11
concurrency). In this setting, while our modalities will be indexed by page table roots, it is possible to modify the address translation for an address
space with root $r$ --- thus changing the interpretation of a modality, and even whether a modality is valid --- \emph{while assertions with that modality are active}.
\add{This is essential to updating page mappings for the current address space in use by a CPU.}
\looseness=-1

\definecolor{dkgreen}{rgb}{0,0.6,0}
\definecolor{ltblue}{rgb}{0,0.4,0.4}
\definecolor{dkviolet}{rgb}{0.3,0,0.5}

% lstlisting coq style (inspired from a file of Assia Mahboubi)
\lstdefinelanguage{Coq}{ 
    % Anything betweeen $ becomes LaTeX math mode
    mathescape=true,
    % Comments may or not include Latex commands
    texcl=false, 
    % Vernacular commands
    morekeywords=[1]{Section, Module, End, Require, Import, Export,
        Variable, Variables, Parameter, Parameters, Axiom, Hypothesis,
        Hypotheses, Notation, Local, Tactic, Reserved, Scope, Open, Close,
        Bind, Delimit, Definition, Let, Ltac, Fixpoint, CoFixpoint, Add,
        Morphism, Relation, Implicit, Arguments, Unset, Contextual,
        Strict, Prenex, Implicits, Inductive, CoInductive, Record,
        Structure, Canonical, Coercion, Context, Class, Global, Instance,
        Program, Infix, Theorem, Lemma, Corollary, Proposition, Fact,
        Remark, Example, Proof, Goal, Save, Qed, Defined, Hint, Resolve,
        Rewrite, View, Search, Show, Print, Printing, All, Eval, Check,
        Projections, inside, outside, Def},
    % Gallina
    morekeywords=[2]{forall, exists, exists2, fun, fix, cofix, struct,
        match, with, end, as, in, return, let, if, is, then, else, for, of,
        nosimpl, when},
    % Sorts
    morekeywords=[3]{Type, Prop, Set, true, false, option},
    % Various tactics, some are std Coq subsumed by ssr, for the manual purpose
    morekeywords=[4]{pose, set, move, case, elim, apply, clear, hnf,
        intro, intros, generalize, rename, pattern, after, destruct,
        induction, using, refine, inversion, injection, rewrite, congr,
        unlock, compute, ring, field, fourier, replace, fold, unfold,
        change, cutrewrite, simpl, have, suff, wlog, suffices, without,
        loss, nat_norm, assert, cut, trivial, revert, bool_congr, nat_congr,
        symmetry, transitivity, auto, split, left, right, autorewrite},
    % Terminators
    morekeywords=[5]{by, done, exact, reflexivity, tauto, romega, omega,
        assumption, solve, contradiction, discriminate},
    % Control
    morekeywords=[6]{do, last, first, try, idtac, repeat},
    % Comments delimiters, we do turn this off for the manual
    morecomment=[s]{(*}{*)},
    % Spaces are not displayed as a special character
    showstringspaces=false,
    % String delimiters
    morestring=[b]",
    morestring=[d],
    % Size of tabulations
    tabsize=3,
    % Enables ASCII chars 128 to 255
    extendedchars=false,
    % Case sensitivity
    sensitive=true,
    % Automatic breaking of long lines
    breaklines=false,
    % Default style fors listings
    basicstyle=\small,
    % Position of captions is bottom
    captionpos=b,
    % flexible columns
    columns=[l]flexible,
    % Style for (listings') identifiers
    identifierstyle={\ttfamily\color{black}},
    % Style for declaration keywords
    keywordstyle=[1]{\ttfamily\color{dkviolet}},
    % Style for gallina keywords
    keywordstyle=[2]{\ttfamily\color{dkgreen}},
    % Style for sorts keywords
    keywordstyle=[3]{\ttfamily\color{ltblue}},
    % Style for tactics keywords
    keywordstyle=[4]{\ttfamily\color{dkblue}},
    % Style for terminators keywords
    keywordstyle=[5]{\ttfamily\color{dkred}},
    %Style for iterators
    %keywordstyle=[6]{\ttfamily\color{dkpink}},
    % Style for strings
    stringstyle=\ttfamily,
    % Style for comments
    commentstyle={\ttfamily\color{dkgreen}},
    %moredelim=**[is][\ttfamily\color{red}]{/&}{&/},
    literate=
    {\\forall}{{\color{dkgreen}{$\forall\;$}}}1
    {\\exists}{{$\exists\;$}}1
    {<-}{{$\leftarrow\;$}}1
    {=>}{{$\Rightarrow\;$}}1
    {==}{{\code{==}\;}}1
    {==>}{{\code{==>}\;}}1
    %    {:>}{{\code{:>}\;}}1
    {->}{{$\rightarrow\;$}}1
    {<->}{{$\leftrightarrow\;$}}1
    {<==}{{$\leq\;$}}1
    {\#}{{$^\star$}}1 
    {\\o}{{$\circ\;$}}1 
    {\@}{{$\cdot$}}1 
    {\/\\}{{$\wedge\;$}}1
    {\\\/}{{$\vee\;$}}1
    {++}{{\code{++}}}1
    {~}{{\ }}1
    {\@\@}{{$@$}}1
    {\\mapsto}{{$\mapsto\;$}}1
    {\\hline}{{\rule{\linewidth}{0.5pt}}}1
}[keywords,comments,strings]
\section{Machine State and Semantics}
% \section{Machine State \& Syntax}
\label{sec:syntax}
To develop our core logical ideas, we instantiate \iris with a simple language for streams of instructions
and a logical machine model corresponding to execution of x86-64 assembly instructions with virtual memory enabled on the
CPU.
%
% \subsection{Registers and Memory}
% Programs we demonstrate in this paper requires accessing two types of computer resource: registers and memory.
A register identifier, $\reg$, is chosen from a fixed finite set of register identifiers, $\regset$.\footnote{\add{$\reg$ abbreviates \emph{general register}, which includes integer registers (e.g., \lstinline|rax|),
control registers (e.g., \lstinline|cr3|), and segmentation registers (which we do not discuss in detail,
but are still used in a limited way in 64-bit mode on modern x86-64 processors).
}} 
We use these identifiers $\reg$ for register names such as \lstinline|rax|, \lstinline|r8|, or \lstinline|cr3|. Our model includes
all x86-64 integer registers (including stack and instruction pointers), as well as \lstinline|cr3| (for page table roots) and \lstinline|rflags| (for
flags set by comparison operations and inspected by conditional jumps).
% \begin{figure}[t]
% \newcommand{\commentary}[1]{ & \text{\small\it #1} \\}
% \[
%   \begin{array}{r@{\;}c@{\;}l}
%     \loc & \in & \Loc \\
%   \end{array}
%   \begin{array}{r@{\;}c@{\;}l}
%     \reg & \in & \regset \\
%   \end{array}
%   \begin{array}{r@{\;}c@{\;}l}
%     \regval & \in & \regvaltype \\
%     % \val & ::= & \vunit
%   \end{array}
% \quad
%   \begin{array}{r@{\;}c@{\;}l}
%     \instrs & ::= &
%     \begin{array}[t]{@{}l@{\hspace{10mm}}l@{}}
%     \begin{array}[t]{@{}ll@{}}
%       \iskip
%                    \commentary{no-op}
%       \iseq\instr\instrs
%                    \commentary{sequencing}
%       % \ising\instr
%                    % \commentary{executing}             
%     \end{array}
%     %&
%     %\begin{array}[t]{@{}ll@{}}
%      % \ialloc\lval\allocsize
%       %             \commentary{heap allocation}
%     %\end{array}
%     \end{array}
%     % \\
%     % \ectx & ::= &
%     %   \hole \mid
%     %   \iseq\ectx\instrs 
%     % \\
%   \end{array}
% \]
% \caption{Syntax}
% \Description{Syntax}
% \label{fig:syntax}
% \end{figure}
For clarity and ease of representation, we use machine words, $\loc \in \Loc$, with the subscripts showing the number of bits in a word,
for memory addresses, values, and offsets, rather than distinct location types that wrap machine words.
For example, $\kw{w}_{12}$ is a 12-bit word, which can be obtained for example truncating away 52 bits of a 64-bit word ($\kw{w}_{64}$).
% \subsection{State}
\label{sec:state}
We represent the machine state mainly as a finite map of registers to register values and a map of word-aligned physical memory addresses 
to 64-bit physical memory values. 
Thus our states $\sigma$ include register maps $\sigma.\mathcal{R}: \kw{greg} \rightarrow_{\textrm{fin}} \kw{regval} $ and
memory maps $\sigma.\mathcal{M}: \Locft \rightharpoonup_{\textrm{fin}} (\Loctw \rightharpoonup_{\textrm{fin}} \Locsf )$
\add{which segment memory into page-sized increments}.
Of particular note, \lstinline|cr3|, the page table register, is included in the machine state.
\looseness=-1

% \subsection{Instructions}
\label{sec:instructions}

% \begin{figure}[t]
% \begin{lstlisting}[language=Coq,mathescape]
%  Definition translate (rtv: word 64) ($\sigma$: state) (vaddr: word 64): word 64 + MemFail :=
%       l4e $\leftarrow$ translate_top_level $\sigma.\mathcal{M}$ rtv vaddr;
%       l3e $\leftarrow$ translate_from_l4e $\sigma.\mathcal{M}$ l4e vaddr;
%       l2e $\leftarrow$ translate_from_l3e $\sigma.\mathcal{M}$ l3e vaddr;
%       l1e $\leftarrow$ translate_from_l2e $\sigma.\mathcal{M}$ l2e vaddr;
%       addr $\leftarrow$ translate_from_l1e $\sigma.\mathcal{M}$ l1e vaddr;
%       return concat (shift_dropn addr 3 three_lt64) low3.
% \end{lstlisting}
% \todo[inline]{I don't think we need this given a more compact version of Figure \ref{fig:pagetables}}
% \vspace{-1em}
% \caption{\coq implementation of address translation (slightly simplified).}
% \label{fig:coq_addr_translation}
% \end{figure}

Programs in our logic are instruction sequences \instrs, which are formed by either a basic instruction \iskip, or prefixing an existing instruction
sequence with an additional instruction (\iseq\instr\instrs).
We model (and later, give program logic rules for) instructions for register loads and stores, and reading and writing memory.
The latter require page table walks.
The most important instructions that we model are memory-accessing variants of the x86-64 \lstinline|mov| instruction, which we format in Intel syntax
(destination on the left, source on the right).
Thus, a store to memory is $\textsf{mov}~[r_m]~r_r$, and a load from memory is $\textsf{mov}~r_r~[r_m]$.
Operationally, each first translates the virtual address stored in the register $r_m$ to a physical address \add{(a page table walk)},
then either updates the memory at that physical location with the contents of $r_r$ (for store)
or updates $r_r$ with the contents of that memory (for load).
Our formalization includes
additional \lstinline|mov| variants (e.g., accessing memory at constant offsets from the base register, or moves between registers),
basic integer and bitwise operations (\lstinline|add|,
\lstinline|and|, bit shifts, etc.) with their effects on \lstinline|rflags|, jumps and some (not all) conditional jumps, \lstinline|call|, \lstinline|ret|, \lstinline|push|,
and \lstinline|pop|.
% The semantics model updates to the instruction pointer, though we elide that register (\lstinline|rip|)
% from specifications in the paper for clarity.
\looseness=-1

\section{Program Logic for Location Virtualization}
\label{sec:logic}
% The predicate gen_heap_interp.
\newcommand{\gammaPred}{\delta}
\newcommand{\gammaPreds}{\delta\textsf{s}}
\newcommand{\rtv}{\textsf{rtv}}
\newcommand{\qone}{\texttt{q1}}
\newcommand{\qtwo}{\texttt{q2}}
\newcommand{\qthree}{\texttt{q3}}
\newcommand{\qfour}{\texttt{q4}}

\newcommand{\sumwalkabs}[3]{
  \ownGhost\gammaPred{\authfrag{\singletonMap{#1}{(#2, #3)}}}
}

\newcommand{\sumapaces}[2]{
  \ownGhost\gammaPreds{\authfrag{\singletonMap{#1}{#2}}}
}
\newcommand{\ptableabswalk}[1]{\mathcal{A}\textsf{bsPTableWalk}(#1)}
\newcommand{\ptablestore}{\theta}

\add{
  A program logic for reasoning about code which may work with (and possibly update) multiple address spaces requires dealing with
  several key challenges. It must ensure that reasoning about memory accesses only depends on assertions that hold
  in the active address space at the time of access. It must allow invariants for code or data structures to refer
  to other address spaces. These constraints mean it must also support reasoning about when the active address space
  changes, as this affects which memory assumptions are usable or not for memory access and thus which data structures are
  immediately directly accessible.
}
\add{
  Many approaches could handle these problems in principle, such as tagging pointers with the relevant address space.
  But such approaches introduce other complexities.
  Most code, even in an operating system kernel, only works with a single address space (the current one).
  Explicitly plumbing address space information through a simple linked list specification, simply because some other part of the program
  may manipulate other address spaces, adds significant specification burden.
  Specifications which need to talk about a particular invariant holding in a specific other address space (or multiple other address spaces) would need to quantify
  over \emph{functions} from address space identity to assertions rather than just assertions.
}

\add{
  Using a modal approach resolves all of these challenges cleanly and uniformly.
  Specifications that are not \emph{about} address space manipulation need not mention address spaces in assertions.
  One can use standard separation logic data structure specifications without restructuring or
  adding explicit address space
  tracking,
  but every assertion can be still stated for either the current or specific other address spaces as needed.
  In short, modalities make it possible for specifications to mention address spaces when
  it is important to the code, and not when it is unimportant to the code.
  \looseness=-1
}

We describe a program logic (a separation logic) along the lines suggested {above}, where every assertion is relative
to an address space in which it is interpreted, allowing us to define \emph{virtual points-to} assertions that make claims
about memory locations in a particular address space. Virtual addresses, and even virtual points-to assertions, 
are not tagged with their address spaces in any way. Memory access in this logic is validated through the use
of virtual points-to assertions in preconditions, which guarantee that address translations succeed.
This supports rules for updating not only typical data in memory that happens to be subject to address translation, but \add{also}
manipulation of the page tables themselves via virtual addresses (as demanded by all modern hardware) {and}
via virtual points-to assertions.
To support specifications that deal with multiple address spaces, our logic incorporates a hybrid-style modality
$[r](P)$ to state that an assertion is true in another (assertion-specified) address space rather than the address space
currently active in hardware, which is not only useful for virtual memory manager invariants, but \add{also} critical to reasoning
about change of address space.
By developing this within the \iris framework, we obtain additional features (e.g., fractional permissions) that allow us to verify
some of the most subtle and technically challenging instruction sequences in an OS kernel (Section \ref{sec:experiment}).

To support making assertions depend on a choice of address space, we work entirely in a pointwise lifting of \iris's base BI logic,
essentially working with separation logic assertions indexed by a choice of page table root as a $\mathcal{W}_{64}$, which we call $\textsf{vProp }\Sigma$:\footnote{
  \iris experts may notice our \lstinline|-b>| resembles another pointwise lifting already in  \iris~\cite{dang2019rustbelt,dang2022compass}. 
  This similarity is real, but the existing lifting does not appear to work with indexed \coq types like our \lstinline|word n| as a domain.
}
\lstinline[language=Coq]|Definition vProp  $\Sigma$ : bi := word 64 -b> iPropI  $\Sigma$|.
This is the (\rocq) type of assertions in our logic.
Most constructs in \iris's base logic are defined with respect to any BI-algebra (of \coq type \lstinline|bi|), so \add{they} automatically
carry over to our derived logic.
However, we must still build up from existing \iris primitives to provide new primitives that depend on the address space --- primarily the notion
of virtual points-to.
To define and use virtual points-to assertions, we require two basic assertions that ignore
the current address space:

\paragraph{Register points-to} 
The assertion $\textsf{r}\;\mapsto_{\textsf{r}}\{q\}\;\textsf{rv}$ ensures the ownership of the register $\rg$ containing the
value of the register $\rv$.
The fraction $\qfrac$ with value 1 asserts the unique ownership of the register mapping and grants update permission {to} it;
otherwise, any value $0 < \qfrac <1$ represents partial ownership, granting read-only permission on the mapping.\footnote{
\add{We adopt the standard naming convention of $\qfrac$-related names representing fractional permission, with fractions
sometimes appearing in braces or as subscripts in various asertions.}}

\paragraph{Physical memory  points-to} The soundness proofs for our logic's rules largely center around
proving that page-table-walk accesses as in Figure \ref{fig:pagetables} succeed, which requires assertions
dealing with physical memory locations.
We have two notions of physical points-to facts. The primitive notion closest to our machine model is captured by an assertion
$ \textsf{pfn} \ \sim \ \textsf{pageoff} \mapsto_{\textsf{a}} \; \{\textsf{q}\} \; \textsf{v} $, where \textsf{pfn} (a $\mathcal{W}_{52}$ \emph{page frame number}) essentially selects a 4KB page of physical memory,
and \textsf{pageoff} (a $\mathcal{W}_{12}$) is an offset within that page.
% could be an 52-bits masked address to level 4 table 
% ($\textsf{w1 } =( \textsf{ l4M52 maddr cr3val}) $),
%and, expectedly \textsf{w2} is an address computed by page-offset computation (e.g. $\textsf{l4off maddr cr3val}$). 
From this we can derive a more concise physical points-to when the split is unimportant:
% Giving a raw 64-bits memory pointsto assertion becomes
{$\textsf{w} \mapsto_{\textsf{p}} \{q\} \textsf{ v} \stackrel{\triangle}{=} (\textsf{drop 12}~w) \ \sim \ (\textsf{bottom 12}~w)\mapsto_{\textsf{a}} \; \{\textsf{q}\} \textsf{ v} $}

 %I put a newline in between the following two Definitions as the second Definition seems not proper without the newline
\begin{figure*}
  \begin{lstlisting}[language=Coq,escapeinside={(*}{*)}]
 Definition $\vaddr\mapsto_{\textsf{t}}\{\textsf{q}\}\; \vpage$ : vProp $\Sigma$ := 
  $\exists_{\textsf{l4e l3e l2e l1e}} \ldotp$  $\ulcorner$ aligned $\vaddr \urcorner \ast$ L4_L1_PointsTo($\vaddr$ l4e l3e l2e l1e paddr) $\ast$ paddr $\mapsto_{p}\{\mathsf{q}\} \vpage$.
 Definition L4_L1_PointsTo (maddr l4e l3e l2e l1e paddr :word 64) : vProp $\Sigma$ := $\lambda$ cr3val.
  $\ulcorner$ entry_present l4e $\land$ entry_present l3e $\land$ entry_present l2e $\land$ entry_present l1e$\urcorner$ $\ast$
  (l4M52 maddr cr3val) $\sim$ (l4off maddr cr3val) $\mapsto_{a}$ {q1}  l4e  $\ast$(*\label{line:l4pointsto}*)
  (l3M52 maddr l4e) $\sim$ (l3off maddr l4e)  $\mapsto_{a}$ {q2}  l3e $\ast$(*\label{line:l3pointsto}*) 
  (l2M52 maddr l3e) $\sim$ (l2off maddr l3e) $\mapsto_{a}$ {q3}  l2e $\ast$(*\label{line:l2pointsto}*)
  (l1M52 maddr l2e) $\sim$ (l1off maddr l2e) $\mapsto_{a}$ {q4}  l1e $\ast$ $\ulcorner$ addr_L1(va,l1e) = paddr $\urcorner$.(*\label{line:l1pointsto}*)
\end{lstlisting}
\vspace{-1em}
\caption{A Strong Virtual Points-to Relation
}
  \label{fig:strongvirtualpointsto}
\vspace{-1em}
\end{figure*}

\subsection{An Overly-Restrictive Definition for Virtual Memory Addressing}
\label{sec:overly-restrictive}
A natural definition for a virtual points-to
asserting that virtual address \textsf{va} points to a value \textsf{v}
would 
% require that in order for a virtual address \textsf{va} to point to a value \textsf{v}, the assertion 
contain
partial ownership of the physical memory involved in the page table walk that would translate \textsf{va} to
its backing physical location --- with locations existentially quantified since a virtual points-to should not assert
\emph{which} locations are accessed in a page table walk, as in Figure \ref{fig:strongvirtualpointsto}.
It asserts the existence of four page-table entries, one at each translation level, and via \lstinline|L4_L1_PointsTo|
asserts that the physical page table walk (per Figure \ref{fig:pagetables}) succeeds in reaching the L1 entry,
which points to the page holding the physical memory backing the virtual address, which contains value \textsf{v}.
Most of the definition lives directly in \textsf{vProp}, using the separation logic structure lifted from \iris's \textsf{iProp}.
\looseness=-1

\lstinline|L4_L1_PointsTo| works by
chaining together the entries for each level, using the sequence of table offsets from the address being translated to index
each table level, and using the physical page address embedded in each entry.\footnote{
  The fractions \lstinline|q1| through \lstinline|q4| represent the fractional ownership of each entry based on how many
  word-aligned addresses might need to share the entry ---  $(\frac{1}{512})^n$ for each level $n$.
}
For example, the first-level address translation to get the L4 entry (\lstinline|l4e|) 
  uses the masks \textsf{l4M52} with the current \lstinline|cr3| to get the physical address of the start of the L4 table
  and \textsf{l4off} with the virtual address being translated to compute the correct \add{byte} offset within that table \add{just as in the first translation
  step of Figure \ref{fig:pagetables}}.\footnote{\add{Note the offsets mentioned in Figure \ref{fig:pagetables} are 9-bit indexes into the 512 entries; the byte offset is that times 8.}}
    Thus Line \ref{line:l4pointsto} asserts that the physical address built from the table base and offset points to the L4 entry \textsf{l4e}.
  Subsequent levels of the page table walk \add{assertion (Lines \ref{line:l3pointsto}--\ref{line:l1pointsto})} work similarly.
The statement of these assertions is simplified by the use of our split physical points-to assertions, since
each level of tables is page-sized. \footnote{We do not address superpages and hugepages in this paper.}
This helper definition is also more explicit in \textsf{vProp} which binds a value to \lstinline|cr3| and uses it to start the translation process.
\looseness=-1

This solution is in fact very close to that of \citet{kolanski08vstte}, who define a separation logic from scratch in \textsc{Isabelle/HOL},
where the semantics of all assertions are functions from pairs of heaps and page table root values to booleans.\footnote{
  This was a typical explicit construction at the time; their work significantly predates \iris.
}
Our solution in the next subsection improves on theirs, removing some restrictions in this definition by further abstracting the handling of address translation.
\looseness=-1

\subsection{Aliasing/Sharing Physical Pages}
  \label{sec:sharingpages}  
  The virtual points-to definition shown in Figure \ref{fig:strongvirtualpointsto} 
  is too strong to specify some operations that a virtual memory manager may need to do, such as move one level of the page table to a different physical location while preserving all virtual-to-physical mappings. %\footnote{
  %   x86-64 hardware, like other architectures, includes a feature (which we do not formalize assertions for) to
  %   replace an L1 page table address in an L2 entry with a pointer to a \emph{larger} 2MB page (called super-pages), 
  %   or replace an L2 page table address in an L3 entry with a pointer to a 1GB page (called huge-pages).
  % }
  The use of $\textsf{L}_{4}\_\textsf{L}_{1}\_\textsf{PointsTo}$ in Figure \ref{fig:strongvirtualpointsto}'s
  virtual points-to definition stores knowledge of the page table walk details with ownership of the backing physical memory.
  Updating any of these mappings (e.g., moving the page tables in physical memory, as in coalescing for superpages or hugepages)
  would require explicitly collecting all virtual points-to facts that traverse affected entries.
  It is preferable to permit the page tables themselves to be updated independently of the virtual points-to assertions,
  so long as those updates preserve the same virtual-to-physical translations.
  But this is not possible with Figure \ref{fig:strongvirtualpointsto}'s definition, which ties ownership of particular pieces of page table memory to the virtual points-to.

  % iris.sty lacks nice syntax for the ghost maps
  \newcommand{\ghostmaptoken}[3]{\ensuremath{#2\hookrightarrow^{#1}#3}}
  \newcommand{\fracghostmaptoken}[4]{\ensuremath{#2\hookrightarrow^{#1}_{#4}#3}}

\newcommand{\vale}{\textsf{val}}
\begin{figure*}
\footnotesize
\centerline{$
\begin{array}{l}
    \vaddr\mapsto_{\textsf{v}}\{\textsf{q}\}\;\vale : \mathsf{vProp}~\Sigma \stackrel{\triangle}{=} 
    \exists \paddr\ldotp
    \exists \delta\ldotp
    % \underbrace{(\lambda\mathit{cr3val}\ldotp\sumapaces{\mathit{cr3val}}\delta)}_\text{Find addr.\;space invariant} \ast 
    \underbrace{(\lambda\mathit{cr3val}\ldotp\exists \epsilon\ldotp\fracghostmaptoken{\delta{}s}{cr3val}{\delta}{\epsilon})}_\text{Find addr.\;space invariant} \ast 
  % \underbrace{\sumwalkabs\vaddr\qfrac\paddr }_\text{Ghost translation}\ast 
  \underbrace{\fracghostmaptoken{\delta}{\mathsf{va}}{\mathsf{pa}}{\mathsf{q}} }_\text{Ghost translation}\ast 
  \underbrace{\paddr \mapsto_{\mathsf{p}}\{\textsf{q}\}\; \vale}_\text{Physical location}
\end{array}
$}
\vspace{-1em}
\caption{Virtual-Points-to for Sharing Pages}
  \label{fig:virtualpointstosharing}
\end{figure*}  

  % Intuitively, the definition in Figure \ref{fig:strongvirtualpointsto} is too strong because the virtual points-to
  % assertion there tracks too much information: when writing programs that access memory via virtual addresses,
  % most code does not care \emph{which physical memory locations are involved in address translation}: it only cares
  % that virtual address translation would succeed. The necessary information about the physical page table walk
  % must still be tracked, but can be tracked separately from the virtual points-to assertion itself.
  % In practice the decisions about which virtual addresses are valid rest not with code posessing a virtual address, but with
  % the virtual memory manager --- and its invariants.

  \begin{figure}
\footnotesize
\vspace{-1em}
  \begin{mathpar}
  \inferrule*[right=\footnotesize GhostMapUpdate]{ }{\mathsf{GhostMap}(\gamma,\theta) \ast \ghostmaptoken{\gamma}{\mathsf{pa}}{\mathsf{va}} \vs 
  \mathsf{GhostMap}(\gamma,\theta[\mathsf{va}\mapsto\mathsf{pa'}]) \ast \ghostmaptoken{\gamma}{\mathsf{va}}{\mathsf{pa}'} 
  }
  \and
  \inferrule*[right=\footnotesize GhostMapLookup]{ }{\mathsf{GhostMap}(\gamma,\theta) \ast \ghostmaptoken{\gamma}{\mathsf{va}}{\mathsf{pa}} \wand \ulcorner \theta(\mathsf{va})=\mathsf{Some}(\mathsf{pa}) \urcorner }
  \end{mathpar}
\vspace{-2em}
  \caption{\iris rules for ghost maps}
  \label{fig:ghostmaps}
  \end{figure}

  We separate the physical page-table walk from the virtual points-to relation, replacing it with a ghost state that merely guarantees that the address translation would succeed.
  \iris includes a \emph{ghost map} construction, which we use to track mappings from virtual addresses to the physical addresses they translate to as a piece of ghost state.
  The map includes, for each key in the map (i.e.,
  each virtual address), a token $\ghostmaptoken{\gamma}{k}{v}$ whose ownership is required to update that key-value pair in the ghost map named $\gamma$. The existence of such a token implies that the actual map $\theta$ tracked by a corresponding $\mathsf{GhostMap}(\gamma,\theta)$
  resource indeed maps $k$ to $v$. These properties are captured by key \iris rules in Figure \ref{fig:ghostmaps}.\footnote{\iris ghost maps lack established notation\add{;}
   the syntax we use captures the details of \texttt{iris.base\_logic.lib.ghost\_map}.}
  There are other rules, but these two are most important for explaining ghost maps.
  \textsc{GhostMapUpdate} says that ownership of the actual ghost map with ghost name $\gamma$ and map contents $\theta$,
  and a token witnessing that $\theta$ maps \textsf{pa} to \textsf{va} permits an update to the ghost map's state,
  changing the map and replacing the token to represent the new value.
  \textsc{GhostMapLookup} allows using the same information to simply conclude that the mapping indicated by the token is true.
  
  The \emph{virtual memory manager's invariant} ensures that for each $\ghostmaptoken{\gamma}{\vaddr}{\paddr}$ mapping in this map, there are \emph{physical} resources sufficient to ensure that the address translation for $\vaddr$
will resolve on the hardware to $\paddr$ --- via $\textsf{L}_{4}\_\textsf{L}_{1}\_\textsf{PointsTo}$.
  \add{This kernel invariant turns out to be a key ingredient in supporting proofs of VMM functionality:
  in Section \ref{sec:experiment} we will see that separating the logical and physical virtual-to-physical mappings
  is what allows stating the global kernel invariants needed for software page traversals, which prior work did not (and could not) pursue.}
  \looseness=-1

  %Thus the specification of \emph{which} physical addresses support translation is separated from the virtual points-to.
  % But these physical resources, which specify \emph{which} physical locations
  %are involved in the page table walk, are now separated from but consistent with
  %the knowledge that such resources exist (which is embodied by the token for $va$, which tracks that $va$ maps to $pa$
  %in the ghost map). Thus we can store the token which summarizes the translation and ensures it exists in the virtual
  %points-to, and keep the ghost map and the invariant that every mapping in the ghost map has corresponding physical resources
  %for translation in a separate global invariant for each address space.

  % In Iris this is realized by using an authoritative resource algebra: there is a single \emph{authoritative} global copy of the (ghost)
  % map caching virtual-to-physical address translations, and for each entry a read only \emph{partial} ownership of that key-value pair.
  % The resource algebra itself is instantiated as:
  % \[\mathcal{A}\textsf{bsPTableWalk} \stackrel{\triangle}{=} \textsc{Auth} (\; \mathcal{W}_{64} \;\rightarrow_{\textrm{fin}} \;  ( (\textsc{Frac }, \mathord{+}) \times (\textsc{Agree } \Loc,\mathord{=}) ))\]
For clarity, we refer to the specific ghost map summarizing virtual-to-physical translations by 
\mbox{$\mathcal{A}\textsf{bsPTableWalk}(\delta,\theta) \stackrel{\triangle}{=} \mathsf{GhostMap}(\delta,\theta)$}
(omitting $\delta$ for brevity when only one is in scope)
and keep this in a per-address-space invariant described shortly.
We then replace the physical traversal $\textsf{L}_{4}\_\textsf{L}_{1}\_\textsf{PointsTo}$ in Figure \ref{fig:strongvirtualpointsto}'s virtual points-to definition
with ownership of the token \ghostmaptoken{\delta}{\vaddr}{\paddr}, %ghost-map ($ \sumwalkabs\vaddr\qfrac\paddr$),
yielding Figure \ref{fig:virtualpointstosharing}'s definition.
This new definition guarantees that the ghost map \add{$\theta$} maps the virtual address ($\vaddr$) to a physical address ($\paddr$),
and thus that the per-address-space invariant \add{described next} will contain the physical resources that guarantee \add{that} the hardware resolves the translation.
\looseness=-1

  \begin{figure*}
  \footnotesize
\vspace{-1em}
\centerline{$
\begin{array}{l}
   I\textsf{ASpace}(\ptablestore,m)\stackrel{\triangle}{=} \textsf{ASpace\_Lookup}(\ptablestore,m) \ast 
  \bigast{(\vaddr, \textsf{pa})\in \ptablestore}{\exists\;(\textsf{l4e l3e l2e, l1e, pa})\ldotp \textsf{L}_{4}\_\textsf{L}_{1}\_\textsf{PointsTo}(\vaddr\textsf{, l4e, l3e, l2e, l1e, pa})} \\
  \textsf{ where } 
   \textsf{ASpace\_Lookup} (\ptablestore,m) \stackrel{\triangle}{=} \lambda\textsf{ cr3val} \ldotp \; \exists \gammaPred \; \ldotp \ulcorner m \; !!\; \textsf{cr3val} = \textsf{Some } \gammaPred \urcorner \ast
    %\ownGhost\gammaPred{\authfull{\ptableabswalk\ptablestore}}
    \ptableabswalk{\delta,\theta}
\end{array}
$}
\vspace{-1em}
\caption{Per-address-space invariant with a fixed global map of address space names $m$}
  \label{fig:peraspaceinvariant}
\vspace{-1em}
  \end{figure*}

We place the authorative ownership of the ghost \add{map} translation $\mathcal{A}\textsf{PTableWalk}$ in a per-address-space invariant
$ I$\textsf{ASpace} (Figure \ref{fig:peraspaceinvariant}), 
{allowing} changes to the page tables
that preserve overall virtual-to-physical translations \add{in isolation},
\add{and also allowing changes to specific the virtual-to-physical translations}
when combined with the
{token} stored in the \add{relevant} virtual points-to (Figure \ref{fig:virtualpointstosharing}).
\looseness=-1

% \todo[inline,color=cyan]{Explain $\delta{s}$ (Iris ghost name) vs $m$ (logical contents of ghost map) in next 2 paragraphs}
% \todo[inline]{This next paragraph below explains details, but should first explain the big picture: the current address space
% is identified by a paddr, so to state the invariant for the address space named by that paddr we need to look up
% what invariants should hold, then assert that those invariants do hold.}
\add{We must also ensure that different address spaces can have independent ghost maps ---
which we resolve with an additional unique global ghost map (with ghost name $\delta{s}$) from address-space identifiers (page table roots
whose values are manipulated by the kernel code) to
the \textsc{Iris} ghost name for that address space's ghost map.
In Figure \ref{fig:virtualpointstosharing}, the extra ghost map token for $\delta{s}$ asserts that $\delta$ --- which is exisentially
quantified --- is the correct ghost name for the current address space. That is then the ghost map named
in the ghost virtual-to-physical translation token of Figure \ref{fig:virtualpointstosharing}.
}
\add{
Just as the ghost name $\delta$ names the ghost map with contents $\theta$,
$\delta{s}$ names a ghost map, whose contents appear as $m$ in Figure \ref{fig:peraspaceinvariant} (the association of $\delta{s}$ to $m$
is a global invariant not shown).
}
$ I\textsf{ASpace}(\theta,m)$ then \add{performs 3 roles:
it associates the current address space's root with an appropriate \textsc{Iris} ghost name $\delta$;
it tracks authoritatively that $\delta$'s logical contents match $\theta$; and it}
stores the physical resources for the current address space mappings \add{(via the iterated $\textsf{L}_{4}\_\textsf{L}_{1}\_\textsf{PointsTo}$)}.

\subsection{Address Space Management}
\label{sec:aspacemanagement}
% So far, we have introduced logical abstractions for a single address space, but VMMs
Real VMMs must
 handle more than one address space.
Doing so requires a way to talk about other address spaces, and means to switch address spaces.
\begin{figure}
\footnotesize
% \[\begin{array}{c}
\centerline{$
    \mbox{$[r](\mathsf{P}) \;: \mathsf{vProp} \; \Sigma \; \stackrel{\triangle}{=} \; \lambda \_,\; \mathsf{P}~\mathsf{r} \vspace{0.5em}$}
    \qquad
    \mbox{$\textsf{Fact P} \stackrel{\triangle}{=} \;\forall \; ,r \; r' \ldotp  \textsf{P r} \dashv\vdash \textsf{P r'}$}
    \qquad
    \inferrule*{ }{\textsf{Fact}\;[r](\textsf{P}) }
    \qquad
    \inferrule*{ }{\textsf{Fact}\;(\textsf{r}\;\mapsto_{\textsf{r}}\{q\}\;\textsf{rv})}
  $}
\vspace{0.5em}
\centerline{$
    \inferrule{ }{\textsf{Fact}\;(\textsf{w} \mapsto_{\textsf{p}} \{q\} \textsf{ v} )}
    \qquad
    \inferrule{ }{(\textsf{P} \vdash \textsf{Q}) \vdash  ([r](\textsf{P}) \vdash  [r](\textsf{Q}))}
    \qquad
    \inferrule{ }{[r](\textsf{P} \ast \textsf{Q}) \dashv\vdash ([r](\textsf{P}) \ast [r](\textsf{Q})}\vspace{0.5em}
  $}
\vspace{0.5em}
\centerline{$
    \inferrule{ }{[r](\textsf{P}\land \textsf{Q}) \dashv\vdash [r](\textsf{P})\land [r](\textsf{Q})}
    \qquad
    \inferrule{ }{[r](\textsf{P}\lor \textsf{Q}) \dashv\vdash [r](\textsf{P})\lor [r](\textsf{Q})}
    \qquad
    \inferrule{ }{\textsf{Fact P} \vdash  [r](\textsf{P}) \dashv\vdash (\textsf{P})} %(\textsf{vProp }\Sigma))\\
$}
% \end{array}\]
\vspace{-1em}
  \caption{Other-space Modality and Its Laws}
  \label{fig:modaldef}
\vspace{-1em}
  \end{figure}
Figure \ref{fig:modaldef} gives the definition of our modal operator for asserting the truth of a modal
(address-space-contingent) assertion \emph{in another address space}, which we call
the \emph{other-space} modality. The definition itself is not
particularly surprising --- as our modal assertions are semantically predicates on a page table root (physical)
address, the assertion $[r](P)$ is a modal assertion that ignores the (implicit) current page table root,
and evaluates the truth of $P$ as if $r$ were the page table root. 
The novelty here is not in the details of the definition, but in recognizing that this is the right way to deal with
multiple address spaces, and working out how to support interaction of multiple address spaces (discussed in the next section).
\add{The modal assertions, together with the other-space modality, mean we can give generic definitions
of data structure assertions (e.g., linked lists, etc.) which do not need to track information
about their own address space. In fact, \emph{only} assertions that explicitly deal with multiple address
spaces need to mention address spaces at all (via the other-space modality).
}

We can prove that this modality follows certain basic laws, showing that its truth is independent of the address
space in which it is considered, \add{that} it distributes over various logical connectives, and \add{that} it follows the rule of
consequence.
We call \textsf{vProp} assertions whose truth is independent of the current address space
\textsf{Fact}s; these include other-space assertions, physical memory points-tos, and register assertions.
\textsf{Fact}s can \add{freely} move in and out of other address space modalities.
\looseness=-1

In general, per-address-space invariants should be collected in a larger
VMM invariant, with individual address spaces' invariants pulled out as needed, such as when proving
soundness of an individual virtual memory access.
However, such larger invariants would contain many kernel-specific properties that are orthogonal
to the fundamental reasoning principles that are the focus of this paper.
We leave such kernel-specific reasoning to future work, but our verification of task switching
(Section \ref{sec:experiment}) demonstrates support for managing multiple address spaces.
\looseness=-1

\subsection{Selected Logical Rules}
\label{sec:selected_rules}
% Per the discussion in Section \ref{sec:issues}, w
As common for assembly-level verification~\add{\cite{Ni2006codeptrs,ni2007contexts}}, we define our logic using Hoare doubles:%\footnote{This
% omits some low-level Iris details (stuckness, observations) that play no meaningful
% role in our development.}
\\\centerline{$
  \begin{array}{l}
    %\textsf{wpd\_def e s E1 } \Phi \;\mathsf{ rtv } : \textsf{iProp }\Sigma := \\
    \{ \Phi \}_\mathsf{ rtv }\;\textsf{e} : \textsf{iProp }\Sigma := 
   % \qquad
   ((\textsf{cr3} \mapsto_{\textsf{r}} \textsf{rtv} \ast \Phi) \textsf{ rtv}) \wand \textsf{WP e } \{\_, \textsf{True} \}
    \end{array}
$}\\
Our Hoare doubles $\{\Phi\}_\textsf{rtv}\;\textsf{e}$ state that the expression (i.e., sequence of instructions)
\textsf{e} are safe to execute (will not fault)
when executed with \textsf{vProp} precondition $\Phi\ast\textsf{cr3}\mapsto_{\textsf{r}} \textsf{rtv}$.
\textsf{WP} is \iris's own weakest precondition modality, unmodified~\cite{jung2018iris}.
Making \textsf{rtv} a parameter to the double (vs.\ a simple register assertion)
makes it possible to ensure ownership of the \lstinline|cr3| register and its value is accounted for
while avoiding some technical headaches with trying to enforce that $\Phi$ itself contains that.
\looseness=-1
% solves a technical problem with ensuring that the page table root used to evaluate
% the \textsf{vProp} (i.e., evaluating the assertion in the \emph{current}) address space
% is feasible.
% \footnote{Consider the difficulty of selecting the correct page table root value from an arbitrary
% opaque $\Phi$, which may even existentially quantify the page table root. An alternative is to
% require $\Phi$ to have a syntactic form where we can directly extract the value of \lstinline|cr3|,
% but this makes using Iris Proof Mode (IPM)~\cite{Krebbers:2017:IPH:3009837.3009855} with \textsf{vProps}
%   difficult; IPM works for any type matching the signature of an Iris \lstinline|bi|, which includes
%   \textsf{vProp}s, but manually guiding IPM to put an assertion in a specific position over and over adds
%   significant proof burden.
% }

The rest of this section describes specifications of three key \textsf{AMD64} instructions 
in our logic. 
These rules and others (e.g., including accessing memory at an instruction-specified offset from a register
value, which is common in most ISAs)
can be found in our artifact.
% Each rule in Figure \ref{fig:wpdamd}, 
% is annotated with a root (i.e., \lstinline|cr3|) address value (\textsf{rtv}), 
% under which the resources mentioned in the specification are valid.
In general, we use metavariables $\textsf{r}_s$ and $\textsf{r}_d$ to specify source and destination registers
for each instruction, and prefix various register value variables with \textsf{rv}.
We sometimes use $\textsf{r}_a$ to emphasize when a register is expected to hold an address used
for memory access, though the figure also uses typical assembler conventions of specifying
memory access operands by bracketing the register holding the memory address.
Standard for Hoare doubles, there is a frame resource $P$ in each rule for passing resources
not used by the first instruction in sequence through to subsequent instructions.
Our rules include tracking of each instruction's memory address to track \lstinline|rip| updates, which is critical
for control transfer instructions. Our development also includes handling of the \lstinline|rflags| register updates from arithmetic instructions.
Most rules are otherwise standard (e.g., \lstinline|mov| between registers, etc.), with Figure \ref{fig:wpdamd} showing the rules
most unique to our development.
\add{As a reminder, in systems of Hoare doubles, an instruction's precondition appears in the conclusion of the rule,
and an instruction's ``postcondition'' appears as the precondition to subsequent instructions in the
antecedent of the rule.}
\looseness=-1

\begin{figure}
  \footnotesize
  \makeatletter % allow us to mention @-commands
\def\arcr{\@arraycr}
\makeatother
\begin{mathpar}
%  \inferrule[AddRegImm]{
%  \{P \ast r_d \mapsto_{r} \textsf{rvs} \ast r_s \mapsto_{r}\{q\} \textsf{ rvs} \}_{\textsf{rtv}}\;\overline{ is}
%}{
%  \{P \ast r_d \mapsto_{r} \textsf{rvd} \ast r_s \mapsto_{r}\{q\} \textsf{ rvs} \}_{\textsf{rtv}}
%  \textsf{ mov}~\textsf{r}_d,~\textsf{r}_s;\;\overline{is}
%  \lstinline|mov %cr3, r| 
%}
%\\
% \inferrule[WriteToPhysMemFromReg]{
%   \{P \ast r_s \mapsto_{r}\{q\}  \textsf{rvs}  \ast r_a \mapsto_{r} \{q\} \textsf{ vaddr} \ast \textsf{vaddr} \mapsto_{\textsf{t}} \textsf{v} \}_{\textsf{rtv}}\;\overline{ is}  
% }{
%   \{P \ast r_s \mapsto_{r}\{q\}  \textsf{rvs}   \ast r_a \mapsto_{r}\{q\} \textsf{ vaddr} \ast \textsf{vaddr} \mapsto_{\textsf{t}} \textsf{rvs} \}_{\textsf{rtv}}
%   \textsf{ mov}~[\textsf{r}_a],~\textsf{r}_s;\;\overline{is}
% }
% \\
 % \inferrule[AddRegImm]{
 %   \{P \ast\textsf{rip} \mapsto_{\textsf{r}} \textsf{iv+7} \ast  \ast (r_d \mapsto_{r}  \textsf{rvd+imm} \lor  r_d \mapsto_{r} 0 ) \ast r_a \mapsto_{r} \{q\} \textsf{ vaddr} \ast \textsf{vaddr} \mapsto_{\textsf{t}} \textsf{v} \}_{\textsf{rtv}}\;\overline{is}
 % }{
 %   \left\{
 %   \begin{array}{l}
 %   P \ast \textsf{rip} \mapsto_{\textsf{r}} \textsf{iv} \ast \textsf{rflags} \mapsto_{\textsf{r}} \textsf{flgs} \ast \ulcorner (\textsf{rvd + imm} > 0 \lor \textsf{rvd + imm} = 0 ) \urcorner \ast \arcr
 %   r_d \mapsto_{r}  \textsf{rvd} \ast r_a \mapsto_{r} \{q\} \textsf{ vaddr} \ast \textsf{vaddr} \mapsto_{\textsf{t}} \textsf{v}
 % \end{array}\right\}_{\textsf{rtv}}  \;
 %   \textsf{add}~\textsf{r}_d,~[\textsf{imm}];\;\overline{is}
 % }
 % \\
\inferrule[\footnotesize WriteToRegFromVirtMem]{
  \left\{ \begin{array}{l}
    P \ast  I\textsf{ASpace}\ast  \textsf{rip} \mapsto_{\textsf{r}} \; {\textsf{iv} + \textsf{MovLen}(\textsf{r}_d,[\textsf{r}_a])} \ast \arcr
    r_d \mapsto_{r}  \textsf{ v} \ast r_a \mapsto_{r} \; \{q\} \textsf{ vaddr} \ast \textsf{vaddr} \mapsto_{\textsf{v}} \;\{q\} \textsf{ v}
  \end{array} \right\}_{\textsf{rtv}}\;\overline{is}
}{
  \{P \ast  I\textsf{ASpace}\ast \textsf{rip} \mapsto_{\textsf{r}} \textsf{ iv} \ast r_d \mapsto_{r}  \textsf{ rvd} \ast r_a \mapsto_{r} \;\{q\} \textsf{ vaddr} \ast \textsf{vaddr} \mapsto_{\textsf{v}} \textsf{ v} \}_{\textsf{rtv}}\;
  \textsf{mov}~\textsf{r}_d,~[\textsf{r}_a];\;\overline{is}
}
\\
\inferrule[\footnotesize WriteToVirtMemFromReg]{
  \{P \ast  I\textsf{ASpace}\ast \textsf{rip} \mapsto_{\textsf{r}} \textsf{ iv} + \textsf{MovLen}(\textsf{r}_d,[\textsf{r}_a])  \ast r_s \mapsto_{r}\;\{q_1\}  \textsf{ rvs}  \ast r_a \mapsto_{r} \; \{q_2\} \textsf{ vaddr} \ast \textsf{vaddr} \mapsto_{\textsf{v}} \textsf{ v} \}_{\textsf{rtv}}\;\overline{ is}  
}{
  \{P \ast  I\textsf{ASpace}\ast \textsf{rip} \mapsto_{\textsf{r}} \textsf{ iv} \ast r_s \mapsto_{r} \;\{q_1\}  \textsf{ rvs}   \ast r_a \mapsto_{r} \; \{q_2\} \textsf{ vaddr} \ast \textsf{vaddr} \mapsto_{\textsf{v}} \textsf{ rvs} \}_{\textsf{rtv}}
  \textsf{ mov}~[\textsf{r}_a],~\textsf{r}_s;\;\overline{is}
}
\\
%\inferrule[WriteToRegCtlFromReg]{
%  \{P \ast r_s \mapsto_{r}\{q\}  \textsf{ rvs}  \}_{\textsf{rvs}} \overline{is}
%}{
%  \{P \ast r_s \mapsto_{r}\{q\}  \textsf{ rvs}   \}_{\textsf{rtv}}
%  \textsf{mov}~\textsf{cr3},~r_s;\;\overline{is}
  %\lstinline|mov %cr3, r| 
%}\\
\inferrule[\footnotesize WriteToRegCtlFromRegModal]{
  \{[\textsf{rtv}](P \ast  I\textsf{ASpace})\ast \textsf{rip} \mapsto_{\textsf{r}} \textsf{ iv + 4} \ast  I\textsf{ASpace} \ast R\ast r_s \mapsto_{r} \;\{q\}  \textsf{ rvs}  \}_{\textsf{rvs}} \;\overline{is}
}{
  \{P \ast  I\textsf{ASpace} \ast \textsf{rip} \mapsto_{\textsf{r}} \textsf{ iv} \ast [\textsf{rvs}](R \ast I\textsf{ASpace})\ast r_s \mapsto_{r}\; \{q\}  \textsf{ rvs}   \}_{\textsf{rtv}} \;
  \textsf{mov}~\textsf{cr3},~r_s;\;\overline{is}
  %\lstinline|mov %cr3, r| 
}
% \\
% \inferrule[WriteToRegFromRegCtl]{
%   \{P \ast  \textsf{rip} \mapsto_{\textsf{r}} \textsf{iv + 4} \ast r_d \mapsto_{r} \textsf{rvs} \ast r_s \mapsto_{r}\{q\} \textsf{ rvs} \}_{\textsf{rtv}}\;\overline{ is}
% }{
%   \{P \ast  \textsf{rip} \mapsto_{\textsf{r}} \textsf{iv} \ast r_d \mapsto_{r} \textsf{rvd} \ast r_s \mapsto_{r}\{q\} \textsf{ rvs} \}_{\textsf{rtv}}
%   \textsf{ mov}~\textsf{r}_d,~\textsf{r}_s;\;\overline{is}
%   %\lstinline|mov %cr3, r| 
% }
\end{mathpar}
\vspace{-1em}
\caption{Proof rules for selected \textsf{AMD64} instructions}
\label{fig:wpdamd}
\vspace{-1em}
\end{figure}

\subsubsection{Accessing Virtual Addresses}
Figure \ref{fig:wpdamd} includes two  rules for accessing memory at an address stored in a register $r_a$. 
\add{Setting aside $P$, $ I\textsf{ASpace}$, and the instruction pointer \lstinline|rip|,}
\TirNameStyle{WriteToRegFromVirtMem} and \TirNameStyle{WriteToVirtMemFromReg}
are nearly-standard (assembly) separation logic rules for memory accesses~\add{\cite{Chlipala2013Bedrock,ni2007contexts}}.
\add{For example, \textsc{WriteToRegFromVirtMem}'s specification
reflects that it reads from the (virtual) memory address \textsf{vaddr} stored in the address
register $\textsf{r}_a$ --- and thus requires register points-to and virtual points-to assertions describing
that relationship and the assumed value \textsf{v} in memory in its precondition (the precondition of the rule's
conclusion). 
It reflects the load (\lstinline|mov|) of that memory value into the destination register $\textsf{r}_d$, with
the updated register points-to in the precondition for $\overline{is}$.
$P$ describes framed resources, which are passed along to the precondition of subsequent instructions,
as in any system of Hoare doubles~\cite{Chlipala2013Bedrock,ni2007contexts}.
\textsc{WriteToVirtMemFromReg} is analogous for writing to memory.
}
\add{There are only two changes specific to our approach.}
\looseness=-1

First,
because we split the physical resources for the page table walk from the
virtual points-to itself ({per the discussion of Section \ref{sec:sharingpages}}), the rule requires $ I\textsf{ASpace}$
for the current address space to be carried through.
The soundness proofs for these rules extract
the token ($\fracghostmaptoken{\delta}{\vaddr}{\paddr}{\qfrac}$) from the virtual points-to,
use that to extract the physical page-table-traversal points-to collection describing
the page table walk for the relevant address ($\textsf{L}_{4}\_\textsf{L}_{1}\_\textsf{PointsTo}$)
from the invariant ($ I\mathsf{ASpace}$), prove that the page table walk succeeds
and that memory or registers are updated appropriately, before re-packing the invariant and virtual points-to resources.
\looseness=-1

Second, the memory access rules --- as with all rules in our logic ---
{increment} the instruction pointer \lstinline|rip| \add{by the length of the encoded instruction.}
\textsf{MovLen} returns how long the instruction encoding for the corresponding \lstinline|mov| is;
x86-64 instruction encodings are often longer for instructions using registers that are absent from the 32-bit
x86 ISA that preceded x86-64.
\looseness=-1

\add{Note that the use of a modal abstraction of address space simplifies these rules.
The antecedents of \textsc{WriteToRegFromVirtMem} and \textsc{WriteToVirtMemFromReg}
only mention the address space in the index of the
Hoare double --- not in $P$, or the (virtual) points-to assertions.
There is no extra condition to discharge that the address being accessed is from
the current address space.
}

\subsubsection{Updating \lstinline|cr3|} 
Unlike other rules, \TirNameStyle{WriteToRegCtlFromRegModal} updates the root address of the 
address space determining the validity of resources, from $\rtv$ before the
\lstinline|mov| to $\textsf{rvs}$ afterwards. The global effects of this rule are reflected in moving
assertions \add{for the current address space ($P$ and $ I\mathsf{ASpace}$)} under an other-space modality for \add{the initial
(outgoing) address space} $\rtv$, and moving the new address space's assertions out of
the corresponding modality \add{(since after the \lstinline|mov|, those will hold in the
new current address space)}.
The \emph{global} aspect is important. A na\"ive frame rule would be unsound for \lstinline|cr3| updates:
one could frame out assertions in one address space, switch address spaces, and bring those assertions from the \emph{old}
address space back into the \emph{new} address space, where they may not hold. 
% Appendix \ref{sec:issues} gives more details.
It is often said that the frame rule (below) is one of the key pieces of separation logic.
\centerline{$
  \mbox{$\inferrule*[right=Frame]{
    \{P\}\;C\;\{Q\}
  }{
    \{P\ast R\}\;C\;\{Q\ast R\}
  }$}
  \qquad\begin{array}{c}\textrm{(unsound with address space changes)}\\\\\end{array}
$}\\
Such a rule is normally recoverable from Hoare doubles (see, e.g., \citet{Chlipala2011Bedrock,Chlipala2013Bedrock}).
However, in the presence of address space changes, the traditional frame rule is unsound.
Consider:\\
\centerline{$
  \inferrule*[right=Frame]{
    \inferrule{\ldots }{
    \{\textsf{Pre}\}
    \texttt{mov}~\texttt{\%cr3},~r%\lstinline|mov %cr3, r| 
    \{\textsf{Post}\}
    }
  }{
    \{a\mapsto_\mathsf{v} x \ast \textsf{Pre}\}
    \texttt{mov}~\texttt{\%cr3},~r%\lstinline|mov %cr3, r| 
    \{a\mapsto_\mathsf{v} x \ast \textsf{Post}\}
  }
$}\\
In this (broken!) hypothetical example,
both the precondition and the postcondition assert that $a\mapsto_\mathsf{v} x$ in the current address space, but
  the new address space may map $a$ to another value. So, this derivation clearly leads to an unsound conclusion. 
The essential problem is that the frame rule is motivated by local reasoning about local updates, but
a switch of address space is a \emph{global} change that may invalidate information about virtual addresses.
Thus, framing around arbitrary \lstinline|cr3| updates is unsound --- hence the \emph{global} nature of \textsc{WriteToRegCtlFromRegModal} ---
though a variant the ensures the same \lstinline|cr3| value is installed before and after the framing
can be recovered.

\subsection{Soundness}
\label{sec:soundness}
Our rules from Figure \ref{fig:wpdamd} are proven to be sound in \iris against an assembly-level hardware model
implementing a fragment of x86-64, including 64-bit address translation with 4-level page tables.
% \todo[inline]{
Our rules for control transfers (\lstinline|jne|, \lstinline|call|, and \lstinline|ret|) are currently
axiomatized (with completely standard specifications~\cite{ni2007contexts,Chlipala2013Bedrock})\footnote{
  \add{An assertion that code at some address is safe to call with a given precondition~\cite{Ni2006codeptrs}
  asserts that the address is mapped and that memory at that address decodes to an instruction sequence
  that is safe with that precondition.
  }
} 
because \iris's built-in machinery does not provide
convenient ways to discard the current continuation; adaptation of others'
approaches~\cite{de2023type} is future work.
Our soundness proofs for all other instructions (including, critically, all memory accesses)
are axiom-free.
% }
% At the moment our proofs do rely on 13 small axioms of properties which should be provable, but
% are challenging to discharge due to some representation choices in our model.\footnote{See \lstinline|srx/x64/machine/current_axioms.v|}
\looseness=-1

%\input{soundness.tex}
%$\assert{\ulcorner \mathsf{aligned maddr} \urcorner \ast \mathsf{r14} \mapsto_{\textsf{r}} \textsf{r14v} \ast \mathsf{r13} \mapsto_{\textsf{r}} \textsf{maddr} \ast \mathsf{rdi} \mapsto_{\textsf{r}} \textsf{rdiv} \ast \mathsf{rax} \mapsto_{\textsf{r}} \textsf{raxv}}$
%$\assert{\ulcorner \ptablestore !! \textsf{maddr} = \textsf{None} \urcorner \ast \ownGhost\gammaPred{\authfull{\ptableabswalk\ptablestore}} \ast \textsf{Pf}} $
\definecolor{main-color}{rgb}{0.6627, 0.7176, 0.7764}
\definecolor{back-color}{rgb}{0.1686, 0.1686, 0.1686}
\definecolor{string-color}{rgb}{0.3333, 0.5254, 0.345}
\definecolor{key-color}{rgb}{0.8, 0.47, 0.196}

\newcommand{\sumwalkabsent}{
  \ownGhost\gammaPred{\authfrag{\singletonMap{\texttt{entry+KERNBASE}}{(\textsf{qfrac}, \textsf{entry})}}}
}

\newcommand{\ventry}{\texttt{entry + KERNBASE}}
\newcommand{\entry}{\texttt{entry}}
\newcommand{\qfraczero}{\textsf{qfrac}}
\newcommand{\true}{\textsf{true}}
\tikzstyle{boxedassert_border} = [sharp corners,line width=0.2pt]
\NewDocumentCommand \boxedassertpv {O{} m o}{%
	\tikz[baseline=(m.base)]{
		%	  \node[rectangle, draw,inner sep=0.8pt,anchor=base,#1] (m) {${#2}\mathstrut$};
		\node[rectangle,inner sep=1.5pt,outer sep=0.2pt,anchor=base] (m) {${\,#2\,}\mathstrut$};
		\draw[#1,boxedassert_border] ($(m.south west)$) rectangle ($(m.north east)$);
	}\IfNoValueF{#3}{^{\,#3}}%
}
\newcommand*{\knowInvpv}[2]{\boxedassertpv{#2}[#1]}
\newcommand*{\ownGhostpv}[2]{\boxedassertpv[dash dot]{#2}[#1]}

\newcommand{\sumpv}[3]{
  \ownGhostpv\gammaPred{\authfrag{\singletonMap{#1}{(#2, #3)}}}
}

\newcommand{\pvmapping}[1]{\mathcal{A}\textsf{P2VMappings}(#1)}

\newcommand{\fpaddr}{\texttt{fpaddr}}
\newcommand{\specline}[1]{{\color{blue}\left\{#1\right\}}}
\newcommand{\sumapacesfull}[2]{
  \ownGhost\gammaPreds{\authfull{\singletonMap{#1}{#2}}}
}
\section{Experiments}
\label{sec:experiment}
%To both validate and demonstrate the value of the modal approach to reasoning about virtual memory management, 
% we study several
% We validate our logic by studying
% distillations of key VMM functionality.
% real concerns of virtual memory managers.
% Recall from Section \ref{sec:logic} that virtual points-to assertions work just like regular points-to assertions, by design.
In this section, we verify several critical and challenging pieces of VMM code.
First, we formally verify a switch into a new address space as part of a task switch,
the first such verification handling both old and new processes' assertions (in different address spaces) at the time of the switch.
Then, in several stages, we work up to mapping a new page in the current address space, addressing significantly more of this process than prior
work that included address translation in its hardware model.
This requires a number of independently challenging substeps: dynamically traversing a page table to find
the appropriate L1 entry to update; inserting additional levels of the page table if necessary (updating
the VMM invariants along the way);
converting the physical addresses found in intermediate entries into the corresponding virtual addresses
that can be used for memory access;
installing the new mapping;
and collecting sufficient resources to form a virtual points-to assertion.
Of these, only the second-to-last step (installing the correct mapping into the
current address space) has previously been formally verified with respect to a machine model with address translation.

\begin{figure}\footnotesize
\begin{lstlisting}[mathescape,escapeinside={(*}{*)}]
$\specline{\textsf{P} \ast  I\texttt{ASpace}(\theta,\Xi,m) \ast [\rtv']( I\texttt{ASpace}(\theta',\Xi',m') \ast \texttt{Pother}) \ast \texttt{rsi}\mapsto_{\textsf{r}} restore \ast \texttt{rdi}\mapsto_{\textsf{r}} save\ast \texttt{rbx}\mapsto_{\textsf{r}} \texttt{rbxv} }_{\rtv}$(*\label{line:start_swtch_pre}*)
$\specline{\texttt{rsp}\mapsto_{\textsf{r}} \texttt{rspv} \ast \texttt{rbp}\mapsto_{\textsf{r}} \texttt{rbpv} \ast \texttt{r12}\mapsto_{\textsf{r}} \texttt{r12v} \ast \texttt{r13}\mapsto_{\textsf{r}} \texttt{r13v} \ast \texttt{r14}\mapsto_{\textsf{r}} \texttt{r14v} \ast \texttt{r15}\mapsto_{\textsf{r}} \texttt{r15v}}_{\rtv}$
$\specline{\mathsf{ContextAt}(save,\_) \ast \mathsf{ContextAt}(restore,[\mathsf{rbxv}',\ldots,\rtv']) }_{\rtv}$(*\label{line:end_swtch_pre}*)
mov 0[rdi], rbx(*\label{line:start_save}*) ... ;; also save rsp, rbp, r12, r13, r14, to offsets 8, 16, ... and 40 from rdi
... mov 48[rdi], r15
$\specline{\textsf{P} \ast  I\texttt{ASpace}(\theta,\Xi,m) \ast [\rtv']( I\texttt{ASpace}(\theta',\Xi',m') \ast \texttt{Pother})  \ast  \texttt{rsi}\mapsto_{\textsf{r}} restore \ast \texttt{rdi}\mapsto_{\textsf{r}} save \ast \texttt{rbx}\mapsto_{\textsf{r}} \texttt{rbxv} }_{\rtv}$
$\specline{ \texttt{rsp}\mapsto_{\textsf{r}} \texttt{rspv} \ast \texttt{rbp}\mapsto_{\textsf{r}} \texttt{rbpv} \ast \texttt{r12}\mapsto_{\textsf{r}} \texttt{r12v} \ast \texttt{r13}\mapsto_{\textsf{r}} \texttt{r13v} \ast \texttt{r14}\mapsto_{\textsf{r}} \texttt{r14v} \ast \texttt{r15}\mapsto_{\textsf{r}} \texttt{r15v}}_{\rtv}$
$\specline{\mathsf{ContextAt}(save,[\mathsf{rbxv,\ldots,\rtv}]) \ast \mathsf{ContextAt}(restore,[\mathsf{rbxv}',\ldots,\rtv']) }_{\rtv}$
mov 56[%rdi], %cr3 $\specline{\ldots \ast \texttt{rdi+56} \mapsto_{\textsf{v}} \rtv \ast \ldots }_{\rtv}$   (*\label{line:end_save}*) 
mov rbx, 0[rsi] (*\label{line:start_restore}*)
mov rsp, 8[rsi] ;; Switch to new stack, which may not be mapped in the current address space!(*\label{line:stack_switch}*)
... ;; load rbp, r12, r13, r14, from offsets 16, 24, 32, and 40 from rsi
mov r15, 48[rsi]
$\specline{\textsf{P} \ast  I\texttt{ASpace}(\theta,\Xi,m) \ast [\rtv']( I\texttt{ASpace}(\theta',\Xi',m') \ast \texttt{Pother})  \ast  \texttt{rsi}\mapsto_{\textsf{r}} restore \ast \texttt{rdi}\mapsto_{\textsf{r}} save \ast \texttt{rbx}\mapsto_{\textsf{r}} \texttt{rbxv}' }_{\rtv}$
$\specline{\texttt{rsp}\mapsto_{\textsf{r}} \texttt{rspv}' \ast \texttt{rbp}\mapsto_{\textsf{r}} \texttt{rbpv}' \ast \texttt{r12}\mapsto_{\textsf{r}} \texttt{r12v}' \ast \texttt{r13}\mapsto_{\textsf{r}} \texttt{r13v}' \ast \texttt{r14}\mapsto_{\textsf{r}} \texttt{r14v}' \ast \texttt{r15}\mapsto_{\textsf{r}} \texttt{r15v}'}_{\rtv}$
$\specline{\mathsf{ContextAt}(save,[\mathsf{rbxv,\ldots,\rtv}]) \ast \mathsf{ContextAt}(restore,[\mathsf{rbxv}',\ldots,\rtv']) }_{\rtv}$
mov cr3, 56[rsi] ;; <-- Switch to the new address space (*\label{line:end_restore}*)
$\specline{ [\rtv](\textsf{P} \ast  I\texttt{ASpace}(\theta,\Xi,m) \ast \mathsf{ContextAt}(save,[\mathsf{rbxv,\ldots,\rtv}]) \ast \mathsf{ContextAt}(restore,[\mathsf{rbxv}',\ldots,\rtv']))  }_{\rtv}$ (*\label{line:modality_switchC}*)
$\specline{  I\texttt{ASpace}(\theta',\Xi',m') \ast \texttt{Pother} \ast  \texttt{rsi}\mapsto_{\textsf{r}} restore \ast \texttt{rdi}\mapsto_{\textsf{r}} save \ast \ldots }_{\rtv'}$
\end{lstlisting}
\vspace{-1em}
\caption{Basic task switch code that switches address spaces.}
\label{fig:swtchC}
\end{figure}

\subsection{Change of Address Space}
A critical piece of \emph{trusted} code in current verified OS kernels is the assembly code to change the current address space; current verified OS kernels currently
lack effective ways to specify and reason about this low-level operation, for reasons outlined in Section \ref{sec:relwork}.

Figure \ref{fig:swtchC} gives simplified code for a basic task switch, the heart of an OS scheduler implementation. This is code that saves the context (registers and stack)
of the running thread, and resumes execution of a previously-suspended thread of execution.
In C this code would be given the signature
\mbox{\lstinline[language=C]|void swtch(context_t* save, context_t* restore)|}.
Saving the context is a straightforward matter of storing each register into the \lstinline|save| context.
Restoring the \lstinline|restore| context is the tricky bit, because both the stack pointer and address space must be restored.
Confusingly, a single dynamic execution of this function begins execution in one thread,
and returns in another thread --- because the execution switches stacks, and thus returns on the second thread's stack.%\footnote{This is the function in UNIX 6th Edition 
% with the infamous ``You are not expected to understand this'' comment~\cite{lions1996lions},
% though Doeppner~\cite{doeppner2010operating} and others, offer detailed explanations.}
Hence this is used, for example, when the scheduler has chosen a new thread to execute for a voluntary (non-preempted)
context switch, and will call
the code with \lstinline|save| pointing to a reserved storage space for the current thread, and
\lstinline|restore| pointing to the context of the next thread to execute.
We will ignore non-integer registers; others are handled similarly.
In Figure \ref{fig:swtchC}, Lines \ref{line:start_save}--\ref{line:end_save} store the callee-save registers (per the System V AMD64 ABI calling conventions) of the calling
thread into the context data structure pointed to by \lstinline|rdi| (at virtual address $save$).
This is justified by the $\textsf{ContextAt}(save,\_)$ assertion in the precondition (Line \ref{line:end_swtch_pre}), which expands
into a full set of full-permission (thus writable) virtual points-to assertions for various offsets from $save$, one for each saved register.
Verification up through line \ref{line:end_save} is straightforward application of \textsc{WriteToVirtMemFromReg} (Figure \ref{fig:wpdamd}).

The code to restore the previously-saved context located at $restore$ (accessed via \lstinline|rsi|) in Lines
\ref{line:start_restore}--\ref{line:end_restore}
is where the proof becomes subtle, though our logic makes the construction of the proof feel similar to typical assembly-level verification
because most instructions are verified with rules that work very similarly to standard proof rules while being proven
sound against a machine model with address translation.
Similar to the precondition for the save context, the restore context has a corresponding $\mathsf{ContextAt}(restore,[\ldots])$
assertion expanding to virtual points-to assertions --- in the caller's adddress space.
The \lstinline|mov| instructions prior to Line \ref{line:end_restore} are each verified with a fixed-register-offset
variant of \textsc{WriteToRegFromVirtMem}, but
Line \ref{line:stack_switch}'s implications are subtle because it switches stacks by updating \lstinline|rsp|.
Because the new stack pointer may only be valid in the address space of the restored context, stack accesses at this point are unsafe.
Prior to Line \ref{line:end_restore}, we can see in code and invariants that the local registers are updated
to match the values populating the restore context, except for the page table root.
Line \ref{line:end_restore} itself is verified with a rule similar to \textsc{WriteToRegCtlFromRegModal} (but obviously
reading from a fixed offset of a register, as needed in Line \ref{line:end_restore}).
This rule also globally moves assertions into and out-of other-space assertions, to reflect that
assertions holding in the outgoing address space \rtv{} generally will not hold in the incoming address space $\rtv'$
and vice versa. Thus the precondition has assertions for the new thread under an other-space modality for the new address space,
and the postcondition has assertions for the old thread under an other-space modality for the old address space.%\footnote{
  % The \textsf{ContextAt} assertions \emph{both} end up under the other-space modality for the old thread.
  % A real kernel would want to transfer both out, but justifying this is highly kernel-specific,
  % and particularly post-Spectre-and-Meltdown is quite varied.
  %\looseness=-1
%}
Both \textsf{ContextAt} assertions end up under the old other-space modality, but in a real kernel would
transfer out based on kernel-specific invariants.
\looseness=-1

The specification above does not directly discuss the relationship between instruction pointers and registers --- and does not need to
because \textsf{P} and \textsf{POther} can be instantiated to capture that relationship with additional information about stack contents.
This code is meant to be called with a return address for the current thread stored on the current stack,
and a return address for the target thread on the target thread's stack.
% But the target thread's precondition is \emph{relative to its address space}, 
% not the address space of the calling thread! This is reflected by
% the other-space modality
% $[\rtv']( I\texttt{ASpace}(\theta,\Xi,m) \ast \texttt{Pother})$
% in the specfication. 
For a given call site, \textsf{P} would be instantiated to require that the initial stack pointer (before \lstinline|rsp| is updated)
has a return address expecting the then-current callee-save register values in the \emph{current} (initial) address space
to (together with other resources used to instantiate \textsf{P}) imply the precondition of the code at that return address.
The situation for the target thread is similar, but using \textsf{POther}, \emph{and using the other-space modality}
because the other thread's stack, code, and other relevant assertions may only be valid in the new address space.
Our logic's rules for updating the page table root, and thus moving assertions into and out of other-space modalities,
neatly manage which assertions are \emph{currently} valid, without the need to explicitly plumb address space labels through
every assertion in the larger proof.

% Immediately after the page table switch, assertions about the saved and restored contexts are
% guarded by a modality for the retiring
% address space \rtv{} (Line \ref{line:modality_switch}), per
% \textsc{WriteToRegCtlFromRegModal} (Figure \ref{fig:wpdamd}),
% because
% there is no guarantee that the data structures of the previous address space are mapped in the new address space.
% The ability to transfer that points-to information out of that modality is specific to a given kernel's design. 
% Kernels that map kernel memory into all address spaces would need invariants
% that justified moving those assertions out of the other-space modality.
% % Following Spectre and Meltdown, this kernel design became less prevalent because speculative execution of accesses to kernel addresses could leak information even if the access did eventually cause a fault (the user/kernel mode permission check was done after fetching data from memory). Thus many modern kernels have reverted to the older kernel design where the kernel inhabits its own unique address space, and user processes have only enough extra material mapped in their address spaces to switch into the kernel (CPUs do not speculate past updates to \texttt{cr3}).
% \looseness=-1

Although prior work has verified context switches within a single address space~\cite{ni2007contexts}, and address space switches
without any code before or after~\cite{syeda2020formal} (that is, not reasoning about the \emph{impact} of address space change
on what data were accessible), this is the first verification that handles both.
\looseness=-1

\subsection{Traversing Live Page Tables}
\label{sec:traversingC}
We build up to the main task of mapping a new page after traversing the page tables in the software.
This algorithm is complex and corresponds to a significant amount of assembly code.
To assist with readability, we present C code for this process, with assertions adjusted slightly to refer to
C program variables rather than registers. The actual verification was carried out on x86-64 assembly
\emph{generated from this source code}.
Listings of the assembly fragments with inline assertions appear in
\ifARXIV
Appendix \ref{sec:experiment_appendix}.
\else
our technical report~\cite{kuru2025modal}.
\fi
Whether in C or assembly,
the page table traversal involved in mapping a new page is very challenging functionality to verify.
Loading the current table root from \lstinline|cr3| is straightforward (a \lstinline|mov| instruction).
However, this produces the \emph{physical} address stored in \lstinline|cr3|, not a \emph{virtual} address the kernel code can use to access that memory.
This problem repeats at each level of the page table: assuming that the code has \emph{somehow} read the appropriate L4 (or L3, or L2) entry, those entries again
yield physical addresses, not virtual.
The only prior work to verify page mapping ignored the traversal and only verified mapping
assuming code \emph{already} had an appropriate virtual address for the L1 entry, where a physical
address could simply be stored. Our proof is the first to additionally deal with the critical code
leading up to that point.

\paragraph{Code Overview}
As described in Section \ref{sec:background}, mapping a new page consists of 
simulating the hardware address translation of Figure \ref{fig:pagetables}, but in software.
The code for this task takes three explicit parameters:
the root pointer (read from \lstinline|cr3| by earlier code),
the page-aligned \emph{virtual} address (\lstinline|va|) at which to make a new piece of memory accessible,
and the \emph{physical} address (\lstinline|fpaddr|) of the memory to map in that location.
The function we ultimately verify, \lstinline|vaspace_mappage| (Figure \ref{fig:mapping_codeC}),
relies primarily on a helper function already shown in Figure \ref{fig:pagetablescode}.
\lstinline|walkpgdir| finds the (virtual) address of the the correct L1 entry to translate \lstinline|va|,
by walking the page tables in software one level at a time.
\lstinline|vaspace_mappage| then uses the result to install the new entry.
\lstinline|walkpgdir| itself relies on its own helper function \lstinline|pte_get_next_table|, also shown in Figure \ref{fig:pagetablescode},
which implements a single-level of traversal from level $n+1$ to level $n$ (and whose specification and proof are therefore
parameterized by page table level), allocating additional levels as needed.

We organize our explanation of the proofs by essentially following execution from the start of \lstinline|walkpgdir|, through
execution of \lstinline|pte_get_next_table|, and out to its callsite in \lstinline|vaspace_mappage|.
While slightly awkward because we start in the middle of the mapping execution,
this ordering allows us to start with the simpler pieces of the proof, and incrementally explain the complexities
of the proofs and kernel invariant, before concluding with the top-level verification.

\begin{figure}\footnotesize
\ifPLDI
\begin{lstlisting}[language=C,mathescape,escapeinside={(*}{*)}]
/* @param entry: virtual address of level n+1 entry which should point to a level-n table base
   @return next: a virtual pointer to the base of a valid level n table */
pte_t *pte_get_next_table(pte_t *entry) {
  pte_t *next;
  $\specline{\textsf{P} \ast  I\texttt{ASpace}_{\textsf{id}}(\theta,\Xi\setminus\{\textsf{entryp}\mapsto\textsf{v}\},m) \ast \ghostmaptoken{\delta{}s}{\rtv}{\delta} \ast \mathsf{entry} \mapsto_r \mathsf{entryp+KERNBASE} \ast \ghostmaptoken{\textsf{id}}{(\mathsf{entryp})}{\textsf{level}}}_{\rtv}$(*\label{line:precondition_entry_out}*)
  $\specline{ \textsf{entryp+KERNBASE} \mapsto_{\textsf{vpte,qfrac}} \textsf{entryp entryv} \ast \ulcorner  \textsf{qfrac} = 1 \leftrightarrow \; \lnot(\textsf{entry\_present entryv})\urcorner}_{\rtv}$ (*\label{line:get_next_vpte_preconditionC}*)
  $\specline{\ulcorner\textsf{entry\_present(entryv)}\urcorner \wand \forall_{i\in\textsf{0..511}} \; \ghostmaptoken{\textsf{id}}{\textsf{table\_root(entryv.pfn) + i * 8}}{\textsf{v-1}} }_{\rtv} $(*\label{line:precondition_conditional}*) 
  /* Reading the page table is entry justified by the virtual pte-points-to */
  if (!entry->present(*\label{line:read_entry_contentsC}*)(*\label{line:conditional_childrenC}*)(*\label{line:check_entry_present_jumpC}*)) {// If the present bit is zero, need to allocate next level (* \label{line:check_entry_presentC} *)(* \label{line:mask_presentC} *) 
    (*\label{line:refined_fractional_ownership}*)(*\label{line:pass_addrof_nextC}*)(*\label{line:alloc_path_startC}*)$\specline{ \textsf{entryp+KERNBASE} \mapsto_{\textsf{vpte,qfrac}} \textsf{entryp entryv} \ast \ulcorner  \textsf{qfrac} = 1 \land \lnot(\textsf{entry\_present entryv})  \urcorner }_{\rtv}$ (*\label{line:after_concluding_qfrac1C}*)
    pte_initialize(entry); // Allocate a zeroed physical page and install in entry(*\label{line:call_to_pte_initializeC}\label{line:page_of_capsC}*)
    entry->present = 1; // Install newly-allocated level n table, mark present(*\label{line:install_new_entryC}*)
    /*Now we know that entry is initialized, and the condition to access children list holds*/(*\label{line:now_we_know}*)
    $\specline{ \textsf{entryp+KERNBASE} \mapsto_{\textsf{vpte,qfrac}}  \textsf{entryp} \; \textsf{pte\_initialized(pfn\_set(entryv nextpaddr))} \ast \ulcorner  \textsf{qfrac} = 1 \urcorner }_{\rtv}$
    $\specline{\forall_{i\in \textsf{0 ... 511} } \ldotp  \ghostmaptoken{\textsf{id}}{((\textsf{table\_root (pte\_initialized (pfn\_set(entryv nextpaddr)))}) + \textsf{i * 8})}{\textsf{v-1}}  }$(*\label{line:available_child_tokens}*) (*\label{line:alloc_path_endC}*)
  } (*\label{line:end_of_allocation_pathC}*)
  $\specline{\textsf{P} \ast  I\texttt{ASpace}_{\textsf{id}}(\theta,\Xi\setminus \{ \mathsf{entryp}\mapsto\textsf{v} \}),m) \ast \textsf{entryp} \mapsto_{\textsf{id}} \textsf{\_} \ast \ghostmaptoken{\delta{}s}{\rtv}{\delta}  }_{\rtv}$
  $\specline{ \textsf{entryp+KERNBASE} \mapsto_{\textsf{vpte,qfrac}} \textsf{entryp} \; \textsf{(pte\_initialized (entryv.pfn))} \urcorner }_{\rtv}$
  $\specline{\forall_{i\in \textsf{0 ... 511} } \ldotp  \ghostmaptoken{\textsf{id}}{((\textsf{table\_root (pte\_initialized (entryv.pfn)))}) + \textsf{i * 8})}{\textsf{v-1}}  }$ (*\label{line:childrenC}*)
  uintptr_t next_phys_addr = PTE_PFN_TO_ADDR(entry->pfn); // Fetch physical addr of next table (*\label{line:finalpieceS}*)(*\label{line:extract_pfn}*)
  uintptr_t next_virt_addr = (uintptr_t) P2V(next_phys_addr); // Convert to virtual address (*\label{line:p2vC}*)
  next = (pte_t *) next_virt_addr;(*\label{line:finalpieceE}*)
  return next;
}
\end{lstlisting}
\else
\begin{lstlisting}[mathescape, basicstyle=\footnotesize,escapeinside={(*}{*)}]
;;pte_t *pte_get_next_table(pte_t *entry) {
... ;; setting up the stack
;; pte_t *next;
$\specline{\textsf{P} \ast  I\texttt{ASpace}_{\textsf{id}}(\theta,\Xi\setminus\{\textsf{entry}\},m)  \ast \texttt{rbp-8} \mapsto_{\textsf{v}} \textsf{entry} \ast  \texttt{rbp-16} \mapsto_{\textsf{v}} \textsf{next} \ast \texttt{r8}  \mapsto_{\textsf{r}} \textsf{\_} \ast \texttt{rdi}  \mapsto_{\textsf{r}} \textsf{\_} \ast \ghostmaptoken{\delta{}s}{\rtv}{\delta}  }_{\rtv}$
$\specline{ \textsf{entry+KERNBASE} \mapsto_{\textsf{vpte,qfrac}} \textsf{entry entry\_val} \ast \ulcorner  \textsf{qfrac} = 1 \leftrightarrow \; \lnot(\textsf{entry\_present entry\_val})\urcorner}_{\rtv}$ (*\label{line:get_next_vpte_precondition}*)
$\specline{\ulcorner\textsf{entry\_present(entry\_val)}\urcorner \wand \forall_{i\in\textsf{0..511}} \; \ghostmaptoken{\textsf{id}}{\textsf{table\_root(entry\_val.pfn) + i * 8}}{\textsf{v-1}} }_{\rtv} $ (*\label{line:conditional_children}*)
mov    -0x8[rbp],rdi
$\specline{\textsf{P} \ast  I\texttt{ASpace}_{\textsf{id}}(\theta,\Xi\setminus\{entry\},m)   \ast \textsf{rbp-8} \mapsto_{\textsf{v}} \textsf{entry} \ast \texttt{r8}  \mapsto_{\textsf{r}} \textsf{\_}  \ast  \texttt{rbp-16} \mapsto_{\textsf{v}} \textsf{next} \ast \texttt{rdi}  \mapsto_{\textsf{r}} \textsf{entry} \ast \ghostmaptoken{\delta{}s}{\rtv}{\delta}  }_{\rtv}$
$\specline{ \textsf{entry+KERNBASE} \mapsto_{\textsf{vpte,qfrac}} \textsf{entry entry\_val} \ast \ulcorner  \textsf{qfrac} = 1 \leftrightarrow \; \lnot(\textsf{entry\_present entry\_val})\urcorner}_{\rtv}$
$\specline{\ulcorner\textsf{entry\_present(entry\_val)}\urcorner \wand \forall_{i\in\textsf{0..511}} \; \ghostmaptoken{\textsf{id}}{\textsf{table\_root(entry\_val.pfn) + i * 8}}{\textsf{v-1}} }_{\rtv} $
mov     rdi, r8
$\specline{\textsf{P} \ast  I\texttt{ASpace}_{\textsf{id}}(\theta,\Xi\setminus \{ \mathsf{entry} \},m)  \ast \textsf{rbp-8} \mapsto_{\textsf{v}} \textsf{entry} \ast \texttt{r8}  \mapsto_{\textsf{r}} \textsf{entry}  \ast  \texttt{rbp-16} \mapsto_{\textsf{v}} \textsf{next} \ast \texttt{rdi}  \mapsto_{\textsf{r}} \textsf{entry} \ast \ghostmaptoken{\delta{}s}{\rtv}{\delta}  }_{\rtv}$
$\specline{ \textsf{entry+KERNBASE} \mapsto_{\textsf{vpte,qfrac}} \textsf{entry entry\_val} \ast \ulcorner  \textsf{qfrac} = 1 \leftrightarrow \; \lnot(\textsf{entry\_present entry\_val})\urcorner}_{\rtv}$
$\specline{\ulcorner\textsf{entry\_present(entry\_val)}\urcorner \wand \forall_{i\in\textsf{0..511}} \; \ghostmaptoken{\textsf{id}}{\textsf{table\_root(entry\_val.pfn) + i * 8}}{\textsf{v-1}} }_{\rtv} $
mov    [r8],rdi (*\label{line:read_entry_contents}*)
$\specline{\textsf{P} \ast  I\texttt{ASpace}_{\textsf{id}}(\theta,\Xi\setminus \{ \mathsf{entry} \},m)  \ast \textsf{rbp-8} \mapsto_{\textsf{v}} \textsf{entry} \ast \texttt{r8}  \mapsto_{\textsf{r}} \textsf{entry}  \ast  \texttt{rbp-16} \mapsto_{\textsf{v}} \textsf{next} \ast \texttt{rdi}  \mapsto_{\textsf{r}} \textsf{entry\_val}  \ast \ghostmaptoken{\delta{}s}{\rtv}{\delta}  }_{\rtv}$
$\specline{ \textsf{entry+KERNBASE} \mapsto_{\textsf{vpte,qfrac}} \textsf{entry entry\_val} \ast \ulcorner  \textsf{qfrac} = 1 \leftrightarrow \; \lnot(\textsf{entry\_present entry\_val})  \urcorner }_{\rtv}$
$\specline{\ulcorner\textsf{entry\_present(entry\_val)}\urcorner \wand \forall_{i\in\textsf{0..511}} \; \ghostmaptoken{\textsf{id}}{\textsf{table\_root(entry\_val.pfn) + i * 8}}{\textsf{v-1}} }_{\rtv} $
and    0x1,rdi (* \label{line:mask_present} *)
mov    rdi,rax
cmp    0x0,rax ;;  if (!entry->present) {(* \label{line:check_entry_present} *)
jne    161 <pte_get_next_table+0xa1> (* \label{line:check_entry_present_jump} *) ;; Jump if the present bit is not zero
$\specline{\textsf{P} \ast  I\texttt{ASpace}_{\textsf{id}}(\theta,\Xi\setminus \{ \mathsf{entry} \},m)  \ast \textsf{rbp-8} \mapsto_{\textsf{v}} \textsf{entry} \ast \texttt{r8}  \mapsto_{\textsf{r}} \textsf{entry} \ast \texttt{rbp-16} \mapsto_{\textsf{v}} \textsf{next} \ast \texttt{rdi}  \mapsto_{\textsf{r}} \textsf{entry\_val \& 0x1} }_{\rtv}$
$\specline{  \textsf{entry} \mapsto_{\textsf{id}} \textsf{v} \ast \ghostmaptoken{\delta{}s}{\rtv}{\delta}  \ast \texttt{rax}  \mapsto_{\textsf{r}} \textsf{entry\_val \& 0x1} \ast \ulcorner\textsf{entry\_val \& 0x1}=\textsf{0x0}\urcorner}_{\rtv}$
$\specline{ \textsf{entry+KERNBASE} \mapsto_{\textsf{vpte,qfrac}} \textsf{entry entry\_val} \ast \ulcorner  \textsf{qfrac} = 1 \leftrightarrow  \lnot(\textsf{entry\_val \& 0x1}=\textsf{0x1})  \urcorner }_{\rtv}$ (*\label{line:before_concluding_qfrac1}*)
mov    rbp,rdi (*\label{line:alloc_path_start}*)
sub    0x10, rdi ;; Store the value of rbp minus 16 bytes (address of (*\textsf{next}*)) into rdi (*\label{line:pass_addrof_next}*)
$\specline{\textsf{P} \ast  I\texttt{ASpace}_{\textsf{id}}(\theta,\Xi\setminus \{ \mathsf{entry} \},m)  \ast \textsf{rbp-8} \mapsto_{\textsf{v}} \textsf{entry} \ast \texttt{r8}  \mapsto_{\textsf{r}} \textsf{entry} \ast \texttt{rbp-16} \mapsto_{\textsf{v}} \textsf{next} \ast \texttt{rdi}  \mapsto_{\textsf{r}} \textsf{rbp - 16} }_{\rtv}$
$\specline{  \textsf{entry} \mapsto_{\textsf{id}} \textsf{v} \ast \ghostmaptoken{\delta{}s}{\rtv}{\delta} }_{\rtv}$
$\specline{ \textsf{entry+KERNBASE} \mapsto_{\textsf{vpte,qfrac}} \textsf{entry entry\_val} \ast \ulcorner  \textsf{qfrac} = 1 \land \lnot(\textsf{entry\_present entry\_val})  \urcorner }_{\rtv}$ (*\label{line:after_concluding_qfrac1}*)
callq  70 <pte_initialize> ;;pte_initialize(entry);(* \label{line:call_to_pte_initialize} *)
$\specline{\textsf{P} \ast  I\texttt{ASpace}_{\textsf{id}}(\theta,\Xi\setminus \{ \mathsf{entry} \},m)  \ast \textsf{rbp-8} \mapsto_{\textsf{v}} \textsf{entry} \ast \texttt{r8}  \mapsto_{\textsf{r}} \textsf{entry}  }_{\rtv}$
$\specline{  \textsf{entry} \mapsto_{\textsf{id}} \textsf{v} \ast \ghostmaptoken{\delta{}s}{\rtv}{\delta} }_{\rtv}$
$\specline{ \textsf{entry+KERNBASE} \mapsto_{\textsf{vpte,qfrac}} \textsf{entry entry\_val} \ast \ulcorner  \textsf{qfrac} = 1 \land \lnot(\textsf{entry\_present entry\_val})  \urcorner }_{\rtv}$
$\specline{\ulcorner  \textsf{qfrac} = 1 \land \lnot(\textsf{entry\_present (pfn\_set (entry\_val nextpaddr))}  \urcorner   }_{\rtv}$
$\specline{ \texttt{rbp-16} \mapsto_{\textsf{v}} \textsf{pfn\_set(entry\_val nextpaddr)}   }_{\rtv}$
$\specline{ \begin{array}{l}\textsf{entry\_present (pte\_initialized (pfn\_set(entryv nextpaddr)))} \wand  \\ \;\;\;\;\;\;\;\;\  \forall_{i\in \textsf{0 ... 511} } \ldotp  \ghostmaptoken{\textsf{id}}{((\textsf{table\_root (pte\_initialized (pfn\_set(entry\_val nextpaddr)))}) + \textsf{i * 8})}{\textsf{v-1}} \end{array}  }$ (*\label{line:page_of_caps}*)
... ;;entry value updates: entry->pfn = nextpaddr; entry->present = 1; (*\label{line:install_new_entry}*)
... ;;now we know that entry is initialized, so we satisfy the condition to access children list
$\specline{\textsf{P} \ast  I\texttt{ASpace}_{\textsf{id}}(\theta,\Xi\setminus \{ \mathsf{entry} \},m)  \ast \textsf{rbp-8} \mapsto_{\textsf{v}} \textsf{entry} \ast \texttt{r8}  \mapsto_{\textsf{r}} \textsf{entry}  }_{\rtv}$
$\specline{  \textsf{entry} \mapsto_{\textsf{id}} \textsf{v} \ast \ghostmaptoken{\delta{}s}{\rtv}{\delta} }_{\rtv}$
$\specline{ \textsf{entry+KERNBASE} \mapsto_{\textsf{vpte,qfrac}}  \textsf{pte\_initialized(pfn\_set(entryv nextpaddr))} \ast \ulcorner  \textsf{qfrac} = 1 \land \lnot(\textsf{entry\_present entry\_val})  \urcorner }_{\rtv}$
$\specline{ \texttt{rbp-16} \mapsto_{\textsf{v}} \textsf{pte\_initialized(pfn\_set(entry\_val nextpaddr))}  \ast rax \mapsto_{r} \textsf{table\_root (pte\_initialized (pfn\_set(entry\_val nextpaddr)))}  }_{\rtv}$
$\specline{\forall_{i\in \textsf{0 ... 511} } \ldotp  \ghostmaptoken{\textsf{id}}{((\textsf{table\_root (pte\_initialized (pfn\_set(entry\_val nextpaddr)))}) + \textsf{i * 8})}{\textsf{v-1}}  }$ (*\label{line:alloc_path_end}*)
;;} (*\label{line:end_of_allocation_path}*)
... ;; Code after conditional continued in Figure (*\ref{fig:p2v}*)
\end{lstlisting}
\fi
% ;;uintptr_t next_virt_addr = (uintptr_t) P2V(next_phys_addr);
% movabs KERNBASE,rcx
% add    rcx,rax
% ...
% ;;next = (pte_t *) next_virt_addr;
% ;;clean up the stack and return next
% \end{lstlisting}
\vspace{-1em}
\caption{Ensuring \textsf{entry} points to a valid next-level table, allocating if necessary, returning its virtual address.}
\label{fig:calltopteinitializeC}
\vspace{-1em}
\end{figure}

\subsubsection{From L$n+1$ Entries to L$n$ Tables}
We discuss access to the level 4 table later (Section \ref{wlkpgdirC}). However, for subsequent levels, the base address of level $n$ must be
fetched from the appropriate entry in the table of level $n+1$.
This is the role of \lstinline|pte_get_next_table| (originally Figure \ref{fig:pagetablescode}, with proof details in Figure \ref{fig:calltopteinitializeC}).
It is passed the virtual address of the page table entry in level $n+1$, and should return the \emph{virtual} 
address of the \emph{base} of the level $n$ table
indicated by that entry.
If the entry is empty (i.e., this is a sparse part of the page table representation),
the code also allocates a page for the level $n$ table, installs it in the level $n+1$ entry, and establishes appropriate invariants.
Figure \ref{fig:calltopteinitializeC} presents the function with proof annotations that we will explain shortly, but we first explain the functionality.
\lstinline|pte_get_next_table| accepts a \emph{virtual} address \lstinline|entry| which points to a level-$n+1$
table entry.
\looseness=-1

On Line \ref{line:read_entry_contentsC}, the code checks the present bit of the entry.
If the bit is unset, there is no level-$n$ table, so one must be allocated via \lstinline|pte_initialize| (explained shortly,
but it essentially
allocates a fresh physical page, and initializes the memory pointed to by \lstinline|entry| with that physical address) and marked present.
By Line \ref{line:finalpieceS} the entry is known to be valid and contain the physical address of
the base of a level $n$ table. That address is then extracted (Line \ref{line:extract_pfn}),
converted to a virtual address (Line \ref{line:p2vC}), and returned to the caller.
We can now discuss \lstinline|pte_get_next_table|'s proof of correctness.
While at first glance this code may look like its subtlety is mostly care to distinguish physical and virtual addresses,
it has a highly nontrivial correctness argument, which depends critically on detailed invariants on how access to page table
entries is shared between parts of the kernel. No prior work has engaged with this problem.
\looseness=-1

For this C presentation of what is really an assembly-level proof, we abuse notation and
use our register points-to for C-level program variables. So on Line \ref{line:precondition_entry_out},
$\mathsf{entry} \mapsto_r \mathsf{entryp+KERNBASE}$ means that the \emph{register representing} the C program variable
\lstinline|entry| (per the calling convention, \lstinline|rdi|) holds the sum on the right (a constant offset added to the physical address $\mathsf{entryp}$ of the entry).
That particular value is one subtlety of the proof related to the aforementioned kernel invariant, and is explained in Section \ref{sec:p2vC}.
The \emph{virtual pte-points-to} from that virtual address (Line \ref{line:get_next_vpte_preconditionC}) indicates that it points to a value
$\mathsf{entryv}$, a (possibly-unpopulated) page table entry.
A virtual pte-points-to is defined just like the normal virtual points-to of Figure \ref{fig:virtualpointstosharing},
except the physical address (\textsf{entryp} on Line \ref{line:get_next_vpte_preconditionC}) is explicit in the assertion
rather than existentially quantified:\\
\centerline{$
    \vaddr\mapsto_{\textsf{vpte,q}} \; \paddr \; \vpage : \mathsf{vProp}~\Sigma \stackrel{\triangle}{=} 
    \exists \delta\ldotp
	(\lambda \mathit{cr3val}\ldotp
	\ghostmaptoken{\delta{}s}{\mathit{cr3val}}{\delta}) \ast 
  \fracghostmaptoken{\delta}{\vaddr}{\paddr}{\qfrac} \ast \paddr \mapsto_{\mathsf{p}} \vpage
$}\\
This supports memory access rules much like Figure \ref{fig:wpdamd}'s rules (which are proven
sound using the virtual pte-points-to rules as lemmas!)
while exposing the physical location being modified.
This is useful for page table modifications, which require knowing the physical location being changed.
They are used throughout the software page table walk because entries in any level may be initialized.
\looseness=-1

\subsubsection{Address Space Invariant: Identity Mappings and Conditional Page Table Ownership}
\label{subsec:identitymappingsC}
Assembly-level verification of compiler output from Figure \ref{fig:calltopteinitializeC} is verbose, but largely
similar to other assembly-level verification thanks to Section \ref{sec:logic}'s logic (including virtual pte-points-to
assertions),
but only after resolving two critical challenges.
Two key challenges stand out and end up affecting both the pre- and post-conditions, neither of which has been addressed by prior work.
First, the update to the memory at (virtual) address \lstinline|entry| depends on subtle ownership invariants:
if the entry is present then its fractional ownership is shared with a large number of $\textsf{L}_{4}\_\textsf{L}_{1}\_\textsf{PointsTo}$ assertions
from the address space invariant (Figure \ref{fig:peraspaceinvariant}),
but if the entry is absent the proof requires full ownership to update it. We resolve this by extending the address space invariant
to make the owned fraction of the entry's memory \emph{dependent on its own contents}.
Second, the conversion of physical addresses into a corresponding virtual address that can be used to modify the specific
physical location relies on subtle, never-before-formalized kernel invariants.
% \looseness=-1
%
% These two factors percolate to the precondition (for conditional fractional ownership) and postcondition
% (for physical-to-virtual conversions), as the caller essentially passes output from one call to \lstinline|pte_get_next_table|
% as input to a subsequent call to traverse multiple levels.
Since the key to solving these challenges is to extend the address space invariant, we
first discuss that invariant and the kernel designs it supports, before returning to the subtle details of
verifying lines \ref{line:install_new_entryC} and \ref{line:p2vC}.
The key idea is to establish extra invariants on physical addresses that are part of a page table ---
but to do so in a way that meshes with the existing invariants (in the informal sense) already preserved in most
unverified kernel designs.
Each of the above problems requires its own extension to the invariant, but we will discuss
physical-to-virtual conversion first, both because it dictates the organization of the invariant
and because when Line \ref{line:check_entry_present_jumpC}'s
conditional check is false that is all that is necessary for the proof; correctness of the conditional branch deals with \emph{both}
extensions.

\subsubsection{Physical-to-Virtual Mappings and \textsf{P2V}}
\label{sec:p2vC}
Kernels need to convert between physical and virtual addresses, in both directions.
Traversing the page tables in software is the simplest way to convert a virtual address to a physical address;
this is the context we are working up to.
However, implementing this virtual-to-physical (V2P) translation in software ironically requires physical-to-virtual (P2V) translation,
because the addresses stored in page table entries are physical, but memory accesses issued by the OS code use virtual addresses.
Because VMM operations are performance-critical for many workloads, most kernels
maintain invariants that enable very fast P2V conversions (rather than adding another data structure).
Specifically, many kernels maintain an invariant on their page tables that the virtual address of any page used for a page table 
% lives at a virtual address whose value 
is \emph{a constant offset from the physical address} --- a practice sometimes referred to as \emph{identity mapping} 
(even though the physical-to-virtual translation
is typically not literally the identity function, but adding a nonzero constant offset).\footnote{Some kernels do this for all physical memory on the machine, simplifying interaction
with DMA devices.
On newer platforms like RISC-V, this sometimes truly is an identity mapping ---
x86-64 machines are forced into offsets by backward compatibility with bootloaders that cannot access the full memory space of the
machine.
}
Thus \lstinline|P2V| on line \ref{line:p2vC} of Figure \ref{fig:calltopteinitializeC} is a macro for adding the fixed constant \lstinline|KERNBASE|.

Figure \ref{fig:peraspaceinvariant_with_p2v_extensionC} extends the per-address-space invariant  to also track which
addresses we can perform a P2V conversion on by adding a constant offset (i.e., the set of physical addresses which participate in page tables).
$\Xi$ is another ghost map, from physical addresses to the level of the page table they represent (1--4).
\emph{Only} physical addresses in $\Xi$ can undergo P2V conversion. 
Section \ref{sec:p2vC} describes the verification of an actual conversion,
but this invariant must be \emph{established} when adding a new page table level (notably on Line \ref{line:call_to_pte_initializeC},
hence the comment of Line \ref{line:now_we_know}).

\begin{figure*}
\footnotesize
\[
\begin{array}{l}
   I\textsf{ASpace}_{\textsf{id}}(\ptablestore,\Xi,m)\stackrel{\triangle}{=} \textsf{ASpace\_Lookup}_{\textsf{id}}(\ptablestore,\Xi,m) \ast\\  %\mathsf{GhostMap}(\mathsf{id},\Xi)\ast\\
  \left(\bigast{(\vaddr, \textsf{paddr})\in \ptablestore}{\exists\;(\textsf{l4e, l3e, l2e, l1e, paddr})\ldotp \textsf{L}_{4}\_\textsf{L}_{1}\_\textsf{PointsTo}(\vaddr\textsf{, l4e, l3e, l2e, l1e, paddr})}\right)\ast \\
  \bigast{(\paddr,\mathsf{level}) \in \Xi}{\exists\; (\textsf{qfrac, q, val,}\vaddr) \ldotp \ulcorner \vaddr = \paddr + \textsf{KERNBASE} \; \textsf{level} > 1\urcorner \ast  \underbrace{\fracghostmaptoken{\delta}{\vaddr}{\paddr}{\qfrac} }_\text{\textcircled{1} Ghost translation} \ast \underbrace{\paddr \mapsto_{\mathsf{p}}\{\textsf{qfrac}\}\; \vale}_\text{\textcircled{2} Physical location}} \ast\\
   \qquad\underbrace{ \ulcorner \textsf{qfrac} = 1 \leftrightarrow \; \lnot\textsf{entry\_present }(\vale) \urcorner}_\text{\textcircled{3} Entry validity}\ast 
    \underbrace{\left(\ulcorner\textsf{present\_L}(\vale,\mathsf{level})\urcorner \wand \forall_{\textsf{i}\in\textsf{0..511}} \ldotp \ghostmaptoken{\textsf{id}}{((\mathsf{entry\_page}\;\vale) + \textsf{i * 8})}{\textsf{level-1}}\right)}_{\text{\textcircled{4} Indexing into next level of tables}} \\
  \textsf{ where } \quad
  \textsf{present\_L}(\vale,\mathsf{level})\stackrel{\triangle}{=} \mathsf{entry\_present}(\vale)\land \mathsf{level} > 1\\
   \textsf{ASpace\_Lookup}_{\textsf{id}}(\ptablestore,\Xi,m) \stackrel{\triangle}{=} \lambda\textsf{ cr3val} \ldotp \; \exists \gammaPred,\gammaPred'\ldotp \ulcorner m \; !!\; \textsf{cr3val} = \textsf{Some } \gammaPred \urcorner \ast
   % \ownGhost\gammaPred{\authfull{\ptableabswalk\ptablestore}} \ast  \ownGhostpv\gammaPred{\authfull{\pvmapping\Xi}}
   \ptableabswalk{\delta,\ptablestore} \ast \pvmapping{\delta',\Xi}
\end{array}
\]
\vspace{-1.5em}
\caption{Address space invariant of Figure \ref{fig:peraspaceinvariant} extended with a ghost map bookkeeping identity mappings. }
  \label{fig:peraspaceinvariant_with_p2v_extensionC}
\vspace{-1em}
\end{figure*}

For each $\paddr\mapsto \textsf{v} \in\Xi$, the invariant contains a virtual points-to justifying that virtual address
$\paddr+\textsf{KERNBASE}$ maps to physical address $\paddr$
(\textcircled{1} in Figure \ref{fig:peraspaceinvariant_with_p2v_extensionC});
fractional ownership of the physical memory for that page table entry (\textcircled{2}, which together with \textcircled{1} is equivalent
to a virtual points-to);
and for valid entries (with the present bit set) above L1, ghost map tokens for $\Xi$ for every entry in the table pointed to by the entry, which can be used
to repeat the process one level down (\textcircled{4}). 
% (L1 entries point to data pages, whose physical memory ownership resides in some virtual points-to).
\textcircled{4} becomes part of the precondition to \lstinline|pte_get_next_table|:
Line \ref{line:precondition_conditional}) says that if the entry is valid (points to a next-level table)
then there are tokens for accessing $\Xi$ for every entry in the next-level table.
By Line \ref{line:finalpieceS} the entry is guaranteed to be valid so all tokens for converting the next table level's physical addresses to virtual
are available (in the form expressed by the assertion on Line \ref{line:available_child_tokens}).
\looseness=-1

As noted above, for the code path where the conditional does not execute (there was already a valid entry), this is all we need of the new
invariant to verify the end of the function from Line \ref{line:finalpieceS} onward.
By that point the invariant holds and applies to the definitely-valid entry,
so we can the physical address of the next-level table to a corresponding virtual address via the identity mappings just described.
Line \ref{line:extract_pfn} simply retrieves the physical address.
Line \ref{line:p2vC} is the critical piece, and arguably corresponds to the most subtle verification of an \lstinline|add| instruction
(\lstinline|add rax, KERNBASE|)
that we are aware of, and something no prior work on verified OS kernels has dealt with.

After Line \ref{line:extract_pfn}, it is already known that the present bit is set in the entry;
Line \ref{line:childrenC}'s assertion reflects that the tokens for $\Xi$ exist for each word-aligned
physical address in the next-level table.
However, note that no argument to this function specifies which virtual address is being accessed,
so \lstinline|pte_get_next_table| does not know which entry in the next table to retrieve.
Even if that address were passed, this function is used for each step-down, so the slice of the
virtual address (per Figure \ref{fig:pagetables}) is not fixed.
Thus Line \ref{line:p2vC} computes the virtual address of the \emph{base} of the next-level table,
and the postcondition includes a renamed version of the assertion on line \ref{line:childrenC},
for the \emph{caller} --- \lstinline|walkpgdir| (discussed next) to perform the conversion for
The caller determines which entry in that table must actually
be accessed --- by selecting the appropriate index into the 512 ghost map tokens returned in the postcondition,
and using the ghost translation and physical location portions of the invariant to assemble a vpte-pointsto
that justifies the caller's subsequent access to a particular entry of the returned table.
The postcondition also passes back the per-address-space invariant with the
identity mapping resources for \lstinline|entry| still pulled out (it was removed by the caller).
\looseness=-1

\subsubsection{Self-Conditional Fractional Ownership and Installing a New Table}
\label{sec:selfconditional}
The fractional ownership of the entry's physical memory is subtle.
As noted above, a \emph{valid} entry must coexist with the fractional ownership from
$\textsf{L}_{4}\_\textsf{L}_{1}\_\textsf{PointsTo}$ and therefore have less than full ownership,
but in the case where the entry is \emph{invalid}, Line \ref{line:call_to_pte_initializeC} must have full permissions in order
to populate the entry (i.e., to install a reference to a next-level table).
Fortunately, the entry can only be in use if its valid bit is set; if the valid bit is not set, we know
that no virtual points-to entry in $\delta$ or $\theta$ holds any partial ownership.
But determining this requires reading the very memory whose ownership is being determined.
We use the invariant portion annotated as ``Entry validity'' (\textcircled{3}) in Figure \ref{fig:peraspaceinvariant_with_p2v_extensionC} to capture this:
if the entry is invalid the invariant holds full ownership of the entry, so it can be updated;
while if the entry is valid, the invariant owns only a constant nonzero fraction sufficient to read but not modify the entry.
Since the fractional ownership is always non-zero, Line \ref{line:read_entry_contentsC} in Figure \ref{fig:calltopteinitializeC} can read the entry
(using a rule similar to \textsc{WriteToRegFromVirtMem}, tailored to virtual PTE-points-to assertions),
and if the entry is dynamically found to be invalid, the invariant is refined (Line \ref{line:refined_fractional_ownership})
to indicate full ownership, allowing updates.
Note that the caller is responsible for providing this conditional ownership, having pulled it out of the invariant earlier.
This is why the precondition (Line \ref{line:precondition_entry_out}) explicitly excludes the entry's physical address from the invariant ($\Xi\setminus\{\mathsf{entry}\}$) ---
its relevant assertions have already been borrowed by the caller.
\looseness=-1

\begin{figure}\footnotesize
  \begin{lstlisting}[language=C,mathescape,escapeinside={(*}{*)}]
/* @param entry: virtual address of non-present entry to initialize with a new physical page*/
$\specline{I\texttt{ASpace}_{\textsf{id}}(\theta,\Xi,m) \ast \lnot(\textsf{entry\_present entryv}) \ast \textsf{entryp+KERNBASE} \mapsto_{\textsf{vpte,1}} \textsf{entryp entryv} \ast \ghostmaptoken{\textsf{id}}{(\mathsf{entryp})}{\textsf{level}} }$
void pte_initialize(pte_t *entry) {
  pte_t *local = kalloc(); // Allocate a full zeroed page for 512 8-byte entries(*\label{line:call_to_kallocC}*)
  entry->pfn = PTE_ADDR_TO_PFN((uintptr_t) local);(*\label{line:kalloc_install}*)
  $\specline{ I\texttt{ASpace}_{\textsf{id}}(\theta,\Xi,m) \ast \lnot(\textsf{entry\_present entryv}) \ast \textsf{entryp+KERNBASE} \mapsto_{\mathsf{vpte,1}} \mathsf{entryp} \; \textsf{pfn\_set(entryv nextpaddr)}}$
  $\specline{\ghostmaptoken{\delta{}s}{\rtv}{\delta} \ast \forall_{i\in\textsf{0..511}}\ldotp ()\textsf{table\_root (pte\_initialized (pfn\_set (entry\_val nextpaddr))) + i * 8}) \mapsto_{\textsf{id}} \textsf{level-1}   }_{\rtv}$(*\label{line:initialize_post}*)
}
\end{lstlisting}
\vspace{-2em}
\caption{Allocating a physical page }
\label{pteinitializespecC}
\vspace{-1em}
\end{figure}

If the entry is not set, \lstinline|pte_initialize|  
allocates a physical page for use as the next-level table.
\lstinline|pte_initialize| (Figure \ref{pteinitializespecC}) calls
\lstinline|kalloc| to allocate a physical page (Figure \ref{pteinitializespecC} Line \ref{line:call_to_kallocC}),
and installs it into the entry (Line \ref{line:kalloc_install}).
The page-allocator's \textsf{kalloc}
is the only unverified (trusted) code in our case study.\footnote{
  This is an allocator for regions of pre-zeroed physical memory that is mapped, but not accessed by the allocator itself,
  as is typical for slab allocators~\cite{bonwick1994slab}.
  Its verification would be similar to verifying a usermode \textsf{malloc} verification ~\cite{Chlipala2013Bedrock,wickerson2010explicit},
  just with additional invariants on the memory pool.
} 
Since we are using \textsf{pte\_initialize} for page-table address allocation, we must relate this newly
allocated physical address to the identity mapping map $\Xi$ --- 
see Line \ref{line:page_of_capsC} in Figure \ref{fig:calltopteinitializeC}, where
\texttt{kalloc}'s specification guarantees it has returned memory from a designated memory
pool that is already mapped
\footnote{A reasonable reader might wonder where this pool
initially comes from and how it might grow when needed. Typically an initial mapping subject to this identity mapping
constraint is set up prior to transition to 64-bit kernel code (notably,
a page table must exist \emph{before} virtual memory is enabled during boot, as part of enabling it is setting
a page table root).
Growing this pool later requires cooperation of physical memory range allocation and virtual memory range allocation,
typically by starting general virtual address allocation at the highest physical memory address plus the identity mapping offset.
This reserves the virtual addresses corresponding to all physical addresses plus the offset for later use in this pool,
as needed.
} 
and satisfies the offset invariants (trivially, as the new page is zeroed).
The presence bit of the entry is \emph{not} set during \lstinline|pte_initialize|, but upon
return to \lstinline|pte_get_next_table|, where it will validate the conditional ownership discussed above.
\lstinline|pte_initialize| has a full-permission virtual pte-pointsto in its precondition.
Then the assertions that hold after Line \ref{line:now_we_know} of Figure \ref{fig:calltopteinitializeC}
are enough to establish the same page table invariants which hold in the case where the entry was already valid,
by updating the current address space's entry.

\subsubsection{The Specification of \lstinline|pte_get_next_table|}
Note that the specification does \emph{not} assume a specific page table level and is used
for all three level transitions (4 to 3, 3 to 2, 2 to 1).
The logical parameter \textsf{v} represents the level
of the entry passed as an argument (c.f. the token $\ghostmaptoken{\textsf{id}}{(\mathsf{entryp})}{\textsf{v}}$ witnessing
that \textsf{entryp} is part of the page table invariant on Line \ref{line:precondition_entry_out}).
This comes into play with a key subtlety of \lstinline|pte_get_next_table|'s
specification: its precondition
includes a virtual {pte-points-to} (discussed earlier, Line \ref{line:get_next_vpte_preconditionC})
but its postcondition does not yield new virtual points-to assertions!
It merely computes the base virtual address of the next table, and returns adequate tokens
% (discussed in Section \ref{subsec:identitymappingsC}, explicit on
(Line \ref{line:childrenC})
for the \emph{caller} to construct a vpte-points-to for any entry of the next table level.
\looseness=-1

\subsubsection{Walking The Page Tables: Calling \textsf{pte\_get\_next\_table} for Each Level}
\label{wlkpgdirC}
\begin{figure}\footnotesize
\ifPLDI
\begin{lstlisting}[language=C,mathescape,escapeinside={(*}{*)}]
/* @param l4: the (virtual) address of the L4 page table tree
   @param va: virtual address to be translated and mapped */
pte_t *walkpgdir(pte_t *l4, const void *va) {
  $\specline{\textsf{P} \ast \ulcorner\textsf{l4p}=\textsf{rtv}\urcorner\ast\ulcorner\textsf{rtv+KERNBASE}=\textsf{l4}\urcorner \ast \ghostmaptoken{\textsf{id}}{(\textsf{l4p}+8*\textsf{L4Offset}(\textsf{va}))}{\textsf{4}} \ast  I\texttt{ASpace}_{\textsf{id}}(\theta,\Xi,m) \ast \ghostmaptoken{\delta{}s}{\rtv}{\delta}}_{\rtv}$(*\label{line:walkpgdir_pre}*)
  pte_t *l4_entry = &l4[L4Offset(va)]; // Virtual address of L4 entry(*\label{line:start_l4_calcC}*)
  $\specline{\textsf{P} \ast  I\texttt{ASpace}_{\textsf{id}}(\theta,\Xi\setminus\{\textsf{l4p+8*L4Offset(va)}\},m) \ast \ghostmaptoken{\delta{}s}{\rtv}{\delta}  \ast \ulcorner  \textsf{qf4} = 1 \leftrightarrow \; \lnot(\textsf{entry\_present l4e\_val})\urcorner }_{\rtv}$
  $\specline{\textsf{entry\_present(l4e\_val)} \wand \forall_{i\in\textsf{0..511}} \; \ghostmaptoken{\textsf{id}}{(\textsf{table\_root(l4e\_val)+ i * 8})}{\textsf{3}} }_{\rtv} $
  $\specline{ \textsf{l4p+KERNBASE+L4Offset(va)} \mapsto_{\textsf{vpte,qf4}} \textsf{(l4p+L4Offset(va)) l4e\_val}}_{\rtv}$    
  pte_t *l3 = pte_get_next_table(l4_entry); (*\label{line:first_getnext_callC}*) 
  /*pte_get_next_table may have allocated a new page which updates entry value*/
  $\specline{\textsf{P} \ast  I\texttt{ASpace}_{\textsf{id}}(\theta,\Xi\setminus\{\textsf{l4p+8*L4Offset(va)}\},m)  \ast \ghostmaptoken{\delta{}s}{\rtv}{\delta}  \ast \ulcorner  \textsf{qf4} = 1 \leftrightarrow \; \lnot(\textsf{entry\_present l4e\_val})\urcorner }_{\rtv}$
  $\specline{\forall_{i\in\textsf{0..511}} \ghostmaptoken{\textsf{id}}{(\textsf{table\_root(l4e\_val'.pfn) + i * 8})}{\textsf{3}}  \ast \ulcorner\textsf{l3-KERNBASE}=\textsf{table\_root(l4e\_val'.pfn)}\urcorner} $(*\label{line:l3tokens}*)
  $\specline{ \textsf{l4p+KERNBASE+L4Offset(va)} \mapsto_{\textsf{vpte,qf4}} \textsf{(l4p+L4Offset(va)) l4e\_val}}_{\rtv}$(*\label{line:first_pte_pointstoC}*)
  pte_t *l3_entry = &l3[L3Offset(va)]; /* Virtual address of L3 entry*/(*\label{line:l3offset}*)
  $\specline{\ulcorner\textsf{l3}+\textsf{L3Offset(va)*8} = \textsf{l3p} \land \textsf{table\_root(l4e\_val'.pfn)}  = \textsf{l3} \urcorner }_{\rtv}$
  $\specline{\textsf{P} \ast  I\texttt{ASpace}_{\textsf{id}}(\theta,\Xi\setminus\{\textsf{l4p+8*L4Offset(va)},\textsf{l3p+8*L3Offset(va)}\},m)  \ast \ghostmaptoken{\delta{}s}{\rtv}{\delta}  }_{\rtv}$
  $\specline{\forall_{i\in\textsf{0..511}\setminus\{\textsf{L3Offset(va)}\}} \; \ghostmaptoken{\textsf{id}}{(\textsf{table\_root(l4e\_val.pfn}) \textsf{ + i * 8})} {\textsf{3}} }_{\rtv}$
  $\specline{ \textsf{l4p+KERNBASE+L4Offset(va)} \mapsto_{\textsf{vpte,qf4}} \textsf{(l4p+L4Offset(va)) l4e\_val} \ast \ulcorner  \textsf{qf4} = 1 \leftrightarrow \; \lnot(\textsf{entry\_present l4e\_val})\urcorner}_{\rtv}$(*\label{line:ex_l4_vpteC}*)
  $\specline{ \textsf{l3p+KERNBASE+L3Offset(va)} \mapsto_{\textsf{vpte,qf3}} \textsf{(l3p+L3Offset(va)) l3e\_val} \ast \ulcorner  \textsf{qf3} = 1 \leftrightarrow \; \lnot(\textsf{entry\_present l3e\_val})\urcorner}_{\rtv}$(*\label{line:ex_l3_vpteC}*)
  $\specline{\textsf{entry\_present(l3e\_val)} \wand \forall_{i\in\textsf{0..511}} \;\ghostmaptoken{\textsf{id}}{(\textsf{table\_root(l3e\_val.pfn) + i * 8})}{ \textsf{2}} } _{\rtv} $
  pte_t *l2 = pte_get_next_table(l3_entry);(*\label{line:getnextl3}*)
  pte_t *l2_entry = &l2[L2Offset(va)];(*\label{line:l2offset}*)
  pte_t *l1 = pte_get_next_table(l2_entry);(*\label{line:getnextl2}*)
  return &l1[L1Offset(va)]; // Return virtual addr of L1 entry with a similar index computation(*\label{line:returnl1entryaddr}*)
} $\specline{\textsf{R}_{\textsf{walk}} \ast \textsf{R}_{\textsf{l1e}} }_{\rtv}$
\end{lstlisting}
\else
\todo[inline]{this is dead code, we're using the PLDI flag set to true}
\fi
\vspace{-1em}
\caption{Walking page-table via calls to \textsf{pte\_get\_next\_table} in Figure \ref{fig:calltopteinitializeC}}
\label{walkpgdirC}
\vspace{-1em}
\end{figure}

\begin{figure}\footnotesize
$\begin{array}{l}
\textsf{R}_{\textsf{walk} }=
\left( \begin{array}{l}
  \ulcorner
    \textsf{l3}+8*\textsf{L3Offset(va)} = \textsf{l3\_entry}
    \land \textsf{l2}+8*\textsf{L2Offset(va)} = \textsf{l2\_entry}
    \land \textsf{l1}+8*\textsf{L1Offset(va)} = \textsf{l1\_entry} \\
    \land \textsf{table\_root(l4e\_val}_{\textsf{pfn}})  = \textsf{l3} 
    \land \textsf{table\_root(l3e\_val}_{\textsf{pfn}})  = \textsf{l2} 
    \land \textsf{table\_root(l2e\_val}_{\textsf{pfn}})  = \textsf{l1} \urcorner  \ast \\
    % \textsf{P} \ast  I\texttt{ASpace}_{\textsf{id}}(\theta,\Xi\setminus\{\textsf{l4p},\textsf{l3p},\textsf{l2p},\textsf{l1p}\},m)  \ast  
    \textsf{P} \ast  I\texttt{ASpace}_{\textsf{id}}(\theta,\Xi\setminus\{\textsf{l4p+8*L4Offset(va)},\textsf{l3p+8*L3Offset(va)},\textsf{l2p+8*L2Offset(va)},\textsf{l1p+8*L1Offset(va)}\},m)  \ast  \\
      \ghostmaptoken{\textsf{id}}{(\mathsf{l4p+8*L4Offset(va)})}{\textsf{4}}\ast \ghostmaptoken{\textsf{id}}{(\mathsf{l3p+8*L3Offset(va)})}{\textsf{3}} \ast \\ \ghostmaptoken{\textsf{id}}{(\mathsf{l2p+8*L2Offset(va)})}{\textsf{2}}  \ast \ghostmaptoken{\textsf{id}}{(\mathsf{l1p+8*L1Offset(va)})}{\textsf{1}} \ast \\
      \ghostmaptoken{\delta{}s}{\rtv}{\delta} \ast (\textsf{l4p+KERNBASE+8*L4Offset(va)}) \mapsto_{\textsf{vpte,qf4}} (\textsf{l4p+8*L4Offset(va)}) \; \textsf{l4e\_val}  \ast \\
      (\textsf{l3p+KERNBASE+8*L3Offset(va)}) \mapsto_{\textsf{vpte,qf3}} \textsf{l3p+8*L3Offset(va)}\;\textsf{l3e\_val}  \ast\\
      (\textsf{l2p+KERNBASE+8*L2Offset(va)}) \mapsto_{\textsf{vpte,qf2}} \textsf{l2p+8*L2Offset(va)} \; \textsf{l2e\_val}
    \end{array} \right)   
    \\
    \textsf{R}_{\textsf{l1e}}=(\textsf{l1p+KERNBASE*8*L1Offset(va)}) \mapsto_{\textsf{vpte,qf1}} \textsf{l1p+8*L1Offset(va)}\;\textsf{l1e\_val} \ast  \ulcorner  \textsf{qf1} = 1 \leftrightarrow \; \lnot(\textsf{entry\_present l1e\_val})\urcorner\end{array} $
\vspace{-1em}
\caption{\textsf{R}$_{\textsf{walk}}$: Resources obtained from invariant during a software page table walk in Figure \ref{fig:calltopteinitializeC}}
\label{fig:rwalkC}
\vspace{-1em}
\end{figure}

Implementing a software page table walk amounts to calling \textsf{pte\_get\_next\_table} for each level as shown in Figure \ref{walkpgdirC}. 
\lstinline|walkpgdir| traverses the page table anchored at \lstinline|l4| (the virtual address of the base of the L4 table)
and returns the virtual address of the L1 entry that should map the virtual address \lstinline|va|.
For each level, \lstinline|walkpgdir| locates the appropriate entry by using the level-specific slice of \lstinline|va| to index into
that table (simulating the hardware translation as in Figure \ref{fig:pagetables}), and passes the virtual address of that entry to
\lstinline|pte_get_next_table| to get the base of the next level down.
For example, Line \ref{line:start_l4_calcC} uses \lstinline|L4Offset| (a bit shifting and masking macro) to extract bits 39--47 of \lstinline|va|,
and uses that to find the address of the L4 entry that would map \lstinline|va| in the address space.
That is then passed to \lstinline|pte_get_next_table| on Line \ref{line:first_getnext_callC}, which
returns the virtual address of the base of the L3 table. This process repeats for 3-to-2 (Lines \ref{line:l3offset}--\ref{line:getnextl3}),
and 2-to-1 (Lines \ref{line:l2offset}--\ref{line:getnextl2}), after which Line \ref{line:returnl1entryaddr} returns
the virtual address of the appropriate L1 entry.

In Figures \ref{walkpgdirC} and \ref{fig:rwalkC}, there are four related concepts for each level.
\textsf{l4p} is the physical address of the L4 table base; \textsf{l4} is the corresponding virtual address
(using the same name for the value and the program variable name for brevity, since the variable is not reassigned);
\textsf{l4\_entry} is the virtual address of the L4 entry used to translate \textsf{va};
and \textsf{l4e\_val} is the value of that table entry.
Other levels are named consistently.
For each of the three level transitions, the main challenges for the proof are to
construct a virtual pte-points-to assertion for the entry in that level's table,
and pass the conditional assertion discussed in Section \ref{sec:p2vC} that if that entry is present then there are
identity map tokens for the the physical address of each entry in the subsequent level's table.
For traversing the L4 table, this proceeds by
exchanging the relevant identity map token provided in the precondition (Line \ref{line:walkpgdir_pre})
and pulling the resources for physical address $\mathsf{l4p}+8*\textsf{L4Offset}(\textsf{va})$
out of the identity map invariant: 
parts \textcircled{1} and \textcircled{2} of Figure \ref{fig:peraspaceinvariant_with_p2v_extensionC}) give a virtual pte-points-to,
and \textcircled{3} and \textcircled{4} satisfy other parts of \lstinline|pte_get_next_table|'s precondition.
This justifies the call on Line \ref{line:first_getnext_callC},
which returns the virtual address of the base of the appropriate L3 table and whose postcondition
includes 512 identity map tokens for each of those entries (anchored to the returned virtuall address minus \textsf{KERNBASE}).
That invariant (Line \ref{line:l3tokens}) is analogous to the one that justified the 4-3 step (Line \ref{line:walkpgdir_pre}), and the next two steps proceed the same way.

Even given the slight adaptation of our assembly-level proof for the C-level presentation in Figure \ref{walkpgdirC},
the proof outline in the figure omits some repeat intermediate assertions for readability.
But by Line \ref{line:ex_l3_vpteC}, it should be clear that the proof accumulates a set of similar assertions for each level.
Figure \ref{fig:rwalkC} expands the abbreviated postcondition to the full set of facts that are accumulated in this way.
$\mathsf{R}_\mathsf{walk}$ together with $\mathsf{R}_{\mathsf{l1e}}$
nearly entail \lstinline|L4_L1_PointsTo| (Figure \ref{fig:strongvirtualpointsto}) within the logic,
forming the basis of the construction of a new virtual points-to for virtual address \textsf{va}.
\lstinline|walkpgdir|'s execution observes most evidence of an address translation for \lstinline|va|,
at least down to a possibly-invalid L1 entry (which \lstinline|walkpgdir|'s caller, \lstinline|vaspace_mappage|, will check). 
Each virtual pte-points-to in $\mathsf{R}_\mathsf{walk}$ internally contains the physical points-to portion of one page table walk step for \lstinline|L4_L1_PointsTo|,
and the pure assertions in $\mathsf{R}_\mathsf{walk}$ ensure the address arithmetic works.
$\mathsf{R}_{\mathsf{l1e}}$ includes the self-conditional fractional ownership of the L1 entry (Section \ref{sec:selfconditional})
for the caller to initialize the entry if it is empty, so the caller can complete a virtual points-to assertions
for a newly-mapped page, as we discuss next.
\looseness=-1

\subsection{Mapping a New Page}
\label{sec:mapnewC}
Finally we come to the top-level routine for mapping a new page of memory into an address space
by updating page tables --- the \lstinline|vaspace_mappage| function in Figure \ref{fig:mapping_codeC}.
Again, our verification was carried out at the assembly level, but presented as on the original C for readability,
and the proof outline omits all relevant facts in favor of the most critial assertions involved in the key parts of the proof.
\lstinline|vaspace_mappage| is typically called by a page fault handler, to map a previously-reserved but lazily-allocated page.
It is passed the virtual address of the L4 table base (\lstinline|l4|), the virtual address to map (\lstinline|va|),
and the \emph{physical} address of an empty memory page which should be used as backing memory for \lstinline|va| and its surrounding page.
It begins by calling \lstinline|walkpgdir| (Line \ref{line:call_walkpgdirC}) to return the virtual address of 
the L1 entry which corresponds to \lstinline|va| (allocating intermediate tables as needed).
It then checks if the entry is already initialized. If not, \lstinline|fpaddr| is installed
into the L1 entry, which is then marked valid (setting the present bit), and the page is mapped.
% To do so, with a given allocated fresh page (\textsf{fpaddr}), then calculate the appropriate
% known-valid page table walks (via \textsf{walkpgdir} Line \ref{line:call_walkpgdirC} in Figure \ref{fig:mapping_codeC}) and update
% the appropriate L1 page table entry (Line 35 in Figure \ref{fig:mapping_codeC});
Unmapping is the reverse of the logic we discuss here.
\looseness=-1
%\lstset{
%  columns=fullflexible,
%  numbers=left,
%  basicstyle=\ttfamily,
%  keywordstyle=\color{blue}\bfseries,
%  morekeywords={mov,add,call},
%  emph={rsp,rdx,rax,rbx,rbp,rsi,rdi,rcx,r8,r9,r10,r11,r12,r13,r14,r15},
%  emphstyle=\color{green},
%  emph={[2]cr3},
%  emphstyle={[2]\color{violet}},
%  morecomment=[l]{;;},
%  mathescape
%}
\begin{figure}\footnotesize
  \begin{lstlisting}[language=C,mathescape,escapeinside={(*}{*)}]
/* @param l4: the virtual address of the root of the page table tree
   @param va: virtual address to be translated and mapped
   @param fpaddr: physical address of a zeroed page to map if va is not already mapped */
void vaspace_mappage(pte_t *l4, void *va,uintptr_t fpaddr ) {
  $\specline{\textsf{P} \ast \ghostmaptoken{\mathsf{\textsf{id}}}{(\textsf{l4p+8*L4Offset(va)})}{4} \ast  I\texttt{ASpace}_{\textsf{id}}(\theta,\Xi,m) \ast \ghostmaptoken{\delta{}s}{\rtv}{\delta} \ast \ulcorner\theta \; !!\; \vaddr = \texttt{None}\urcorner}_{\rtv}$ (*\label{line:mappage_pre}*)
  pte_t *pteaddr = walkpgdir(l4, va); (*\label{line:call_walkpgdirC}*)
  $\specline{ \textsf{R}_{\textsf{walk}} \ast \textsf{R}_{\textsf{l1e}} }$
  if (!pteaddr->present){ (*\label{line:mappage_pte_present_startC}*)
    /*The entry is not present(*\label{line:mappage_pte_present_endC}*)*/
    pteaddr->pfn = PTE_ADDR_TO_PFN(fpaddr); /*Store updated entry back to L1 entry (*\label{line:l1entry_storeC}*)*/
$\specline{ \textsf{l1\_entry+KERNBASE} \mapsto_{\textsf{vpte,1}} \textsf{l1\_entry} \; (\textsf{set\_pfn}(\textsf{l1e\_val}, \textsf{ADDR\_TO\_PFN}(\textsf{fpaddr}))) \ast  \ulcorner \lnot(\textsf{entry\_present l1e\_val}) \urcorner}$
    pteaddr->present = 1;(*\label{line:l1entry_setpresent}*) // Set the present bit in entry
  }
$\specline{ \begin{array}{l}(\textsf{l1\_entry+KERNBASE} \mapsto_{\textsf{vpte,qfrac}} \textsf{l1\_entry} \; \textsf{l1e\_val} \ast \ulcorner (\textsf{entry\_present l1e\_val}) \urcorner) \lor ( \ulcorner \lnot(\textsf{entry\_present l1e\_val}) \urcorner \\
 \ast \textsf{l1\_entry+KERNBASE} \mapsto_{\textsf{vpte,1}} \textsf{l1\_entry} \; (\textsf{set\_pfn}(\textsf{l1e\_val}, \textsf{ADDR\_TO\_PFN}(\textsf{fpaddr}))) )  \end{array} }$
}
\end{lstlisting}
\vspace{-2em}
  \caption{Specification of updating L1 entry to reference a new page (\textsf{fpaddr}).}
\label{fig:mapping_codeC}
\end{figure}

For brevity this example is specialized to the case where $\vaddr$ is known to not be mapped:
the precondition on Line \ref{line:mappage_pre} includes $\theta \; !!\; \vaddr = \texttt{None}$;
generalization to returning an error if it is already mapped is straightforward.
% .\footnote{
%   The generalization to handling either case and returning an error is straightforward.
% }
The precondition on Line \ref{line:mappage_pre}, directly entails
the precondition of the \lstinline|walkpgdir| call.\footnote{The proof of \lstinline|vaspace_mappage|'s caller
would extract this single identity map token for the specific L4 entry from a set of 512 that are
part of the kernel invariant, as \lstinline|walkpgdir|'s proof does for lower levels.
}
\lstinline|walkpgdir|, as just discussed,
returns the virtual address of an allocated L1 entry and its postcondition contains almost all of the information
needed to construct a virtual points-to for \lstinline|va| --- except information about the L1 entry
being present and pointing to a data page.
We already discussed for the upper level page-tables how the entry-present checks are handled, and
Line \ref{line:mappage_pte_present_startC} is similar: $\mathsf{R}_{\textsf{l1e}}$ includes the self-conditional
fractional permission for the L1 entry, so as it is not present, by Line \ref{line:l1entry_storeC}
it is known that full permission is held to update that entry.
Lines \ref{line:l1entry_storeC} and \ref{line:l1entry_setpresent} are verified using the pte-points-to
memory store rule. In the assembly proof, this is a bitwise-and of the word-aligned \lstinline|fpaddr| with 1
(setting the present flag), yielding a single store.
\looseness=-1

% By incorporating verification of the
% \lstinline|ensure_L1| function (see Section \ref{sec:traversing}), our verification also directly handles several subtle aspects which
% were axiomatized in prior work.
\ifPLDI
\else
\subsection{Unmapping a Page}
\todo[inline]{update (esp. line refs) for new mapping code}
The reverse operation, unmapping a designated page that is currently mapped,
would essentially be the reverse of
the reasoning around line 22 above: given the virtual points-to assertions for all 512
machine words of memory that the L1 entry would map,
and information about the physical location, 
full permission on the L1 entry could be obtained, allowing the construction of a
full virtual PTE pointer for it, setting to 0, and reclaiming the now unmapped physical memory.
\fi

\section{Related Work}
\label{sec:relwork}
We will discuss two streams of related work: OS verification, and 
program logics with modalities.

\paragraph{OS Kernel Verification}
There has been relatively little prior work on formal verification of virtual memory.
Most OS verification work minimizes reasoning about virtual memory management.
The original \textsc{Verisoft} project~\cite{alkassar2008verisoft,alkassar2010pervasive,alkassar2008formal,dalinger2005verification,hillebrand2005address,alkassar2008formal,starostin2010formal} 
relied on custom hardware which always ran kernel code with virtual memory disabled, removing the circularity that is a key challenge of verifying
VMM code for real hardware: at that point page tables become a basic partial map data structure to represent user program address translations,
with an idiosyncratic format. Subsequent OS verification work
\add{targets real hardware that uses virtual addressing in the kernel, but unsoundly
uses hardware models that do not.} 
\add{Thus} they \emph{trust}
that the particular page table manipulations do not, for example, unmap kernel code
 (which can crash the machine even if done ``temporarily'').
This is true for \textsc{seL4}~\cite{Klein2009seL4,seL4TOCS,Sewell2013translation}, whose formal machine model omits address translation,
and \textsc{CertiKOS}~\cite{gu15,gu2016certikos,gu2018certikos,chen2016interrupts}, whose refinement proofs rely on
\textsc{CompCert}'s usermode-oriented memory model~\cite{leroy2008formal,leroy2009formally} which assumes
updates to one memory address are independent of updates to another ---
which is not true of page table updates.
Other work on OS verification either never progressed to VMM verification
(\textsc{Verisoft XT}~\cite{cohen2009vcc,cohen2010local,dahlweid2009vcc,cohen2013SOFSEM}),
or uses memory-safe languages
for process isolation instead of address translation
(\textsc{Singularity}~\cite{Fahndrich2006language,Hunt2007singularity,Hunt2007sealing,Barnett2011specsharp}, \textsc{Verve}~\cite{Yang2010Verve},
and \textsc{Tock}~\cite{levy2017multiprogramming}), ensuring memory safety, but ignoring other functional uses of virtual memory
hardware, like swapping~\cite{Denning1970VM} or exploiting copy-on-write techniques for dynamic migration of virtual machines~\cite{clark2005live}.
\looseness=-1

\add{
 These limitations motivate work like ours on reasoning soundly about virtual memory management.
 As discussed earlier, 
}
\citet{kolanski08vstte} are the only other researchers to study VMM verification against a realistic hardware
model, where page table updates are performed through virtual memory accesses (later adding C-level type modeling~\cite{kolanski09tphols}).
As noted in Section \ref{sec:overly-restrictive}, Kolanski and Klein's virtual points-to is similar to that in Figure \ref{fig:strongvirtualpointsto},
with the attendant problems discussed earlier and lifted by our model. Their approach had modal elements,
but
 did not tackle evaluations that would benefit from modality.
 Our work improves significantly on the technical capabilities of ther logic and
 evaluates on kernel code that is more complete and more challenging than theirs.
\looseness=-1

\add{
 Our modal approach makes it possible to specify address space changes cleanly, which their logic cannot do at all.
 Our use of virtual pte-points-to assertions enables nearly the same proof rules
 as standard memory accesses, and constructing virtual points-to information within the logic
 (c.f.\ the logical entailment between $\textsf{R}_\textsf{walk}$ and $\textsf{L}_{4}\_\textsf{L}_{1}\_\textsf{PointsTo}$)
 whereas Kolanski and Klein must reason semantically about when the model state supports new virtual points-to assertions.
}

\add{
Kolanski and Klein verify the critical step of updating an already-located L1 entry to map a new page
(ARM assembly corresponding roughly to Lines \ref{line:l1entry_storeC} and \ref{line:l1entry_setpresent}
in our Figure \ref{fig:mapping_codeC}), but ignore the essential code
preceding that step --- which as our \lstinline|walkpgdir| and \lstinline|pte_get_next_table| verifications
demonstrate, side-steps a significant amount of complexity and critical reasoning tasks.
We have verified the entirety of the software page table walk up to mapping a new page, aside from a trusted physical memory allocator resembling
\lstinline|malloc|~\cite{Chlipala2013Bedrock,wickerson2010explicit}.
 As a consequence of tackling this larger verification challenge, our work is the first to formally specify large
 segments of the self-referential portion of an OS kernel's virtual memory management invariants (per Sections \ref{sec:p2vC} and \ref{sec:selfconditional}),
 and to reason about converting from physical addresses to virtual addresses efficiently.
\looseness=-1
}

Due to a lack of fractional permissions in their formalism, they incidentally pick up other limitations orthogonal to their
foundational focus: by requiring
a virtual points-to have \emph{full} ownership of the page table walk memory, they limit their system to having only as many
virtual points-to assertions as there are entries in the top-level table ({512}) because they cannot share access to entries.
We inherit fractional permission support form \iris, and use it extensively (the overly-restrictive
Figure \ref{fig:strongvirtualpointsto} is already an improvement in this way).
\add{While we cannot claim credit for \textsc{Iris}'s extensive feature set, the fact that the model of our
assertions is based on a classic algebraic tool (pointwise lifting) makes our approach compatible with
other logical bases as well, such as \textsc{PulseCore}~\cite{ebner2025pulsecore} or \textsc{VST}~\cite{appel2014program}.
}
\looseness=-1
 
Kolanski and Klein prove that updates to any memory locations that are not part of the page tables
do not affect the interpretation of other memory addresses, just like on real hardware.
This implies that programs that do not modify memory mappings could be reasoned about without
concern for mappings.
 An analagous result should hold of our model (though we have not proven it).
Informally this is visible in the rules for
\texttt{mov} instructions,
which are nearly identical to rules in a VM-ignorant logic~\cite{Chlipala2013Bedrock,ni2007contexts}.
  In principle both our approach and Kolanski and Klein's could
 be extended to account for pageable points-to assertions by adding 
disjunctions to an extended points-to definition\add{,} \add{which would be}
 the appropriate next step to extend reasoning to usermode programs running with a kernel that may demand-page the program's
memory.
\looseness=-1

As noted in Section \ref{sec:backgroundonmachinemodel}, we do not formally model or reason about \add{translation lookaside buffers} \add{(}TLBs\add{)}.
 TLB flushes are necessary when a page is \emph{un}mapped, or when switching address spaces.
This occurs in few places in uniprocessor kernels (in some, only 3 locations), \add{so} full verified
kernels including \textsc{seL4}~\cite{Klein2009seL4,seL4TOCS} and \textsc{CertiKOS}~\cite{gu15,gu2016certikos}
trust TLB management. Neither of the aforementioned systems has a hardware model including a TLB, so neither is able
to verify TLB management in any form --- they \emph{must} trust its operation.
 This is true of multicore verified kernels as well~\cite{von2013clustered,gu2016certikos}, though there
 the situation becomes much more challenging: when unmapping pages, the running CPU must invalidate the relevant TLB entries locally,
 but also send an inter-procesor interrupt (IPI) to all other cores to ensure they also invalidate affected entries on their TLBs.
 No formal hardware model currently exists for IPIs on any architecture, or even for the memory-mapped IO
 used to trigger those IPIs.

The only work to tackle TLB code verification was
\citet{syeda2018program,syeda2020formal},
who are also the only prior work on verifying address-space-aware context switching.
However, they verified only the 4 instructions to switch address spaces and update the TLB on an ARM processor, in isolation
(i.e., not the full context switch including changing stacks with address spaces).
The specification they proved for those instructions did not address program invariants that may be valid in one address space and not the other,
so is not flexible enough to extend directly to a full context switching primitive as in Figure \ref{fig:swtchC}.
\add{Their logic adds an assertion tracking known-inconsistent addresses (i.e., recently-unmapped by a table update or change in page table root)
and their memory access rules require accessed memory to lie outside that set.
% However their logic is proven sound against a generalized semantics where
% any page table change adds \emph{all}
% addresses to the inconsistent set,
% which is too imprecise for finer-grained unmapping.
However in their logic,
reasoning about updates to the inconsistent set (from a page table update)
requires direct reasoning
directly about the physical memory heap, which limits modularity.
}
The right general solution would be to combine our work
and \add{an extension of} theirs, \add{a substantial project} which we leave to future work.
\add{
 No other prior work has considered address space changes during context switching. \textsc{XCAP}~\cite{ni2007contexts}
 and Bedrock~\cite{Chlipala2011Bedrock,Chlipala2013Bedrock,Chlipala2015webapp} deal
 only with usermode threading (in a single process). \textsc{CertiKOS} and \textsc{seL4} trust assembly primitives for
 context switches, and do  not model address translation for executing code.
 \looseness=-1
}

\paragraph{Program Logics with Modalities}
Modalities have long been a staple of program logics, at least {since} Dijkstra's weakest precondition calculus~\cite{dijkstra1975guarded}
and Pratt's observation the Hoare triples
could be decomposed using the weakest-precondition modality of dynamic logic~\cite{pratt1976semantical},
in a form quite similar to what \iris uses today~\cite{jung2018iris}.
Variants of Nakano's later modality~\cite{nakano2000modality} have long been used to deal with step-indexing 
for impredicative and recursive features of logics and type systems~\cite{Appel2007,hobor2010theory,birkedal2011step,birkedal2013intensional,jung2018iris}.
\looseness=-1

As noted earlier, our other-space modality derives from hybrid logic~\cite{areces2001hybrid,blackburn1995hybrid,gargov1993modal,goranko1996hierarchies},
where modalities are indexed by \emph{nominals} which are names for specific individual states in a Kripke model.
\add{Readers mostly familiar with modalities in prominent program verification approaches~\cite{Appel2007,birkedal2013intensional,birkedal2011step,hobor2010theory,jung2018iris,nakano2000modality,pnueli1977temporal,pratt1976semantical}
may not recognize hybrid logics, but as we discuss in Section \ref{sec:relwork}, they (like temporal logics) trace their roots back to Arthur Prior in the 1950s.
}
Little prior work combines these ideas with program logics. \citet{brotherston2014parametric} show that traditional
nominals extends the expressive power of separation logic. \citet{gordon2019modal}
uses nominals to refer to states of other actors in an actor language.
In parallel with our work, \citet{wagner2024realistic} use a hybrid-logic-inspired modality
to abstract reference-counting specifics from specifications of a low-level application binary interface (ABI)
--- their $@_l(P)$ indicates that $l$ is the location of a reference count for resources satisfying $P$.
% As noted throughout the paper, the inspiration for our other-space modality comes from hybrid logic~\cite{areces2001hybrid,blackburn1995hybrid,gargov1993modal,goranko1996hierarchies},
% where modalities are indexed by \emph{nominals} which are names for specific individual states in a Kripke model.
% We are aware of only two prior works combining hybrid logics with program logics specifically. 
% Brotherston and Villard~\cite{brotherston2014parametric} demonstrated that may properties true of various 
% separation logics are not definable in boolean \BI (\BBI), and showed that a hybrid extension \HyBBI allows
% most such properties to be defined (e.g., the fact that separating conjunction is cancellative is unprovable 
% in boolean \BI, but provable in \HyBBI). There, nominals named resources 
% (roughly, but not exactly, heap fragments). 
% Gordon~\cite{gordon2019modal} described a use of hybrid logic in the verification of actor programs, 
% where nominals named the local state of individual actors (with such assertions stabilized with a 
% rely/guarantee approach). 
Beyond these, there is limited work on the interaction of hybrid logic with general substructural logics, in restricted forms
that do not affect expressivity~\cite{despeyroux2014hybrid,chaudhuri2019hybrid}.
\looseness=-1
% Primarily there is a line of work on hybrid linear logic (\HyLL)~\cite{despeyroux2014hybrid}, 
% originally used as a way to more conveniently express aspects of transition systems in linear logic. 
% However, \HyLL's proof rules offer no non-trivial interactions with multiplicative connectives 
% (every \HyLL proof can in fact be embedded into regular linear logic~\cite{chaudhuri2019hybrid}, 
% unlike Brotherston and Villard's \HyBBI, which demonstrably increases expressive power over its base \BBI.

% In both \HyLL and \HyBBI, nominals denote worlds with monoidal structure (as worlds in Kripke semantics
% for either LL or \BBI necessarily have monoidal structure). Our nominals, by contrast, 
% do not name worlds in the same sense with respect to Iris's CMRAs, 
% but in fact \emph{classes} of worlds, because the names are locations 
% (a means of \emph{selecting} resources) rather than resources.  
% A key difference is that the use of nominals in those logics corresponds specifically to hypothetical 
% reasoning about resources (until a nominal is connected to a current resource, in which case conclusions 
% can be drawn about the current resource), which means the modalities themselves do not ``own'' resources. 
% Instead, assertions under our other-space modality can and do
% have resource footprints.
% Pleasantly, we sidestep most of the metatheoretical complexity of those other substructural hybrid
% systems by building our logic within a substructural metatheory (\iris).

Some logics for weak memory models~\cite{dang2019rustbelt,dang2022compass} have been formalized
in \iris using pointwise lifting, parameterizing by thread-local views of events (an operationalization of the release-acquire + nonatomic
portion of the repaired C11 memory model~\cite{lahav2017repairing}). 
There modalities $\Delta_\pi(P)$ and $\nabla_\pi(P)$
represent that $P$ held before or will hold after certain memory fences by thread $\pi$.
The definitions of those specific modalities existentially quantify over other views, related to the ``current'' view (the one where
the current thread's assertions are evaluated), and evaluate $P$ with respect to those other views. This approach to parameterizing
assertion semantics by a point of evaluation, and evaluating modalized assertions at other points quanfied in the definition of a modality,
is the classic notion of modal assertions, 
whereas hybrid logics expose the choice of evaluation point in assertions,
allowing statements of more properties. 
In these weak memory examples this additional expressive power would not be useful,
because any relevant points of evaluation (thread views) are intimately tied to memory fences performed by the program, whereas
for virtual memory management the kernel must be able to choose or construct arbitrary other address spaces.
\looseness=-1

%   is what it means
% to have a modality at all.
% It is \emph{not}, however, an instance of hybrid logic, which is specifically demarcated by an assertion language where
% \emph{assertions}, not their semantics, choose and name the evaluation points for modal assertions.
% A hybrid extension of the aforementioned logics would include assertions which named specific views at which to evaluate
% $P$, in the syntax of the assertion (e.g., $\Delta_\pi^v(P):=\lambda\_\ldotp (P\;v)$) rather than the 
% $\Delta_\pi(P):= \lambda v\ldotp (\exists v_{rel}\ldotp \ownGhost{\pi}{\mathsf{RelV}(v_{rel})\;v} \ast (P\;v_{rel})))$ actually used.
% Note the hybrid version takes the place to evaluate $P$ as a parameter, and therefore allows the \emph{derived} (modal) logic to explicitly
% reason in terms of evaluation points, rather than hiding all points of evaluation in the internal definitions of modalities.
% This prior \iris-based work also uses modalities where the interpretations are fixed a priori by logic designers. In contrast, our
% address space modalities' interpretations can be changed by program behaviors via page table updates; the equivalent in the prior work would
% be if programs could directly manipulate the buffers used to model weak memory behaviors,
% which they cannot.\footnote{Note that the views modelled modally in this prior work are abstractions of a wide range of hardware, so such
% manipulation of corresponding hardware resources is in fact impossible,
% while the address space mappings in our work reflect known hardware components present in a variety of CPUs.}

\section{Conclusions}
This paper advances the state of the art in formal verification of programs manipulating virtual memory mappings.
We treat assertions about virtual memory locations explicitly as assertions in a modal logic, where the notion of context
is a particular address space, named by the page table root.
We improved the modularity of our reasoning about virtual address translation and virtual points-to assertions
to permit page table modifications that
preserve mappings without collecting all affected virtual points-to assertions.
To specify of code involving other address spaces, we adapt 
hybrid logic's notion of \add{modalities} explicitly naming alternative conditions.
We implemented these ideas in a derived separation logic within \iris, and proved soundness of
the rules for essential memory- and address-space-change-related x86-64 instructions 
against a hardware model of 64-bit 4-level address translation.
Finally, we used our rules to verify the correctness of key VMM instruction sequences,
including the first assembly-level proof of correctness for a change
of address space expressing which assertions hold in which address space, 
the first physical-to-virtual translation proof,
\add{and the first verification of a software page table walk, all of which
are beyond reach of prior work}.
\looseness=-1

\begin{acks}
This work was supported in part by US NSF Award \#CCF-1844964.
\end{acks}

\bibliographystyle{ACM-Reference-Format}
\bibliography{vmm}

%\ifARXIV
\appendix
\section{\add{Assembly-level Verification for Page Table Traversal and Mapping}}
\label{sec:experiment_appendix}
%To both validate and demonstrate the value of the modal approach to reasoning about virtual memory management, 
% we study several
% We validate our logic by studying
% distillations of key VMM functionality.
% real concerns of virtual memory managers.
% Recall from Section \ref{sec:logic} that virtual points-to assertions work just like regular points-to assertions, by design.

\replace{
In this section we verify several critical and challenging pieces of VMM code.
First, in several stages, we work up to mapping a new page in the current address space.
This requires a number of independently challenging substeps: dynamically traversing a page table to find
the appropriate L1 entry to update; inserting additional levels of the page table if necessary (updating
the VMM invariants along the way);
converting the physical addresses found in intermediate entries into the corresponding virtual addresses
that can be used for memory access;
installing the new mapping;
and collecting sufficient resources to form a virtual points-to assertion.
Of these, only the second-to-last step (installing the correct mapping into the
current address space) has previously been formally verified with respect to a machine model with address translation.
Second, we formally verify a switch into a new address space as part of a task switch,
the first such verification handling both old and new processes' assertions (in different address spaces) at the time of the switch.
}{
While our logic was developed and proven sound for x86-64 assembly,
Section \ref{sec:traversingC} described verification of software page table walking code (\lstinline|pte_get_next_table| and \lstinline|walkpgdir|)
as if at the level of C for improved readability.
This appendix describes the actual assembly-level verification carried out in Rocq.
Careful readers of both Section \ref{sec:traversingC} and this appendix will notice
strong similarities in the assertions and and reasoning, for good reason:
The C code in Section \ref{sec:traversingC} was the original kernel code that was compiled
(with no optimizations) to x86-64 assembly and verified with our logic, and the proof outlines
in that section largely back-port the assembly proofs back to C.
\looseness=-1
}

\add{
 This section describes the assembly proofs without reference to the C outlines given in Section \ref{sec:traversingC}.
 The main additional details of note at the assembly level are:
 \begin{itemize}
 \item Accurate treatment of register management (particularly the AMD64 System V calling convention) leading to more direct correspondence
       with our logic
 \item The assembly is naturally more verbose than the C, so the proof outlines are relatively more sparse, with assertions written
       only for key updates.
 \item Bitwise manipulations of page table entries are harder to follow than C's bitfield access support.
       Multiple manipulations which are each explicit in C become adjacent (or sometimes non-adjacent) bitwise operations.
       The critical ones are commented in the assembly figures.
 \item And compared to the C-based presentation earlier, there are differences in logical variable names. For example,
       the assembly proofs use \textsf{entry} as the name for the \emph{physical} address of the entry modified by
       \lstinline|pte_get_next_table| in Figure \ref{fig:calltopteinitialize}, whereas to make sense of the C code
       in Figure \ref{fig:calltopteinitializeC} we used \textsf{entry} consistently with the C variable name and introduced
       separate logical names for physical addresses. This propagates to figures presenting larger invariants separately,
       as they also refer to the logical names from the proofs.
 \end{itemize}
}

\subsection{Traversing Live Page Tables}
\label{sec:traversing}
We build up to the main task of mapping a new page after traversing page tables in software.
The mapping operation of Figure \ref{fig:mapping_code} assumes an operation \textsf{walkpgdir} which must traverse the page tables
in order to locate the address of the L1 entry to update --- 
% possibly allocating tables for levels 3, 2, and 1 in the process,
% installing them into levels 4, 3, and 2, along the way.
possibly allocating new L3, L2, and L1 tables as necessary.
Traversing the page tables is itself challenging functionality to verify: loading the current table root from \lstinline|cr3| is straightforward
(a \lstinline|mov| instruction), however this produces the physical address of \lstinline|cr3|, not the virtual address the kernel code would use to access that memory.
This problem repeats at each level of the page table: assuming the code has \emph{somehow} read the appropriate L4 (or L3, or L2) entry, those entries again
yield physical addresses, not virtual.

\subsubsection{Loading Page-Table Address Value}
We will discuss access to the level 4 table later (Section \ref{wlkpgdir}). But for subsequent levels, the base address of level $n$ must be
fetched from the appropriate entry in the level $n+1$ table.
This is the role of \lstinline|pte_get_next_table| (Figures \ref{fig:calltopteinitialize} and \ref{fig:p2v}):
it is passed the virtual address of the page table entry in level $n+1$, and should return the \emph{virtual} 
address of the \emph{base} of the level $n$ table
indicated by that entry.
If the entry is empty (i.e., this is a sparse part of the page table representation),
the code also allocates a page for the level $n$ table, installs it in the level $n+1$ entry, and establishes appropriate invariants.
Figure \ref{fig:calltopteinitialize} presents the initial part of the function, which performs the allocation if necessary.
Figure \ref{fig:p2v} (discussed in Section \ref{sec:p2v}) deals with the cases where no allocation is necessary \emph{or} the allocation has already
been performed by the code in this figure.
\looseness=-1

Note that the specification does \emph{not} assume a specific page table level --- logical parameter \textsf{v} represents the level
of the entry passed as an argument, and this code
is used for all three level transitions when traversing page tables (4 to 3, 3 to 2, 2 to 1).
This comes into play with a subtlety of the specification of \lstinline|pte_get_next_table| that we will
revisit several times: \lstinline|pte_get_next_table|'s specification
assumes it is given a virtual \emph{vpte-pointsto}
(a virtual points-to exposing the underlying physical address instead of existentially quantifying it;
 see Section \ref{sec:mapnew}) granting access to the specified entry,
but its postcondition does not yield new virtual points-to assertions!
Instead it merely computes the base virtual address of the next table, and returns adequate capabilities (discussed in Section \ref{subsec:identitymappings})
for the \emph{caller} to construct a vpte-pointsto for any entry of the next table level (if this is not an L1 entry ---
the caller knows which level of the table this is for).
\looseness=-1

Within \textsf{get\_next\_table}, after a standard function prologue, the code 
loads the entry pointed to by the argument (logical variable \textsf{entry} in the proof outline).
This is a page table entry: a 64-bit word divided into bit-fields for
the physical address of the next table, and control bits like the valid bit, as discussed in 
Section \ref{sec:backgroundonmachinemodel}.

\ifPLDI
Line \ref{line:mask_present} checks % In the condensed figure, it's all on one line
\else
Lines \ref{line:mask_present}--\ref{line:check_entry_present} check
\fi
if the entry's ``present'' bit is set.
If it is zero, a new page must be allocated for the next level of the table --- which is done by the fall-through
from Line \ref{line:check_entry_present_jump}'s conditional jump. Otherwise the code jumps ahead to
the case for the next level already existing, which is discussed in Section \ref{sec:p2v} and Figure \ref{fig:p2v}.
First, we must discuss another refinement of the address space invariant, establishing
enough structure on the page tables themselves to allow the traversal.
The code for allocating a new level of the page table must establish this extended invariant.

%wshiftll (wshiftll (natToWord 64 entry) (WordImpl.concat (WordImpl.zero 56) (WordImpl.from_nat 8 12 ^& WordImpl.concat (WordImpl.zero 2) WO~1~1~1~1~1~1)) ^& constf)
%(WordImpl.concat (WordImpl.zero 56) (natToWord 8 12 ^& WordImpl.concat (WordImpl.zero 2) WO~1~1~1~1~1~1))
%
%wshiftll
 %      (wshiftll
%          ((((natToWord 64 entry ^& WordImpl.concat (WordImpl.zero 32) consta ^| WordImpl.concat (WordImpl.zero 32) (natToWord 32 2))
%             ^& WordImpl.concat (WordImpl.zero 32) constb ^| WordImpl.concat (WordImpl.zero 32) (natToWord 32 4)) ^& constd
%            ^| wshiftll
%                 (wshiftll (nextpaddr ^+ ^~ (natToWord 64 KERNBASE))
%                    (WordImpl.concat (WordImpl.zero 56) (WordImpl.from_nat 8 12 ^& WordImpl.concat (WordImpl.zero 2) WO~1~1~1~1~1~1))
%                  ^& constf)
%                 (WordImpl.concat (WordImpl.zero 56) (WordImpl.from_nat 8 12 ^& WordImpl.concat (WordImpl.zero 2) WO~1~1~1~1~1~1)))
%           ^& WordImpl.concat (WordImpl.zero 32) conste ^| wone 64)
%          (WordImpl.concat (WordImpl.zero 56) (WordImpl.from_nat 8 12 ^& WordImpl.concat (WordImpl.zero 2) WO~1~1~1~1~1~1)) ^& constf)
%       (WordImpl.concat (WordImpl.zero 56) (natToWord 8 12 ^& WordImpl.concat (WordImpl.zero 2) WO~1~1~1~1~1~1)) 
\begin{figure}%\footnotesize
\ifPLDI
\begin{lstlisting}[mathescape, basicstyle=\footnotesize,escapeinside={(*}{*)}]
;;pte_t *pte_get_next_table(pte_t *entry) { ... // setting up the stack for a pte_t *next elided
$\specline{\textsf{P} \ast  I\texttt{ASpace}_{\textsf{id}}(\theta,\Xi\setminus\{\textsf{entry}\},m)  \ast \texttt{rbp-8} \mapsto_{\textsf{v}} \textsf{entry} \ast  \texttt{rbp-16} \mapsto_{\textsf{v}} \textsf{next} \ast \texttt{r8}  \mapsto_{\textsf{r}} \textsf{\_} \ast \texttt{rdi}  \mapsto_{\textsf{r}} \textsf{\_} \ast \ghostmaptoken{\delta{}s}{\rtv}{\delta}  }_{\rtv}$
$\specline{ \textsf{entry+KERNBASE} \mapsto_{\textsf{vpte,qfrac}} \textsf{entry entry\_val} \ast \ulcorner  \textsf{qfrac} = 1 \leftrightarrow \; \lnot(\textsf{entry\_present entry\_val})\urcorner}_{\rtv}$ (*\label{line:get_next_vpte_precondition}*)
$\specline{\ulcorner\textsf{entry\_present(entry\_val)}\urcorner \wand \forall_{i\in\textsf{0..511}} \; \ghostmaptoken{\textsf{id}}{\textsf{table\_root(entry\_val.pfn) + i * 8}}{\textsf{v-1}} }_{\rtv} $ (*\label{line:conditional_children}*)
mov    -0x8[rbp],rdi $\specline{\ldots \ast \texttt{rdi}  \mapsto_{\textsf{r}} \textsf{entry} \ast \ldots }$
mov     rdi, r8 $\specline{\ldots \ast \texttt{r8}  \mapsto_{\textsf{r}} \textsf{entry} \ast \ldots }$
mov    [r8],rdi (*\label{line:read_entry_contents}*) $\specline{\ldots \ast \texttt{rdi}  \mapsto_{\textsf{r}} \textsf{entry\_val}  \ast\ldots}$ ;; reading page table entry Justified by virtual pte-pointsto
and    0x1,rdi;  mov    rdi,rax; cmp    0x0,rax ;;  if (!entry->present) {(* \label{line:check_entry_present} *)(* \label{line:mask_present} *)
jne    161 <pte_get_next_table+0xa1> (* \label{line:check_entry_present_jump} *) ;; Jump if the present bit is not zero; no need to allocate next level
$\specline{\ldots \ast \texttt{rdi}  \mapsto_{\textsf{r}} \textsf{entry\_val \& 0x1} }_{\rtv}$
mov    rbp,rdi; sub    0x10, rdi ;; Store the value of rbp minus 16 bytes (address of (*\textsf{next}*)) into rdi (*\label{line:pass_addrof_next}*)(*\label{line:alloc_path_start}*)
$\specline{ \textsf{entry+KERNBASE} \mapsto_{\textsf{vpte,qfrac}} \textsf{entry entry\_val} \ast \ulcorner  \textsf{qfrac} = 1 \land \lnot(\textsf{entry\_present entry\_val})  \urcorner }_{\rtv}$ (*\label{line:after_concluding_qfrac1}*)
callq  70 <pte_initialize> ;;pte_initialize(entry);(* \label{line:call_to_pte_initialize} *)
$\specline{ \ldots \ast \ulcorner  \textsf{qfrac} = 1 \land \lnot(\textsf{entry\_present entry\_val})  \urcorner \ast \ulcorner  \textsf{qfrac} = 1 \land \lnot(\textsf{entry\_present (pfn\_set (entry\_val nextpaddr))}  \urcorner   }_{\rtv}$
$\specline{ \texttt{rbp-16} \mapsto_{\textsf{v}} \textsf{pfn\_set(entry\_val nextpaddr)}   }_{\rtv}$
$\specline{ \begin{array}{l}\textsf{entry\_present (pte\_initialized (pfn\_set(entryv nextpaddr)))} \wand  \\ \;\;\;\;\;\;\;\;\  \forall_{i\in \textsf{0 ... 511} } \ldotp  \ghostmaptoken{\textsf{id}}{((\textsf{table\_root (pte\_initialized (pfn\_set(entry\_val nextpaddr)))}) + \textsf{i * 8})}{\textsf{v-1}} \end{array}  }$ (*\label{line:page_of_caps}*)
... ;;entry value updates: entry->pfn = nextpaddr; entry->present = 1; (*\label{line:install_new_entry}*)
... ;;now we know that entry is initialized, so we satisfy the condition to access children list
$\specline{ \textsf{entry+KERNBASE} \mapsto_{\textsf{vpte,qfrac}}  \textsf{pte\_initialized(pfn\_set(entryv nextpaddr))} \ast \ulcorner  \textsf{qfrac} = 1 \land \lnot(\textsf{entry\_present entry\_val})  \urcorner }_{\rtv}$
$\specline{ \texttt{rbp-16} \mapsto_{\textsf{v}} \textsf{pte\_initialized(pfn\_set(entry\_val nextpaddr))}  \ast rax \mapsto_{r} \textsf{table\_root (pte\_initialized (pfn\_set(entry\_val nextpaddr)))}  }_{\rtv}$
$\specline{\forall_{i\in \textsf{0 ... 511} } \ldotp  \ghostmaptoken{\textsf{id}}{((\textsf{table\_root (pte\_initialized (pfn\_set(entry\_val nextpaddr)))}) + \textsf{i * 8})}{\textsf{v-1}}  }$ (*\label{line:alloc_path_end}*)
;;} (*\label{line:end_of_allocation_path}*)
... ;; Code after conditional continued in Figure (*\ref{fig:p2v}*)
\end{lstlisting}
\else
\begin{lstlisting}[mathescape, basicstyle=\footnotesize,escapeinside={(*}{*)}]
;;pte_t *pte_get_next_table(pte_t *entry) {
... ;; setting up the stack
;; pte_t *next;
$\specline{\textsf{P} \ast  I\texttt{ASpace}_{\textsf{id}}(\theta,\Xi\setminus\{\textsf{entry}\},m)  \ast \texttt{rbp-8} \mapsto_{\textsf{v}} \textsf{entry} \ast  \texttt{rbp-16} \mapsto_{\textsf{v}} \textsf{next} \ast \texttt{r8}  \mapsto_{\textsf{r}} \textsf{\_} \ast \texttt{rdi}  \mapsto_{\textsf{r}} \textsf{\_} \ast \ghostmaptoken{\delta{}s}{\rtv}{\delta}  }_{\rtv}$
$\specline{ \textsf{entry+KERNBASE} \mapsto_{\textsf{vpte,qfrac}} \textsf{entry entry\_val} \ast \ulcorner  \textsf{qfrac} = 1 \leftrightarrow \; \lnot(\textsf{entry\_present entry\_val})\urcorner}_{\rtv}$ (*\label{line:get_next_vpte_precondition}*)
$\specline{\ulcorner\textsf{entry\_present(entry\_val)}\urcorner \wand \forall_{i\in\textsf{0..511}} \; \ghostmaptoken{\textsf{id}}{\textsf{table\_root(entry\_val.pfn) + i * 8}}{\textsf{v-1}} }_{\rtv} $ (*\label{line:conditional_children}*)
mov    -0x8[rbp],rdi
$\specline{\textsf{P} \ast  I\texttt{ASpace}_{\textsf{id}}(\theta,\Xi\setminus\{entry\},m)   \ast \textsf{rbp-8} \mapsto_{\textsf{v}} \textsf{entry} \ast \texttt{r8}  \mapsto_{\textsf{r}} \textsf{\_}  \ast  \texttt{rbp-16} \mapsto_{\textsf{v}} \textsf{next} \ast \texttt{rdi}  \mapsto_{\textsf{r}} \textsf{entry} \ast \ghostmaptoken{\delta{}s}{\rtv}{\delta}  }_{\rtv}$
$\specline{ \textsf{entry+KERNBASE} \mapsto_{\textsf{vpte,qfrac}} \textsf{entry entry\_val} \ast \ulcorner  \textsf{qfrac} = 1 \leftrightarrow \; \lnot(\textsf{entry\_present entry\_val})\urcorner}_{\rtv}$
$\specline{\ulcorner\textsf{entry\_present(entry\_val)}\urcorner \wand \forall_{i\in\textsf{0..511}} \; \ghostmaptoken{\textsf{id}}{\textsf{table\_root(entry\_val.pfn) + i * 8}}{\textsf{v-1}} }_{\rtv} $
mov     rdi, r8
$\specline{\textsf{P} \ast  I\texttt{ASpace}_{\textsf{id}}(\theta,\Xi\setminus \{ \mathsf{entry} \},m)  \ast \textsf{rbp-8} \mapsto_{\textsf{v}} \textsf{entry} \ast \texttt{r8}  \mapsto_{\textsf{r}} \textsf{entry}  \ast  \texttt{rbp-16} \mapsto_{\textsf{v}} \textsf{next} \ast \texttt{rdi}  \mapsto_{\textsf{r}} \textsf{entry} \ast \ghostmaptoken{\delta{}s}{\rtv}{\delta}  }_{\rtv}$
$\specline{ \textsf{entry+KERNBASE} \mapsto_{\textsf{vpte,qfrac}} \textsf{entry entry\_val} \ast \ulcorner  \textsf{qfrac} = 1 \leftrightarrow \; \lnot(\textsf{entry\_present entry\_val})\urcorner}_{\rtv}$
$\specline{\ulcorner\textsf{entry\_present(entry\_val)}\urcorner \wand \forall_{i\in\textsf{0..511}} \; \ghostmaptoken{\textsf{id}}{\textsf{table\_root(entry\_val.pfn) + i * 8}}{\textsf{v-1}} }_{\rtv} $
mov    [r8],rdi (*\label{line:read_entry_contents}*)
$\specline{\textsf{P} \ast  I\texttt{ASpace}_{\textsf{id}}(\theta,\Xi\setminus \{ \mathsf{entry} \},m)  \ast \textsf{rbp-8} \mapsto_{\textsf{v}} \textsf{entry} \ast \texttt{r8}  \mapsto_{\textsf{r}} \textsf{entry}  \ast  \texttt{rbp-16} \mapsto_{\textsf{v}} \textsf{next} \ast \texttt{rdi}  \mapsto_{\textsf{r}} \textsf{entry\_val}  \ast \ghostmaptoken{\delta{}s}{\rtv}{\delta}  }_{\rtv}$
$\specline{ \textsf{entry+KERNBASE} \mapsto_{\textsf{vpte,qfrac}} \textsf{entry entry\_val} \ast \ulcorner  \textsf{qfrac} = 1 \leftrightarrow \; \lnot(\textsf{entry\_present entry\_val})  \urcorner }_{\rtv}$
$\specline{\ulcorner\textsf{entry\_present(entry\_val)}\urcorner \wand \forall_{i\in\textsf{0..511}} \; \ghostmaptoken{\textsf{id}}{\textsf{table\_root(entry\_val.pfn) + i * 8}}{\textsf{v-1}} }_{\rtv} $
and    0x1,rdi (* \label{line:mask_present} *)
mov    rdi,rax
cmp    0x0,rax ;;  if (!entry->present) {(* \label{line:check_entry_present} *)
jne    161 <pte_get_next_table+0xa1> (* \label{line:check_entry_present_jump} *) ;; Jump if the present bit is not zero
$\specline{\textsf{P} \ast  I\texttt{ASpace}_{\textsf{id}}(\theta,\Xi\setminus \{ \mathsf{entry} \},m)  \ast \textsf{rbp-8} \mapsto_{\textsf{v}} \textsf{entry} \ast \texttt{r8}  \mapsto_{\textsf{r}} \textsf{entry} \ast \texttt{rbp-16} \mapsto_{\textsf{v}} \textsf{next} \ast \texttt{rdi}  \mapsto_{\textsf{r}} \textsf{entry\_val \& 0x1} }_{\rtv}$
$\specline{  \textsf{entry} \mapsto_{\textsf{id}} \textsf{v} \ast \ghostmaptoken{\delta{}s}{\rtv}{\delta}  \ast \texttt{rax}  \mapsto_{\textsf{r}} \textsf{entry\_val \& 0x1} \ast \ulcorner\textsf{entry\_val \& 0x1}=\textsf{0x0}\urcorner}_{\rtv}$
$\specline{ \textsf{entry+KERNBASE} \mapsto_{\textsf{vpte,qfrac}} \textsf{entry entry\_val} \ast \ulcorner  \textsf{qfrac} = 1 \leftrightarrow  \lnot(\textsf{entry\_val \& 0x1}=\textsf{0x1})  \urcorner }_{\rtv}$ (*\label{line:before_concluding_qfrac1}*)
mov    rbp,rdi (*\label{line:alloc_path_start}*)
sub    0x10, rdi ;; Store the value of rbp minus 16 bytes (address of (*\textsf{next}*)) into rdi (*\label{line:pass_addrof_next}*)
$\specline{\textsf{P} \ast  I\texttt{ASpace}_{\textsf{id}}(\theta,\Xi\setminus \{ \mathsf{entry} \},m)  \ast \textsf{rbp-8} \mapsto_{\textsf{v}} \textsf{entry} \ast \texttt{r8}  \mapsto_{\textsf{r}} \textsf{entry} \ast \texttt{rbp-16} \mapsto_{\textsf{v}} \textsf{next} \ast \texttt{rdi}  \mapsto_{\textsf{r}} \textsf{rbp - 16} }_{\rtv}$
$\specline{  \textsf{entry} \mapsto_{\textsf{id}} \textsf{v} \ast \ghostmaptoken{\delta{}s}{\rtv}{\delta} }_{\rtv}$
$\specline{ \textsf{entry+KERNBASE} \mapsto_{\textsf{vpte,qfrac}} \textsf{entry entry\_val} \ast \ulcorner  \textsf{qfrac} = 1 \land \lnot(\textsf{entry\_present entry\_val})  \urcorner }_{\rtv}$ (*\label{line:after_concluding_qfrac1}*)
callq  70 <pte_initialize> ;;pte_initialize(entry);(* \label{line:call_to_pte_initialize} *)
$\specline{\textsf{P} \ast  I\texttt{ASpace}_{\textsf{id}}(\theta,\Xi\setminus \{ \mathsf{entry} \},m)  \ast \textsf{rbp-8} \mapsto_{\textsf{v}} \textsf{entry} \ast \texttt{r8}  \mapsto_{\textsf{r}} \textsf{entry}  }_{\rtv}$
$\specline{  \textsf{entry} \mapsto_{\textsf{id}} \textsf{v} \ast \ghostmaptoken{\delta{}s}{\rtv}{\delta} }_{\rtv}$
$\specline{ \textsf{entry+KERNBASE} \mapsto_{\textsf{vpte,qfrac}} \textsf{entry entry\_val} \ast \ulcorner  \textsf{qfrac} = 1 \land \lnot(\textsf{entry\_present entry\_val})  \urcorner }_{\rtv}$
$\specline{\ulcorner  \textsf{qfrac} = 1 \land \lnot(\textsf{entry\_present (pfn\_set (entry\_val nextpaddr))}  \urcorner   }_{\rtv}$
$\specline{ \texttt{rbp-16} \mapsto_{\textsf{v}} \textsf{pfn\_set(entry\_val nextpaddr)}   }_{\rtv}$
$\specline{ \begin{array}{l}\textsf{entry\_present (pte\_initialized (pfn\_set(entryv nextpaddr)))} \wand  \\ \;\;\;\;\;\;\;\;\  \forall_{i\in \textsf{0 ... 511} } \ldotp  \ghostmaptoken{\textsf{id}}{((\textsf{table\_root (pte\_initialized (pfn\_set(entry\_val nextpaddr)))}) + \textsf{i * 8})}{\textsf{v-1}} \end{array}  }$ (*\label{line:page_of_caps}*)
... ;;entry value updates: entry->pfn = nextpaddr; entry->present = 1; (*\label{line:install_new_entry}*)
... ;;now we know that entry is initialized, so we satisfy the condition to access children list
$\specline{\textsf{P} \ast  I\texttt{ASpace}_{\textsf{id}}(\theta,\Xi\setminus \{ \mathsf{entry} \},m)  \ast \textsf{rbp-8} \mapsto_{\textsf{v}} \textsf{entry} \ast \texttt{r8}  \mapsto_{\textsf{r}} \textsf{entry}  }_{\rtv}$
$\specline{  \textsf{entry} \mapsto_{\textsf{id}} \textsf{v} \ast \ghostmaptoken{\delta{}s}{\rtv}{\delta} }_{\rtv}$
$\specline{ \textsf{entry+KERNBASE} \mapsto_{\textsf{vpte,qfrac}}  \textsf{pte\_initialized(pfn\_set(entryv nextpaddr))} \ast \ulcorner  \textsf{qfrac} = 1 \land \lnot(\textsf{entry\_present entry\_val})  \urcorner }_{\rtv}$
$\specline{ \texttt{rbp-16} \mapsto_{\textsf{v}} \textsf{pte\_initialized(pfn\_set(entry\_val nextpaddr))}  \ast rax \mapsto_{r} \textsf{table\_root (pte\_initialized (pfn\_set(entry\_val nextpaddr)))}  }_{\rtv}$
$\specline{\forall_{i\in \textsf{0 ... 511} } \ldotp  \ghostmaptoken{\textsf{id}}{((\textsf{table\_root (pte\_initialized (pfn\_set(entry\_val nextpaddr)))}) + \textsf{i * 8})}{\textsf{v-1}}  }$ (*\label{line:alloc_path_end}*)
;;} (*\label{line:end_of_allocation_path}*)
... ;; Code after conditional continued in Figure (*\ref{fig:p2v}*)
\end{lstlisting}
\fi
% ;;uintptr_t next_virt_addr = (uintptr_t) P2V(next_phys_addr);
% movabs KERNBASE,rcx
% add    rcx,rax
% ...
% ;;next = (pte_t *) next_virt_addr;
% ;;clean up the stack and return next
% \end{lstlisting}
\vspace{-1em}
\caption{Ensuring \textsf{entry} points to a valid next table, allocating if necessary.}
\label{fig:calltopteinitialize}
\vspace{-1em}
\end{figure}

\subsubsection{Identity Mappings}
\label{subsec:identitymappings}
Kernels need to convert between physical and virtual addresses, in both directions.
Traversing the page tables in software is the simplest way to convert a virtual address to a physical address; this is the context we are working up to.
However, implementing this virtual-to-physical (V2P) translation in this way ironically requires physical-to-virtual (P2V) translation,
because the addresses stored in page table entries are physical, but memory accesses issued by the OS code use virtual addresses.
% There is no universal way to convert physical addresses to virtual --- doing so relies on the kernel maintaining careful invariants or
% additional data structures to enable P2V translation.
\looseness=-1

Because VMM operations are performance-critical for many workloads, most kernels 
maintain invariants that enable very fast P2V conversions (rather than adding another data structure).
Most kernels maintain an invariant on their page tables that the virtual address of any page used for a page table 
% lives at a virtual address whose value 
is \emph{a constant offset from the physical address} --- a practice sometimes referred to as \emph{identity mapping} 
(even though the physical-to-virtual translation
is typically not literally the identity function, but adding a non-zero constant offset).\footnote{Some kernels do this for all physical memory on the machine, simplifying interaction
with DMA devices.
On newer platforms like RISC-V, this sometimes truly is an identity mapping ---
x86-64 machines are forced into offsets by backwards compatibility with bootloaders that cannot access the full memory space of the
machine.
}

For this reason we extend the per-address-space invariant as in Figure \ref{fig:peraspaceinvariant_with_p2v_extension}, to also track which
addresses we can perform a P2V conversion on by a adding a constant offset.
$\Xi$ is another ghost map, from physical addresses to the level of the page table they represent (1--4).
\emph{Only} physical addresses in $\Xi$ can undergo P2V conversion. 
Section \ref{sec:p2v} describes the actual conversion,
but we describe the invariant here 
because adding new level 3/2/1 tables must maintain the invariant.

\begin{figure*}
\footnotesize
\[
\begin{array}{l}
   I\textsf{ASpace}_{\textsf{id}}(\ptablestore,\Xi,m)\stackrel{\triangle}{=} \textsf{ASpace\_Lookup}_{\textsf{id}}(\ptablestore,\Xi,m) \ast \mathsf{GhostMap}(\mathsf{id},\Xi)\ast\\
  \left(\bigast{(\vaddr, \textsf{paddr})\in \ptablestore}{\exists\;(\textsf{l4e, l3e, l2e, l1e, paddr})\ldotp \textsf{L}_{4}\_\textsf{L}_{1}\_\textsf{PointsTo}(\vaddr\textsf{, l4e, l3e, l2e, l1e, paddr})}\right)\ast \\
  \bigast{(\paddr,\mathsf{level}) \in \Xi}{\exists\; (\textsf{qfrac, q, val,}\vaddr) \ldotp \ulcorner \vaddr = \paddr + \textsf{KERNBASE} \; \textsf{level} > 1\urcorner \ast  \underbrace{\fracghostmaptoken{\delta}{\vaddr}{\paddr}{\qfrac} }_\text{Ghost translation} \ast \underbrace{\paddr \mapsto_{\mathsf{p}}\{\textsf{qfrac}\}\; \vale}_\text{Physical location}} \ast\\
   \qquad\underbrace{ \ulcorner \textsf{qfrac} = 1 \leftrightarrow \; \lnot\textsf{entry\_present }(\vale) \urcorner}_\text{Entry validity}\ast 
    \underbrace{\left(\ulcorner\textsf{present\_L}(\vale,\mathsf{level})\urcorner \wand \forall_{\textsf{i}\in\textsf{0..511}} \ldotp \ghostmaptoken{\textsf{id}}{((\mathsf{entry\_page}\;\vale) + \textsf{i * 8})}{\textsf{level-1}}\right)}_{\text{Indexing into next level of tables}} \\
  \textsf{ where } \\
   \textsf{ASpace\_Lookup}_{\textsf{id}}(\ptablestore,\Xi,m) \stackrel{\triangle}{=} \lambda\textsf{ cr3val} \ldotp \; \exists \gammaPred \; \ldotp \ulcorner m \; !!\; \textsf{cr3val} = \textsf{Some } \gammaPred \urcorner \ast
   % \ownGhost\gammaPred{\authfull{\ptableabswalk\ptablestore}} \ast  \ownGhostpv\gammaPred{\authfull{\pvmapping\Xi}}
   \ptableabswalk{\delta,\ptablestore} \ast \pvmapping{\delta,\Xi}\\
  \textsf{present\_L}(\vale,\mathsf{level})\stackrel{\triangle}{=} \mathsf{entry\_present}(\vale)\land \mathsf{level} > 0
  
\end{array}
\]
\vspace{-1em}
\caption{Global Address-Space Invariant in Figure \ref{fig:peraspaceinvariant} extended with a ghost map bookkeeping identity mappings }
  \label{fig:peraspaceinvariant_with_p2v_extension}
\vspace{-1em}
\end{figure*}

For each $\paddr\mapsto \textsf{v} \in\Xi$, the invariant tracks a virtual points-to justifying that virtual address $\paddr+\textsf{KERNBASE}$ maps to physical address $\paddr$
(the ``Ghost translation'' in Figure \ref{fig:peraspaceinvariant_with_p2v_extension});
fractional ownership of the physical memory for that page table entry;
and for valid entries (with the present bit set) above L1, ghost map tokens for every entry in the table pointed to by the entry, which can be used
to repeat the process one level down. 
% (L1 entries point to data pages, whose physical memory ownership resides in some virtual points-to).
The assertion on Line \ref{line:conditional_children} of Figure \ref{fig:calltopteinitialize} comes from the invariant one level up; 
if the valid bit is set,
the code can return those child tokens without the conditional guard.
\looseness=-1

The fractional ownership of the entry's physical memory is subtle. Recall that $\textsf{L}_{4}\_\textsf{L}_{1}\_\textsf{PointsTo}$ retains some physical
ownership of each page table entry that is traversed (proportional to how many virtual addresses share the entry).
So in general the invariant cannot keep full permission to the memory in this part of the invariant, or it would overlap the page table walk for virtual points-to
assertions. But in the case where the entry is invalid, we may need to write to it (e.g., to install a reference to a next-level table, as we do in Figure \ref{fig:calltopteinitialize}),
which requires full permission. Fortunately, the entry can only be in use if its valid bit is set; if the valid bit is not set we know
that no virtual points-to entry in $\delta$/$\theta$ holds any partial ownership.
Thus we use the invariant portion annotated as ``Entry validity'' in Figure \ref{fig:peraspaceinvariant_with_p2v_extension} to capture this:
if the entry is invalid the invariant holds full ownership of the entry, so it can be updated; while if the entry is valid,
the invariant owns only a constant non-zero fragment sufficient to read the entry, but not modify it (which would invalidate some virtual points-to assertions):
\begin{equation*}
 \ulcorner \textsf{qfrac} = 1 \leftrightarrow \; \lnot\textsf{entry\_present }(\vale) \urcorner \tag{*}
\end{equation*}
Thus the fractional ownership of the physical location is enough for Line \ref{line:read_entry_contents} in Figure \ref{fig:calltopteinitialize} to access the entry, though in \lstinline|get_next_table|
the caller has pulled that piece of information out of the invariant and passed it for the entry at hand.
This removal appears explicitly in assertions,
as the argument to the invariant is $\Xi\setminus\{\mathsf{entry}\}$ (indexing by the set $\Xi$ allows us to borrow the physical resources
for a specific page table entry out of the invariant, and later put them back).
Line \ref{line:check_entry_present_jump}'s conditional then determines in the fall-through case that the bit is not set, which 
together with other facts entails $\textsf{qfrac} = 1$ at Line \ref{line:after_concluding_qfrac1},
and permits storing a new entry (in ellided code around Line \ref{line:install_new_entry}).
\looseness=-1

This seemingly-simple piece of code has a highly non-trivial correctness argument, which depends critically on detailed invariants on how access to page table
entries is shared between parts of the kernel. No prior work has engaged with this problem.

 \subsubsection{Installing a New Table}
 After obtaining the identity mapping for \textsf{entry}, we are able to load the \textsf{entry\_val} into \textsf{rdi}, and check the presence bit through
\ifPLDI
Line \ref{line:mask_present} % in condensed version, all on same line
\else
Lines \ref{line:mask_present}--\ref{line:check_entry_present} 
\fi
in Figure \ref{fig:calltopteinitialize}.
Accessing the presence bit and checking the value allows us to exploit the condition (*) that was just discussed when verifying the allocation
path (i.e., when the entry is invalid  and Lines \ref{line:alloc_path_start}--\ref{line:alloc_path_end} in Figure \ref{fig:calltopteinitialize}
must allocate the next level of tables).
This operation is subtle. To reiterate: the operation requires that the relevant table entry is readable, but the exact portion of ownership 
returned must be determined by inspecting the valid bit of the value in memory --- so full ownership is returned only for unused entries.
When the bit is not set, that entails full ownership of the entry's memory ($\textsf{qfrac} = 1$) and justifies writing to that memory.
Otherwise, the code jumps past the end of this listing, to the following code at the top of Figure \ref{fig:p2v} (which is also the
continuation of this code).

\begin{figure}\footnotesize
  \begin{lstlisting}[mathescape,escapeinside={(*}{*)}]
$\specline{\textsf{P} \ast  I\texttt{ASpace}_{\textsf{id}}(\theta,\Xi,m)   \ast \texttt{rbp-16} \mapsto_{\textsf{r}} \textsf{\_} \ast \texttt{rdi}  \mapsto_{\textsf{r}} \textsf{entry+KERNBASE}  \ast \texttt{rax} \mapsto_{\textsf{r}} \textsf{\_} \ast \lnot(\textsf{entry\_present entry\_val} }_{\rtv}$
$\specline{\textsf{entry+KERNBASE} \mapsto_{\textsf{vpte,1}} \textsf{entry entryv} \ast \ghostmaptoken{\textsf{id}}{(\mathsf{entry})}{\textsf{level}} }$
;;void pte_initialize(pte_t *entry) { ... Stack frame setup elided    
callq  81 <kalloc> ;;allocate a full zeroed page for 512 8-byte entries(*\label{line:call_to_kalloc}*)
mov    rax,-0x10[rbp] ;; Store into 'local'
;;entry->pfn = PTE_ADDR_TO_PFN((uintptr_t) local);
mov    -0x10[rbp],rax; mov    -0x8[rbp],rdi; mov    rax,[rdi]
$\specline{\textsf{P} \ast  I\texttt{ASpace}_{\textsf{id}}(\theta,\Xi,m)   \ast \texttt{rbp-16} \mapsto_{\textsf{r}} \textsf{\_} \ast \texttt{rdi}  \mapsto_{\textsf{r}} \textsf{entry+KERNBASE} \ast \texttt{rax} \mapsto_{\textsf{r}} \textsf{nextpaddr} \ast \lnot(\textsf{entry\_present entry\_val}) }$
$\specline{ \textsf{entry+KERNBASE} \mapsto_{\mathsf{vpte,1}} \mathsf{entry} \; \textsf{pfn\_set(entryv nextpaddr)}}$
$\specline{\ghostmaptoken{\delta{}s}{\rtv}{\delta} \ast \begin{array}{l} \ulcorner\textsf{entry\_present (pte\_initialize(pfn\_set(entry\_val nextpaddr)),level)}\urcorner \wand \\  \;\;\;\;\;\;\; \forall_{i\in\textsf{0..511}} \; \textsf{table\_root (pte\_initialized (pfn\_set (entry\_val nextpaddr))) + i * 8} \mapsto_{\textsf{id}} \textsf{level-1}    \end{array}  }_{\rtv}$
... ;;clean up the stack, return
\end{lstlisting}
\vspace{-1em}
\caption{Allocating a physical page }
\label{pteinitializespec}
\vspace{-1em}
\end{figure}

If the entry is not set, \textsf{pte\_initialize} (Line \ref{line:call_to_pte_initialize} in Figure \ref{fig:calltopteinitialize}) 
allocates a physical page (internally utilizing the only unverified (trusted) code in our case studies, the page-allocator's \textsf{kalloc},\footnote{
  This is an allocator for regions of pre-zeroed physical memory that is mapped, but not accessed by the allocator itself,
  as is typical for slab allocators~\cite{bonwick1994slab}.
  Its verification would be similar to verifying a usermode \textsf{malloc} verifications~\cite{Chlipala2013Bedrock,wickerson2010explicit},
  just with additional invariants on the memory pool.
} 
on Line \ref{line:call_to_kalloc} in Figure \ref{pteinitializespec}). 
Since we are using \textsf{pte\_initialize} for page-table address allocation, we must relate this newly
allocated physical address to the identity mapping map $\Xi$ --- 
see Line \ref{line:page_of_caps} in Figure \ref{fig:calltopteinitialize}, where
\texttt{kalloc}'s specification guarantees it has returned memory from a designated memory
pool that is already mapped
\ifPLDI
\else
\footnote{A reasonable reader might wonder where this pool
initially comes from, and how it might grow when needed. Typically an initial mapping subject to this identity mapping
constraint is set up prior to transition to 64-bit kernel code (notably,
a page table must exist \emph{before} virtual memory is enabled during boot, as part of enabling it is setting
a page table root).
Growing this pool later requires cooperation of physical memory range allocation and virtual memory range allocation,
typically by starting general virtual address allocation at the highest physical memory address plus the identity mapping offset.
This reserves the virtual addresses corresponding to all physical addresses plus the offset for later use in this pool,
as needed.
} 
\fi
and satisfies the offset invariants.
% \todo[inline,color=blue]{colin frontier.
% Stuck with line 31 onwards in Figure 7. rax holds nextpaddr, but I think that should be entrypfn, and 
% the explicit entrypfn id token assertion should go away, as its covered by the forall assertion.
% then the postcondition for pte-initialize should have a specific level now for the entries,
% like 0, which can be updated in the view shift on line 42.
% }
% Focusing on the specification of \textsf{pte\_initialize} separately in Figure \ref{fig:pteinitializespec}, 
% we right immediately realize that instead of seeing see a physical pointsto for the fresly page-table address 
% (e.g. $\mathsf{nextpaddr} \mapsto_{\mathsf{p}} \mathsf{w64\_0}$) deliberately in the post-conditoin in Lines 15-16,
%  we observe a full-ownership token representing the knowledge that a frame and all the entries indexed from this 
% frame are freshly allocated with full-ownership to be a part of the identity map, $\Xi$. 
The soundness argument of this specification relies on the fact that these freshly allocated resources are part 
of an entry construction that has not been completed yet: the presence bit is set 
(Line \ref{line:install_new_entry} in Figure \ref{fig:calltopteinitialize}) after these freshly allocated resources are incorporated to the 
entry construction via the page-frame portion of the PTE. In other words, the side condition, (*),
 formalizes that any access to the entry with these resources is \textit{invalid} (in the sense of not necessarily
having accompanying resources) until the entry is marked present (and thus the memory returned from \textsf{kalloc}
moves into the page table invariant.

\add{Note that the C presentation in Figure \ref{fig:calltopteinitializeC}
omitted the precondition on the implication of Figure \ref{fig:calltopteinitialize}'s Line \ref{line:page_of_caps},
which is logically equivalent to \textsf{True} since \textsf{entry\_present} checks if the present bit is set in an entry,
and \textsf{pte\_initialize} sets that bit. The actual invariant has this form here, and in the postcondition
of \lstinline|pte_initialize| (Figure \ref{pteinitializespec}), to match the conditional form from earlier in
\lstinline|pte_get_next_table| (which is also provably true when the check of the present bit
determines that the entry was already valid/present).
Our proof discharges the conditional at the join point, rather than eagerly in each branch.
}

\subsubsection{Physical-to-Virtual Conversion with \textsf{P2V}}
\label{sec:p2v}
Once we know the entry refers to a physical address in the identity mapping range ($\Xi$)
(via the branch at Line \ref{line:check_entry_present_jump}, or  by allocating and installing a new entry
as just discussed for Lines \ref{line:check_entry_present_jump}--\ref{line:end_of_allocation_path}), 
we can convert this frame address to a corresponding virtual address via the identity mappings
discussed in Section \ref{subsec:identitymappings} and Figure \ref{fig:peraspaceinvariant_with_p2v_extension}.
in the last lines of \lstinline|pte_get_next_table| shown in Figure \ref{fig:p2v} (the continuation of Figure \ref{fig:calltopteinitialize}).
This is a critical piece of the full page table walk verification.
In our small kernel (Line \ref{line:p2v} in Figure \ref{fig:p2v}), as in larger kernels, the C macro \texttt{P2V} common to many kernels
is actually just addition by the constant offset mentioned in Section \ref{subsec:identitymappings}.
But the correctness of this simple instruction is quite subtle.
%  and cannot be proven 
% without the extended invariant (Figure \ref{fig:peraspaceinvariant_with_p2v_extension})
% worked out Section \ref{subsec:identitymappings}.

\begin{figure}\footnotesize
\begin{lstlisting}[mathescape,escapeinside={(*}{*)}]
... ;; Continued from Figure (*\ref{fig:calltopteinitialize}*); assertions below specialized to non-allocating path for clarity
$\specline{\textsf{P} \ast  I\texttt{ASpace}_{\textsf{id}}(\theta,\Xi\setminus \{ \mathsf{entry} \}),m)  \ast \textsf{rbp-8} \mapsto_{\textsf{v}} \textsf{entry} \ast \texttt{rcx}  \mapsto_{\textsf{r}} \textsf{\_}  \ast \textsf{entry} \mapsto_{\textsf{id}} \textsf{\_} \ast \ghostmaptoken{\delta{}s}{\rtv}{\delta}  }_{\rtv}$
$\specline{ \textsf{entry+KERNBASE} \mapsto_{\textsf{vpte,qfrac}} \textsf{(pte\_initialized (entry\_val.pfn))} \urcorner }_{\rtv}$
$\specline{ \texttt{rbp-16} \mapsto_{\textsf{v}} \textsf{(pte\_initialized (entry\_val.pfn)))} \ast \texttt{rax} \mapsto_{\textsf{r}} \textsf{ table\_root (pte\_initialize(entry\_val.pfn))} }_{\rtv}$
$\specline{\forall_{i\in \textsf{0 ... 511} } \ldotp  \ghostmaptoken{\textsf{id}}{((\textsf{table\_root (pte\_initialized (entry\_val.pfn)))}) + \textsf{i * 8})}{\textsf{v-1}}  }$ (*\label{line:children}*)
;;uintptr_t next_virt_addr = (uintptr_t) P2V(entry.pfn<<12); (*\label{line:p2v}*) 
movabs KERNBASE,rcx $\specline{\ldots \ast  \texttt{rcx}  \mapsto_{\textsf{r}} \textsf{KERNBASE}  \ast \ldots}_{\rtv}$
add    rcx,rax $\specline{ \ldots \ast \texttt{rax} \mapsto_{\textsf{r}} \textsf{ table\_root (pte\_initialize(entry\_val.pfn)) + KERNBASE}  \ast \ldots}_{\rtv}$
... ;;clean up the stack and return the rax value
\end{lstlisting}
\vspace{-1em}
\caption{Converting a physical address of a PTE to a virtual address (w/o instruction pointer or flag updates).
 % Abreviating the other relevant resources (including the lower entries) peeled-out of the invariant as $\textsf{R}_{\textsf{children}}$
\vspace{-1em}
}
\label{fig:p2v}
\end{figure}
Figure \ref{fig:p2v} shows the verification of the end of \lstinline|pte_get_next_table| specialized to the case where 
where no allocation was necessary (i.e., the conditional on Line \ref{line:check_entry_present} of Figure \ref{fig:calltopteinitialize} was taken).
In this case, the true present bit allows access to the child tokens from Line \ref{line:conditional_children} of Figure \ref{fig:calltopteinitialize},
which is then refined to the assertion on Line \ref{line:children} of Figure \ref{fig:p2v}.
The code loads \lstinline|rcx| with the offset value \textsf{KERNBASE}, which gives us the value of the virtual address ($\textsf{entry}_{\textsf{pfn}}$ \textsf{+KERNBASE})
of the \emph{base} of the next level of the page table.
% \todo[inline]{the next sentence depends on having figure 10 updated to reflect the page-worth of tokens}
While we could now convert this address to a virtual points-to, this is not necessarily the correct thing to do.
The caller \lstinline|walkpgdir| (discussed next) uses \lstinline|pte_get_next_table| to retrieve just the base address,
because only the caller knows which entry in the subsequent table will be accessed (it depends on the corresponding bits from the virtual
address being translated). So instead we pass back the per-address-space invariant with the identity mapping resources for \lstinline|entry|
pulled out. The caller determines which entry in that table must actually
be accessed --- by selecting the appropriate index into the 512 ghost map tokens returned in the postcondition,
and using the ghost translation and physical location portions of the invariant to assemble a vpte-pointsto
that justifies the caller's subsequent access to a particular entry of the returned table.
% in the identity map ($\Xi\setminus\{entry\}$) of the kernel invariant.
% the logical update in Specification  Lines 5-10 to 10-14 for obtaining virtual-pointsto resource for the frame 
% ($\textsf{entry}_{\textsf{pfn}}$) by removing it from the ghost map ($\Xi\setminus\{entry\}\cup \{\textsf{entry}_{\textsf{pfn}}) \}$) 
% in Line 5 and compute the identity mapping for this physical frame address in Line 13 in Figure \ref{fig:p2v}).

\subsubsection{Walking Page-Table Tree: Calling \textsf{pte\_get\_next\_table} for Each Level}
\label{wlkpgdir}
\begin{figure}\footnotesize
\ifPLDI
\begin{lstlisting}[mathescape,escapeinside={(*}{*)}]
;;pte_t *walkpgdir(pte_t *pml4, const void *va) {
... ;; Stack setup
$\specline{\textsf{P} \ast \ulcorner\textsf{rtv+KERNBASE}=\textsf{pml4}\urcorner \ast \forall_{i\in\textsf{0..511}}\ghostmaptoken{\textsf{id}}{(\textsf{rtv}+i*8)}{\textsf{4}} \ast  I\texttt{ASpace}_{\textsf{id}}(\theta,\Xi,m) \ast \ghostmaptoken{\delta{}s}{\rtv}{\delta}}_{\rtv}$
;;pte_t *pml4_entry = &pml4[PML4EX(va)]; // Virtual address of L4 entry(*\label{line:start_pml4_calc}*)
mov    -0x8[rbp],rsi; mov    -0x10[rbp],rdi; ;; Load pml4 and va
;; Extract 9-bit (0x1ff=511) L4 index from va, multiply by 8 (byte index), add to pml4 for entry address
shr    0x27,rdi; and    0x1ff,rdi; shl    0x3,rdi; add    rdi,rsi(*\label{line:end_pml4_calc}*)
mov    rsi,-0x18[rbp] ;; Store virt. addr. to local var pml4_entry; logical pml4_entry is physical!
$\specline{\textsf{P} \ast  I\texttt{ASpace}_{\textsf{id}}(\theta,\Xi\setminus\{\textsf{pml4\_entry}\},m) \ast \forall_{i\in\textsf{0..511}\setminus\{\textsf{PML4EX(va)}\}}\ghostmaptoken{\textsf{id}}{(\textsf{rtv}+i*8)}{\textsf{4}}\ast \ghostmaptoken{\delta{}s}{\rtv}{\delta}  }_{\rtv}$
$\specline{\textsf{entry\_present(l4e\_val)} \wand \forall_{i\in\textsf{0..511}} \; \ghostmaptoken{\textsf{id}}{\textsf{table\_root(l4e\_val)+ i * 8}}{\textsf{3}} }_{\rtv} $
$\specline{ \textsf{pml4\_entry+KERNBASE} \mapsto_{\textsf{vpte,qfrac}} \textsf{pml4\_entry l4e\_val} \ast \ulcorner  \textsf{qfrac} = 1 \leftrightarrow \; \lnot(\textsf{entry\_present l4e\_val})\urcorner}_{\rtv}$    
;;pte_t *pdp = pte_get_next_table(pml4_entry);
mov    -0x18[rbp],rdi; ...; callq  c0 <pte_get_next_table>(*\label{line:first_getnext_call}*)
;;save the physical next table address in rax
$\specline{\textsf{P} \ast  I\texttt{ASpace}_{\textsf{id}}(\theta,\Xi\setminus\{\textsf{pml4\_entry}\},m)  \ast \ghostmaptoken{\delta{}s}{\rtv}{\delta}  }_{\rtv}$
$\specline{\forall_{i\in\textsf{0..511}} \ghostmaptoken{\textsf{id}}{\textsf{table\_root(l4e\_val'.pfn) + i * 8}}{\textsf{3}}  \ast \ulcorner\textsf{pdp-KERNBASE}=\textsf{table\_root(l4e\_val'.pfn)}\urcorner} $
$\specline{ \textsf{pml4\_entry+KERNBASE} \mapsto_{\textsf{vpte,qfrac}} \textsf{pml4\_entry l4e\_val'} \ast \ulcorner  \textsf{qfrac} = 1 \leftrightarrow \; \lnot(\textsf{entry\_present l4e\_val})\urcorner}_{\rtv}$(*\label{line:first_pte_pointsto}*)
$\specline{\texttt{rax}  \mapsto_{\textsf{r}} \textsf{table\_root(l4e\_val'.pfn)} }_{\rtv}$ ;; pte_get_next_table may have allocated a new page, updating entry
;;pte_t *pdp_entry = &pdp[PDPEX(va)]; // Virtual address of L3 entry
$\specline{\ulcorner\textsf{pdp}+\textsf{PDPEX(va)*8} = \textsf{pdp\_entry} \land \textsf{table\_root(l4e\_val'.pfn)}  = \textsf{pdp} \urcorner }_{\rtv}$
$\specline{\textsf{P} \ast  I\texttt{ASpace}_{\textsf{id}}(\theta,\Xi\setminus\{\textsf{pml4\_entry},\textsf{pdp\_entry}\},m)  \ast \ghostmaptoken{\delta{}s}{\rtv}{\delta}  }_{\rtv}$
$\specline{\forall_{i\in\textsf{0..511}\setminus\{\textsf{PDPEX(va)}\}} \; \textsf{table\_root(pml4\_entry}_{\textsf{pfn}}) \textsf{ + i * 8} \mapsto_{\textsf{id}} \textsf{3} }_{\rtv}$
$\specline{ \textsf{pml4\_entry+KERNBASE} \mapsto_{\textsf{vpte,qfrac4}}\textsf{pml4\_entry}\; \textsf{l4e\_val'} \ast \ulcorner  \textsf{qfrac4} = 1 \leftrightarrow \; \lnot(\textsf{entry\_present l4e\_val'})\urcorner}_{\rtv}$(*\label{line:ex_l4_vpte}*)
$\specline{ \textsf{pdp\_entry+KERNBASE} \mapsto_{\textsf{vpte,qfrac3}}\textsf{pdp\_entry}\; \textsf{l3e\_val} \ast \ulcorner  \textsf{qfrac3} = 1 \leftrightarrow \; \lnot(\textsf{entry\_present l3e\_val})\urcorner}_{\rtv}$(*\label{line:ex_l3_vpte}*)
$\specline{\textsf{entry\_present(l3e\_val)} \wand \forall_{i\in\textsf{0..511}} \;\ghostmaptoken{\textsf{id}}{\textsf{table\_root(pdp\_entry.pfn) + i * 8}}{ \textsf{2}} } _{\rtv} $
... ;; Similar assembly & proofs for levels 3 and 2
... ;;return &pt[PTEX(va)]; // return address of L1 entry, similar index computation
 $\specline{\textsf{R}_{\textsf{walk}} \ast \textsf{R}_{\textsf{l1e}} }_{\rtv}$
\end{lstlisting}

\else
\todo[inline]{this is dead code, we're using the PLDI flag set to true}
\fi
\vspace{-1em}
\caption{Walking page-table directory via calls to \textsf{pte\_get\_next\_table} in Figure \ref{fig:calltopteinitialize}}
\label{walkpgdir}
\vspace{-1em}
\end{figure}

\begin{figure}\footnotesize
  \vspace{-1em}
$\begin{array}{l}\left( \begin{array}{l} \ulcorner \textsf{pdp}+\textsf{PDPEX(va)} = \textsf{pdp\_entry} \land \textsf{table\_root(pml4\_entry}_{\textsf{pfn}})  = \textsf{pdp} \\ \land \textsf{pd}+\textsf{PDEX(va)} = \textsf{pd\_entry} \land \textsf{table\_root(pml3\_entry}_{\textsf{pfn}})  = \textsf{pd} \\ \land \textsf{pt}+\textsf{PDPEX(va)} = \textsf{pt\_entry} \land \textsf{table\_root(pml2\_entry}_{\textsf{pfn}})  = \textsf{pt} \urcorner  \ast \\  \textsf{P} \ast  I\texttt{ASpace}_{\textsf{id}}(\theta,\Xi\setminus\{\textsf{pml4\_entry},\textsf{pdp\_entry},\textsf{pd\_entry},\textsf{pt\_entry}\},m)  \\ \ghostmaptoken{\delta{}s}{\rtv}{\delta} \ast  
      \ghostmaptoken{\textsf{id}}{(\mathsf{pml4\_entry})}{\textsf{4}}\ast \ghostmaptoken{\textsf{id}}{(\mathsf{plm3\_entry})}{\textsf{3}} \ast \ghostmaptoken{\textsf{id}}{(\mathsf{plm2\_entry})}{\textsf{2}}  \ast \\  \textsf{pml4\_entry+KERNBASE} \mapsto_{\textsf{vpte,qfrac}} \textsf{pml4\_entry} \; \textsf{l4e\_val}  \ast \\  \textsf{pdp\_entry+KERNBASE} \mapsto_{\textsf{vpte,qfrac}} \textsf{pdp\_entry}\;\textsf{l3e\_val}  \ast \\  \textsf{pd\_entry+KERNBASE} \mapsto_{\textsf{vpte,qfrac}} \textsf{pd\_entry} \; \textsf{l2e\_val}  \end{array} \right)  =\textsf{R}_{\textsf{walk} } 
    \\ \ast \left(\begin{array}{l} \textsf{rax} \mapsto \textsf{pt\_entry+KERNBASE} \ast  \textsf{pt\_entry+KERNBASE} \mapsto_{\textsf{vpte,qfrac}} \textsf{pt\_entry}\;\textsf{l1e\_val} \ast \\ \ulcorner  \textsf{qfrac} = 1 \leftrightarrow \; \lnot(\textsf{entry\_present l1e\_val})\urcorner \end{array}\right) =\textsf{R}_{\textsf{l1e}}\end{array} $
\caption{\textsf{R}$_{\textsf{walk}}$: Resources obtained from invariant for walking page-table directory with \emph{non-existing entries} via calls to \textsf{l1e\_get\_next\_table} in Figure \ref{fig:calltopteinitialize}}
\label{fig:rwalk}
\vspace{-1em}
\end{figure}
Implementing a software page-table walk amounts to calling \textsf{pte\_get\_next\_table} for each level as shown in Figure \ref{walkpgdir}. 
The key part of the specification and proof for a page table walk is accumulation of memory mappings for the page-table entries 
visited and frame addresses for page-tables. 
For example, Lines \ref{line:ex_l4_vpte} and \ref{line:ex_l3_vpte} in Figure \ref{walkpgdir} show the virtual pte-pointsto assertions for L4 and L3 entries.
In the final post-condition, we expect the accumulation of these resources from each level -- $\textsf{R}_{\textsf{walk}}$ -- 
which allows us to construct and return the path to the L1 entry in the tree to insert a new page.  

This is the code which performs most actual physical-to-virtual conversions using the identity mapping portion of the per-address-space invariant.
\lstinline|walkpgdir| accepts a \emph{virtual} pointer to the base of the L4 table, and the address to translate.
The precondition provides knowledge that the virtual base of the L4 is at the appropriate offset from the current \lstinline|cr3| value,
but does not provide a virtual points-to assertion --- because the function must calculate (Lines \ref{line:start_pml4_calc}--\ref{line:end_pml4_calc})
which entry it needs access to.
Instead the precondition has 512 identity map tokens, guaranteeing that every entry on the page is subject to the identity mapping invariant.
Line \ref{line:end_pml4_calc} calculates the virtual address of the relevant entry, and the subsequent view shift
pulls that entry out of the identity mapping ($\Xi$) and fetches its corresponding resources as
described by Figure \ref{fig:peraspaceinvariant_with_p2v_extension} and Section \ref{subsec:identitymappings}.
The ghost translation and physical location are used to form the virtual pte-pointsto for the L4 entry
(Line \ref{line:first_pte_pointsto}), with the entry validity and next-level indexing
satisfying the rest of the precondition for \lstinline|pte_get_next_table|.
\lstinline|pte_get_next_table| then, as described earlier, checks the valid bit in the indicated
entry and either returns the (unconditional) tokens for the L3 entry physical addresses (if valid), or
allocates into the entry and returns new (also unconditional) tokens for the L3 entry physical addresses.
\lstinline|pte_get_next_table|'s first call (Line \ref{line:first_getnext_call}) returns
the virtual address of the base of the L3 table (a \emph{page directory pointer}, so PDP, in official
x86-64 terminology). Then the situation to move from that pointer to the base of the L2
is just like the process just followed: the proof calculates the address of the relevant
L3 entry, uses the appropriate L3 identity mapping token to construct a virtual pte-pointsto to that entry,
and passes that along with additional resources pulled out of the invariant to another call to
\lstinline|pte_get_next_table|. That call then returns the base of an L2 table, and the process
repeats until the function returns the virtual address of the relevant L1 entry.
That will then be used in the next section by the caller of \lstinline|walkpgdir|
to install a new mapping.

\subsection{Mapping a New Page}
\label{sec:mapnew}
One of the key tasks of a page fault handler in a general-purpose OS kernel is
to map new pages into an address space by writing into an existing page table via a call\\
\centerline{\textsf{vaspace\_mappage(pte\_t *pml4, void *va,uintptr\_t fpaddr)}}\\
in Figure \ref{fig:mapping_code}.
To do so, with a given allocated a fresh page (\textsf{fpaddr}), then calculate the appropriate
known-valid page table walks (via \textsf{walkpgdir} Line \ref{line:call_walkpgdir} in Figure \ref{fig:mapping_code})  and update 
the appropriate L1 page table entry (Line 35 in Figure \ref{fig:mapping_code});
unmapping is the reverse of the logic we discuss here.
\looseness=-1
%\lstset{
%  columns=fullflexible,
%  numbers=left,
%  basicstyle=\ttfamily,
%  keywordstyle=\color{blue}\bfseries,
%  morekeywords={mov,add,call},
%  emph={rsp,rdx,rax,rbx,rbp,rsi,rdi,rcx,r8,r9,r10,r11,r12,r13,r14,r15},
%  emphstyle=\color{green},
%  emph={[2]cr3},
%  emphstyle={[2]\color{violet}},
%  morecomment=[l]{;;},
%  mathescape
%}
\begin{figure}\footnotesize
  \begin{lstlisting}[mathescape,escapeinside={(*}{*)}]
;;complstatus_t vaspace_mappage(pte_t *pml4, void *va,uintptr_t fpaddr ) {
... ;;setting up the stack      
 $\specline{\textsf{P} \ast \textsf{pml4} \mapsto_{\textsf{id}} \textsf{\_} \ast  I\texttt{ASpace}_{\textsf{id}}(\theta,\Xi,m) \ast \ghostmaptoken{\delta{}s}{\rtv}{\delta} \ast \ulcorner\theta \; !!\; \vaddr = \texttt{None}\urcorner}_{\rtv}$      
;; pte_t *pteaddr = walkpgdir(pml4, a);
 mov    -0x8[rbp], %rdi
 mov    -0x16[rbp], rsi
 mov    0x0, rax
 callq  291 <walkpgdir> (*\label{line:call_walkpgdir}*)
 $\specline{ \textsf{R}_{\textsf{walk}} \ast \textsf{R}_{\textsf{l1e}} }$
 mov  rax, rdx
 mov    rdx,-0x28[rbp]
;;if (!pteaddr->present){ (*\label{line:mappage_pte_present_start}*)
 mov    -0x28[rbp], rdx
 mov    [rdx], rcx
 and    0x1, rcx
 mov    rcx, r8
 cmp    0x0, r8
 jne    3e0 <vaspace_mappage+0x190> (*\label{line:mappage_pte_present_end}*)
 ... ;; Not present
;;pteaddr->pfn          = PTE_ADDR_TO_PFN(fpaddr);
 mov    -0x18[rbp], rdx
 shr    0xc, rdx
 mov    -0x28[rbp], rsi
 mov    [rsi], rdi ;; Load current entry in rdi
 movabs 0xffffffffff, rcx
 and    rcx, rdx
 shl    0xc, rdx
 movabs 0xfff0000000000fff, rcx
 and    rcx, rdi
 or     rdx, rdi
 mov    rdi, [rsi] ;; Write back updated entry with new PFN (*\label{line:updatepfn}*)
$\specline{ \textsf{pt\_entry+KERNBASE} \mapsto_{\textsf{vpte,1}} \textsf{pt\_entry} \; (\textsf{set\_pfn}(\textsf{l1e\_val}, \textsf{PTE\_ADDR\_TO\_PFN}(\textsf{fpaddr}))) \ast  \ulcorner \lnot(\textsf{entry\_present l1e\_val}) \urcorner \ast \textsf{P}}$
 ;; Store updated entry back to L1 entry (*\label{line:l1entry_store}*)
;;}
... ;;set present bit in entry, then clean up the stack
... ;;and return either unaccessible pte pointsto or one with fresh page (fpaddr)
 $\specline{ \begin{array}{l}\textsf{pt\_entry+KERNBASE} \mapsto_{\textsf{vpte,qfrac}} \textsf{pt\_entry} \; \textsf{l1e\_val} \ast \ulcorner (\textsf{entry\_present l1e\_val}) \urcorner \ast \textsf{P} \\
 \lor \; \;  \textsf{pt\_entry+KERNBASE} \mapsto_{\textsf{vpte,1}} \textsf{pt\_entry} \; (\textsf{set\_pfn}(\textsf{l1e\_val}, \textsf{PTE\_ADDR\_TO\_PFN}(\textsf{fpaddr}))) \ast  \ulcorner \lnot(\textsf{entry\_present l1e\_val}) \urcorner \ast P \end{array} }$
;;}
\end{lstlisting}
\vspace{-1em}
  \caption{Specification of updating L1 entry (\textsf{pt\_entry}) to reference a new page (\textsf{fpaddr}).}
\label{fig:mapping_code}
\end{figure}

In Figure \ref{fig:mapping_code}, we see an address ($\vaddr$) currently not
mapped to a page ($\theta \; !!\; \vaddr = \texttt{None}$). Mapping a fresh
physical page to back the desired virtual page first requires ensuring
the existence of a memory location for an appropriate L1 table entry.
The code uses a helper function \lstinline{walkpgdir} (discussed again in Section \ref{sec:traversing}).
\textsf{walkpgdir}'s postcondition contains virtual \emph{PTE} pointsto assertions ($\mapsto_{\textsf{vpte}}$)
both for ensuring partial page table walk reaching the
L1 entry (l1e) by asserting that higher levels of the page table exist (R$_{\textsf{walk}}$ in Figure \ref{fig:rwalk}), 
and for allowing access to the memory of the L1 entry via virtual address (R$_{\textsf{l1e}}$ in Figure \ref{fig:rwalk}).

% After obtaining a virtual address \textsf{pte\_addr} in \textsf{rax} backed 
% by the physical memory for the L1 entry that will be used to translate the virtual addresses
% we are mapping, we save it to \textsf{r14} to be updated later in Line 9.

%In the precondition, we see Line 12 allocates a fresh page-aligned, zero-initialized page  (at \textsf{fpaddr}),
%returning a pre-filled PTE entry in \textsf{rax} ($+3$ sets the lower 2 bits).

% , to hold the freshly
% allocated physical page address (\textsf{fpaddr}) in Line X.

We already discussed for the upper level page-tables how the entry-present checks are handled.
However, for L1 entries this check is left to the caller of the 
traversal function \textsf{walkpgdir}. In other words, unlike what we see in R$_{\textsf{walk}}$ for the upper levels where all entry-present
checks have already been performed, the specification in R$_{\textsf{l1e}}$ ensures that page table entry for L1 needs to be checked at the caller site. 
By doing so, as we see in Figure \ref{fig:mapping_code}, the page reference \textsf{fpaddr} is linked to back the virtual address \textsf{va} 
only if it is not already referring to a physical resource (Lines \ref{line:mappage_pte_present_start}--\ref{line:mappage_pte_present_end} in Figure \ref{fig:mapping_code}). 

The crucial step in addition to traversing the page table in Figure \ref{walkpgdir} is actually updating the L1 entry (Line \ref{line:updatepfn} in Figure \ref{fig:mapping_code}),
via the virtual address (\textsf{pt\_entry+KERNBASE}) known to translate to the appropriate physical address, in our example the L1
table entry address ($\textsf{PTE\_ADDR\_TO\_PFN(fpaddr)}$).

Unlike the only prior work verifying analogous code for mapping a new page~\cite{kolanski08vstte,kolanski09tphols}, our proof above
does \emph{not} need to reason directly over the operational semantics,
making this the first verification we know of for mapping a virtual memory page that 
stays entirely at the program logic level.
\looseness=-1
% By incorporating verification of the
% \lstinline|ensure_L1| function (see Section \ref{sec:traversing}), our verification also directly handles several subtle aspects which
% were axiomatized in prior work.
\ifPLDI
\else
\subsection{Unmapping a Page}
\todo[inline]{update (esp. line refs) for new mapping code}
The reverse operation, unmapping a designated page that is currently mapped,
would essentially be the reverse of
the reasoning around line 22 above: given the virtual points-to assertions for all 512
machine words of memory that the L1 entry would map,
and information about the physical location, 
full permission on the L1 entry could be obtained, allowing the construction of a
full virtual PTE pointer for it, setting to 0, and reclaiming the now-unmapped physical memory.
\fi

\end{document}